\documentclass[aps,prd,groupedaddress,preprintnumbers,%
              showpacs,showkeys,floatfix]{revtex4-1}       
\usepackage[latin1]{inputenc}                    
\usepackage{graphicx}                            
\usepackage{epsfig}
\usepackage{epsf}
\usepackage{latexsym}                            
\usepackage{amsfonts}                            
\usepackage{amssymb}                             
\usepackage{amsmath}                             
\usepackage[mathscr]{eucal}                      
\usepackage{dcolumn}                             
\usepackage{multirow}
\usepackage{bm}                                  
\usepackage{hyperref}                            
\usepackage[usenames]{color}
\usepackage{array}
\usepackage{appendix}

\graphicspath{{.}{Graphics/}}                    

\newcommand{\be}{\begin{equation}}
\newcommand{\ee}{\end{equation}}
\newcommand{\bea}{\begin{eqnarray}}
\newcommand{\eea}{\end{eqnarray}}
\newcommand{\tendto}{\mathop{\longrightarrow}}

\def\eq#1{Eq.~(\ref{#1})}
\def\fig#1{Fig. \ref{#1}}
\def\tbl#1{Table \ref{#1}}
\def \3{\ss }

\newcommand{\tr}{\operatorname{Tr}}

\newcommand{\beqn}{\begin{eqnarray}}
\newcommand{\eeqn}{\end{eqnarray}}

\newcommand{\idnty}{\hbox{1$\!\!$1}}

\newcommand{\reci}[1]{\frac{1}{#1}}

\newcommand{\eps}{\epsilon_{abc}}
\newcommand{\con}[3]   {\left( {#1}_a^T C\gamma_5     {#2}_b \right) {#3}_c}
\newcommand{\conm}[3]{\left( {#1}_a^T C\gamma_\mu {#2}_b \right) {#3}_c}
\newcommand{\cone}[3]{\eps \left( {#1}_a^T C\gamma_5 {#2}_b \right) {#3}_c}
\newcommand{\conme}[3]{\eps \left({#1}_a^T C\gamma_\mu {#2}_b \right) {#3}_c}

\def\cyp{a}
\def\cyi{b}
\def\nic{c}
\def\odense{d}

\begin{document}

\begin{titlepage}
  {\vspace{-0.5cm} \normalsize
  \hfill \parbox{60mm}{
DESY 14-096\\
}}\\[10mm]
  \begin{center}
    \begin{LARGE}
      \textbf{Baryon spectrum with $N_f=2+1+1$ twisted
       mass fermions} \\
    \end{LARGE}
  \end{center}

 \vspace{.5cm}

 \vspace{-0.8cm}
  \baselineskip 20pt plus 2pt minus 2pt
  \begin{center}
    \textbf{
      C.~Alexandrou$^{(\cyp, \cyi)}$,
      V.~Drach$^{(\nic,\odense)}$,
      K.~Jansen$^{(\nic)}$,
      C. Kallidonis$^{(\cyi)}$, 
      G. Koutsou$^{(\cyi)}$
}
  \end{center}

  \begin{center}
    \begin{footnotesize}
      \noindent 	
 	$^{(\cyp)}$ Department of Physics, University of Cyprus, P.O. Box 20537,
 	1678 Nicosia, Cyprus\\	
 	$^{(\cyi)}$ Computation-based Science and Technology Research Center, The Cyprus Institute, 20 Kavafi Str., Nicosia 2121, Cyprus \\
      $^{(\nic)}$ NIC, DESY, Platanenallee 6, D-15738 Zeuthen, Germany\\
      $^{(\odense)}$ CP$^3$-Origins and the Danish Institute for Advanced Study DIAS, University of Southern Denmark, Campusvej 55, DK-5230~Odense~M, Denmark\\
     \vspace{0.2cm}
    \end{footnotesize}
  \end{center}
  
  \begin{abstract}

The masses of the low lying baryons are evaluated 
using a total of ten ensembles of dynamical twisted mass fermion gauge configurations.  The simulations are performed using two degenerate flavors of light quarks,  and a strange and a charm quark fixed to approximately their physical values.  The light sea quarks correspond to pseudo scalar masses in the range of about  
210~MeV to 430~MeV. We use the Iwasaki improved gluonic action
at three values of the coupling constant corresponding to 
lattice spacing $a=0.094$~fm, 0.082~fm and 0.065~fm determined from the nucleon mass.
We check for both finite volume and cut-off effects on the baryon masses. 
We examine the issue of isospin symmetry breaking for the octet and
decuplet baryons and its dependence on the lattice spacing.
We show that in the continuum limit isospin breaking is consistent with zero, 
as expected.
We performed a  chiral extrapolation of
the forty baryon masses using SU(2) $\chi$PT.
After taking  the continuum limit and
extrapolating to the physical pion mass our results are in good agreement with experiment. We provide predictions for the mass of the doubly charmed $\Xi_{cc}^*$, as well as of the doubly and triply charmed $\Omega$s that have not yet been determined experimentally.

\begin{center}
\today
\end{center}
 \end{abstract}
\maketitle 
\end{titlepage}

\section{Introduction}

Simulations of QCD defined on  four-dimensional Euclidean lattice using near to physical values of the light quark masses are enabling the reliable extraction  of the masses of the low lying hadrons. This progress in lattice QCD coupled with 
the interest in  charmed-baryon spectroscopy, partly triggered
by the first observation of a family of doubly charmed baryons $\Xi^+_{cc}(3519)$ and $\Xi^{++}_{cc}(3460)$ by the SELEX collaboration~\cite{Mattson:2002vu,Russ:2002bw,Ocherashvili:2004hi}, make the study of the charmed hadron masses particularly timely.  The fact that the observation of $\Xi^+_{cc} (3519)$ or $\Xi^{++}_{cc}(3460)$, 
has not be confirmed by  the BABAR~\cite{Aubert:2006qw} nor the 
BELLE~\cite{Chistov:2006zj}  experiments calls for further attention
into the existence of doubly charmed $\Xi$s.
Even more interesting is the mass splitting  of about 60~MeV
  for this doublet as compared to the splitting of other previously observed 
isospin partners that have mass differences one order of magnitude smaller. Theoretical studies using e.g. the non relativistic~\cite{Roberts:2007ni} and relativistic quark
models~\cite{Martynenko:2007je,Ebert:2002ig}, and QCD sum rules~\cite{Wang:2010hs} predict the $\Xi_{cc}$ mass to be 100-200 MeV higher than that observed by SELEX.
Heavy baryon spectra will be further studied experimentally at the recently upgraded Beijing Electron- Positron Collider (BEPCII) detector, the Beijing Spectrometer (BES-III) and  at the anti Proton Annihilation at DArmstadt (PANDA) at FAIR. 
Lattice QCD calculations can provide theoretical input for these experiments.
 A number of lattice QCD studies have recently looked at the
mass of charmed baryons. Most of these studies employ a mixed action approach using staggered sea quarks.   In Ref.~\cite{Briceno:2012wt}  $N_f=2+1+1$ staggered sea quarks with clover light and strange valence quarks  and a relativistic action for the charm quark are employed and the results are  extrapolated to the continuum limit. In Refs.~\cite{Na:2008hz,Liu:2009jc}  $N_f=2+1$ staggered sea quarks are used with  staggered light and strange~\cite{Na:2008hz} or domain wall~\cite{Liu:2009jc}   valence quarks with a relativistic action for the charm quark.

In this work we extend our previous study on the low-lying spectrum of the baryon 
 octet and decuplet using $N_f=2$  twisted mass fermions~\cite{Alexandrou:2009qu} to $N_f=2+1+1$
twisted mass fermions at maximal twist.
For the valence  strange and charm sector we use Osterwalder-Seiler 
 quarks avoiding mixing between these two sectors.
 The strange and charm valence quark masses are tuned 
using the $\Omega^-$ and $\Lambda_c$ baryon mass, respectively.
We analyze a total of ten $N_f=2+1+1$ ensembles at three different lattice spacings and volumes.
This enables us to take the continuum limit and assess volume effects. Our results  are fully compatible with an ${\cal O}(a^2)$ behavior which is used to  extrapolate to the continuum limit.

The good precision of our results on the baryon masses allows us to perform 
a study of chiral extrapolations to obtain results at 
the physical point. This study shows that one of 
the main uncertainties in predicting the mass at the physical point is
 caused by the chiral extrapolations, which yield the largest systematic error.

An important issue is the restoration of the
explicitly broken isospin symmetry in the continuum limit.
 At finite lattice spacing, baryon masses
display $\mathcal{O}(a^2)$ isospin breaking effects. 
There are, however, theoretical arguments \cite{Frezzotti:2007qv} as well as numerical evidence
\cite{Dimopoulos:2008sy,Jansen:2008vs} that these isospin breaking effects are 
particularly pronounced
 for the neutral pseudo scalar mass, whereas for other quantities studied 
so far by the European Twisted Mass Collaboration (ETMC) they are compatible with zero.   
In this paper, we will corroborate this result also in the baryon sector 
showing that isospin 
breaking
effects are in general small or even compatible with zero. 
For a preliminary account of these
results see Ref.~\cite{Alexandrou:2014yha}.

The paper is organized as follows:
The details of our lattice setup, namely those concerning 
the twisted mass action, 
the parameters of the simulations and the interpolating fields used, 
are given in Section~II. 
Section~III contains the numerical results of the baryon masses computed
 for different 
lattice volumes, lattice spacings and bare quark masses.
Lattice artifacts, including finite volume and discretization errors are also discussed  with special 
emphasis on the $\mathcal{O}(a^2)$ isospin breaking effects inherent
 in the twisted mass formulation
of lattice QCD.
The chiral extrapolations are analyzed in Section~IV.
Section~V  contains a comparison with other existing calculations and
conclusions are finally  drawn in Section~VI. 

\section{Lattice techniques}

\subsection{The lattice action}
In the present work we employ the twisted mass
fermion (TMF) action~\cite{Frezzotti:2000nk} and the Iwasaki improved
gauge action~\cite{Weisz:1982zw}. Twisted mass fermions provide an
attractive formulation of lattice QCD that allows for automatic ${\cal
O}(a)$ improvement, infrared regularization of small eigenvalues and
fast dynamical simulations~\cite{Frezzotti:2003ni}.

The twisted mass Wilson action used for the light degenerate doublet of quarks ($u$,$d$) is given by~\cite{Frezzotti:2003ni,Frezzotti:2000nk}
\be
S_F^{(l)}\left[\chi^{(l)},\overline{\chi}^{(l)},U \right]= a^4\sum_x  \overline{\chi}^{(l)}(x)\bigl(D_W[U] + m_{0,l} + i \mu_l \gamma_5\tau^3  \bigr ) \chi^{(l)}(x)
\label{eq:S_tml}
\ee
with $\tau^3$ the third Pauli matrix acting in the flavour space, $m_{0,l}$ the bare untwisted light quark mass, $\mu_l$ the bare twisted light quark mass and the massless Wilson-Dirac operator given by 
\be \label{eq:wilson_term}
D_W[U] = \frac{1}{2} \gamma_{\mu}(\nabla_{\mu} + \nabla_{\mu}^{*})
-\frac{ar}{2} \nabla_{\mu}
\nabla^*_{\mu} 
\ee
where
\be
\nabla_\mu \psi(x)= \frac{1}{a}\biggl[U^\dagger_\mu(x)\psi(x+a\hat{\mu})-\psi(x)\biggr]
\hspace*{0.5cm} {\rm and}\hspace*{0.5cm} 
\nabla^*_{\mu}\psi(x)=-\frac{1}{a}\biggl[U_{\mu}(x-a\hat{\mu})\psi(x-a\hat{\mu})-\psi(x)\biggr]
\quad .
\ee
The quark fields denoted by $\chi^{(l)}$ in \eq{eq:S_tml} are in the so-called ``twisted basis". The fields in the ``physical basis", $\psi^{(l)}$, are obtained for maximal twist by the simple transformation

\be
\psi^{(l)}(x)=\reci{\sqrt{2}}\left(\idnty+ i \tau^3\gamma_5\right) \chi^{(l)}(x),\qquad
\overline{\psi}^{(l)}(x)=\overline{\chi}^{(l)}(x) \reci{\sqrt{2}}\left(\idnty + i \tau^3\gamma_5\right)
\quad.
\ee

In addition to the light sector, a twisted heavy mass-split doublet $\chi^{(h)} = \left(\chi_c,\chi_s \right)$ for the strange and charm quarks is introduced, described by the action \cite{Frezzotti:2004wz,Frezzotti:2003xj}
\be
S_F^{(h)}\left[\chi^{(h)},\overline{\chi}^{(h)},U \right]= a^4\sum_x  \overline{\chi}^{(h)}(x)\bigl(D_W[U] + m_{0,h} + i\mu_\sigma \gamma_5\tau^1 + \tau^3\mu_\delta  \bigr ) \chi^{(h)}(x)
\label{eq:S_tmh}
\ee
where $m_{0,h}$ is the bare untwisted quark mass for the heavy doublet, $\mu_\sigma$ is the bare twisted mass along the $\tau^1$ direction and $\mu_\delta$ is the mass splitting in the $\tau^3$ direction. The quark fields for the heavy quarks in the physical basis are obtained from the twisted basis through the transformation

\be
\psi^{(h)}(x)=\reci{\sqrt{2}}\left(\idnty+ i \tau^1\gamma_5\right) \chi^{(h)}(x),\qquad
\overline{\psi}^{(h)}(x)=\overline{\chi}^{(h)}(x) \reci{\sqrt{2}}\left(\idnty + i \tau^1\gamma_5\right)
\quad.
\ee 

In this paper, unless otherwise stated, the quark fields will be understood as ``physical fields", $\psi$, in particular when we define the baryonic interpolating fields.

The form of the fermionic action in \eq{eq:S_tml} breaks parity and isospin at non-vanishing lattice spacing. In particular, the isospin breaking in physical observables is a cut-off effect of ${\cal O}(a^2)$~\cite{Frezzotti:2003ni}. 

Maximally twisted Wilson quarks are obtained by setting the untwisted quark mass $m_0$ to its critical value $m_{\rm cr}$, while the twisted quark mass parameter $\mu$ is kept non-vanishing in order to work away from the chiral limit. A crucial advantage of the twisted mass formulation is
the fact that, by tuning the bare untwisted quark mass $m_0$ to its critical value
 $m_{\rm cr}$, all physical observables are automatically 
${\cal O}(a)$ improved \cite{Frezzotti:2003ni,Frezzotti:2003xj}. 
In practice, we implement
maximal twist of Wilson quarks by tuning to zero the bare untwisted current
quark mass, commonly called PCAC (Partially Conserved Axial Current) mass, $m_{\rm PCAC}$ \cite{Boucaud:2008xu,Frezzotti:2005gi}, which is proportional to
$m_0 - m_{\rm cr}$ up to ${\cal O}(a)$ corrections. A convenient way to evaluate $m_{\rm PCAC}$ is through

\be \label{eq:m_pcac}
m_{\rm PCAC} = \lim_{t/a \gg 1} \frac{\sum_{\bf x} \langle \partial_4\tilde{A}_4^b({\bf x},t)\tilde{P}^b(0)\rangle}{\sum_{\bf x} \langle \tilde{P}^b({\bf x},t)\tilde{P}^b(0) \rangle} \qquad b=1,2 \qquad ,
\ee
where $\tilde{A}_\mu^b=\overline{\chi}\gamma_\mu\gamma_5 \frac{\tau^b}{2}\chi$ is the axial vector current and $\tilde{P}^b=\overline{\chi}\gamma_5\frac{\tau^b}{2}\chi$ is the pseudoscalar density in the twisted basis. The large $t/a$ limit is required in order to isolate the contribution of the lowest-lying charged pseudoscalar meson state in the correlators of \eq{eq:m_pcac}. This way of determining $m_{\rm PCAC}$ is equivalent to imposing on the lattice the validity of the axial Ward identity $\partial_\mu\tilde{A}_\mu^b = 2 m_{\rm PCAC} \tilde{P}^b$, $b=1,2$, between the vacuum and the charged zero three-momentum one-pion state. When $m_0$ is taken such that $m_{\rm PCAC}$ vanishes, this Ward identity expresses isospin conservation, as it becomes clear by rewriting it in the physical quark basis. The value of $m_{\rm cr}$ is determined at each $\mu_l$ in our $N_f=2+1+1$ simulations, a procedure that preserves ${\cal O}(a)$ improvement
and keeps ${\cal O}(a^2)$ small~\cite{Boucaud:2008xu,Frezzotti:2005gi}.
The reader can find more details on the twisted mass fermion action in Ref.~\cite{Baron:2010bv}.
Simulating a charm quark may give rise to concerns regarding cut-off effects. An analysis presented in Ref~\cite{Athenodorou:2011zp}
shows that they are surprising small.  
In this work we investigate in detail the 
cut-off effects on the hyperon and charmed baryon masses
using  simulations at our three values of the lattice spacings.
All final results are extrapolated to the continuum limit.

In order to avoid complications due to flavor mixing in the heavy quark sector
we only use Osterwalder-Seiler valence strange and charm quarks.
Since the bare heavy quark masses in the sea were approximately tuned to the mass of the kaon and D-meson, in order to match their masses exactly  tuning would have been required even if we used
twisted mass quarks for the strange and the charm. Since our interest in this work is the baryon spectrum we
choose to use the physical mass of the $\Omega^-$ and the $\Lambda_c$ in order to tune the Osterwalder-Seiler  strange and charm quark masses. This means
that we need to choose a value of strange (charm) quark mass, perform the
computation at several values of the pion mass and then chiral extrapolate the
$\Omega^-$ ($\Lambda_c$) mass and compare with its experimental value.
If our chirally extrapolated results do not reproduce the right mass we
change the strange (charm) quark mass and iterate until we reach agreement
with the experimental value.  
Osterwalder-Seiler fermions are doublets
with $r=\pm1$ like the the {\it u-} and {\it d-} doublet, i.e.
  $\chi^{(s)} = \left(s^+,s^- \right)$ and $\chi^{(c)} = \left(c^+,c^- \right)$, having an action that is the same as for the doublet of light quarks, as given in \eq{eq:S_tml}, but with $\mu_l$ in \eq{eq:S_tml}  replaced with the tuned value of the bare twisted mass of the strange (charm) valence quark. Taking $m_{0}$ to be equal to the critical mass determined in the light sector the $\mathcal{O}(a)$ improvement in any observable still applies. One can equally work with  the upper or the lower component of the strange and charm doublets. In the continuum limit both choices are equivalent. In this work we choose to work with the upper components, namely the $s^+$ and $c^+$. 
  The action for the heavy quarks would then read
\be
S_{OS}^{(h)}\left[\chi^{(h)},\overline{\chi}^{(h)},U \right]= a^4\sum_x \sum_{h=s}^c  \overline{\chi}^{(h)}(x)\bigl(D_W[U] + m_{\rm cr} + i \mu_h \gamma_5  \bigr ) \chi^{(h)}(x)
\label{eq:S_tmh2}
\ee

 The reader interested in the advantage of this mixed action in the mesonic sector is referred to the Refs~\cite{Blossier:2007vv,Blossier:2009bx,Frezzotti:2004wz,AbdelRehim:2006ra,AbdelRehim:2006ve}. We give more details on
 the tuning of the strange and charm quark masses in subsection F.


\subsection{Simulation details}

We summarize the input parameters of the calculations, namely $\beta$, $L/a$, the light quark mass $a\mu$ as well as the value of the pion mass in Table~\ref{Table:params}. A total of ten gauge ensembles at three values of $\beta$ are considered, namely $\beta=1.90$, $\beta=1.95$ and $\beta=2.10$, allowing for an investigation of finite lattice spacing effects and for taking the  continuum limit. The values of the lattice spacings $a$ given in Table \ref{Table:params} are determined using the nucleon mass as explained in subsection E. The pion masses for the simulations span a range from about 210~MeV to 430~MeV, which is close enough to the physical point mass to allow us to perform chiral extrapolations. 
\begin{table}[h]
\begin{center}
\renewcommand{\arraystretch}{1.2}
\renewcommand{\tabcolsep}{5.5pt}
\begin{tabular}{c|lcccc}
\hline\hline
\multicolumn{6}{c}{ $\beta=1.90$, $a=0.0936(13)$~fm  ${r_0/a}=5.231(38)$}\\
\hline
\multirow{4}{*}{$32^3\times 64$, $L=3.0$~fm}  &$a\mu$        & 0.0030  & 0.0040 &  0.0050 &    \\
                                              & No. of Confs        &  200    &   200  &    200  &  \\
                              			     	  &$m_\pi$~(GeV) & 0.261  & 0.298 &  0.332 &  \\
                        					        &$m_\pi L$     &   3.97  &   4.53 &    5.05 &    \\
\hline\hline
\multicolumn{6}{c}{ $\beta=1.95$,  $a=0.0823(10)$~fm, ${r_0/a}=5.710(41)$ }\\
\hline
\multirow{4}{*}{$32^3\times 64$, $L=2.6$~fm} & $a\mu$  &   0.0025	& 0.0035   & 0.0055  & 0.0075     \\
 														 & No. of Confs          &	200		& 200 &  200  &  200       \\
                                             & $m_\pi$~(GeV) & 0.256	& 0.302   &  0.372 & 0.432      \\
				                                 &$m_\pi L$     &   3.42	&  4.03     &  4.97   &	5.77      \\
\hline\hline
\multicolumn{6}{c}{ $\beta=2.10$, $a=0.0646(7)$~fm  ${r_0/a}=7.538(58)$}\\
\hline
\multirow{4}{*}{$48^3\times 96$, $L=3.1$~fm}  &$a\mu$        & 0.0015  & 0.002  & 0.003  &  \\
                                              & No. of Confs        &  196    &  184    &  200   &  \\
                              			     	  &$m_\pi$~(GeV) & 0.213  & 0.246 & 0.298 & \\
                        					        &$m_\pi L$     &   3.35    &  3.86    &  4.69  &  \\ 
\hline \hline
\end{tabular}
\caption{Input parameters ($\beta,L, a\mu$) of our lattice simulations with the corresponding lattice spacing ($a$), pion mass ($m_{\pi}$) as well as the number of gauge configurations analyzed.}
\label{Table:params}
\end{center}
\vspace*{-.0cm}
\end{table} 


\subsection{Two-point correlation functions and effective mass}

In order to extract baryon masses we consider two-point correlation functions at $\vec{p}=\vec{0}$ defined by
\be \label{eq:two-point}
C_X^\pm (t,\vec{p}=\vec{0}) = \sum_{{\bf x}_{\rm sink}-{\bf x}_{\rm source}}\langle\reci{4} {\rm Tr}\left(1\pm \gamma_0 \right)   \mathcal{J}_X\left({\bf x}_{\rm sink},t_{\rm sink}\right)\bar{\mathcal{J}}_X\left({\bf x}_{\rm source},t_{\rm source}\right)\rangle , \qquad t=t_{\rm sink}-t_{\rm source}
\ee
where $\mathcal{J}_X$ is the interpolating field of the baryon state of interest  acting  at the source  $\left({\bf x}_{\rm source},t_{\rm source}\right)$ and the sink, $\left({\bf x}_{\rm sink},t_{\rm sink}\right)$. Space-time reflection symmetries of the action and the anti-periodic boundary conditions in the temporal direction for the quark fields imply, for zero three-momentum correlators, that $C_X^+(t) = -C_X^-(T-t)$. Therefore, in order to decrease errors we average correlators in the forward and backward direction and define
\be
   C_X(t) = C_X^+(t) - C_X^-(T-t) \, .
\ee
In addition, the source location is chosen randomly on the whole lattice for each configuration, in order to decrease correlation among measurements. 

\begin{figure}[!ht]\vspace*{-0.2cm}
\center
\begin{minipage}{8.5cm}
{\includegraphics[width=\textwidth]{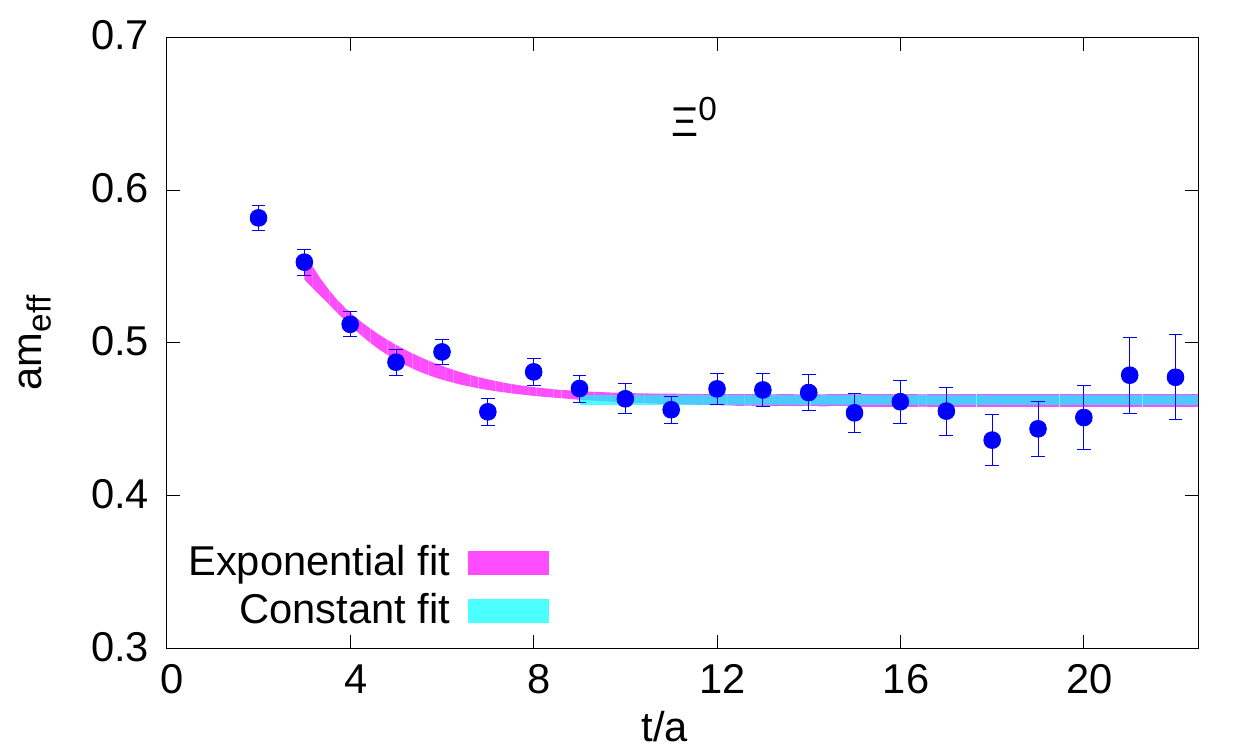}}
\end{minipage}\hfill
\begin{minipage}{8.5cm}
{\includegraphics[width=\textwidth]{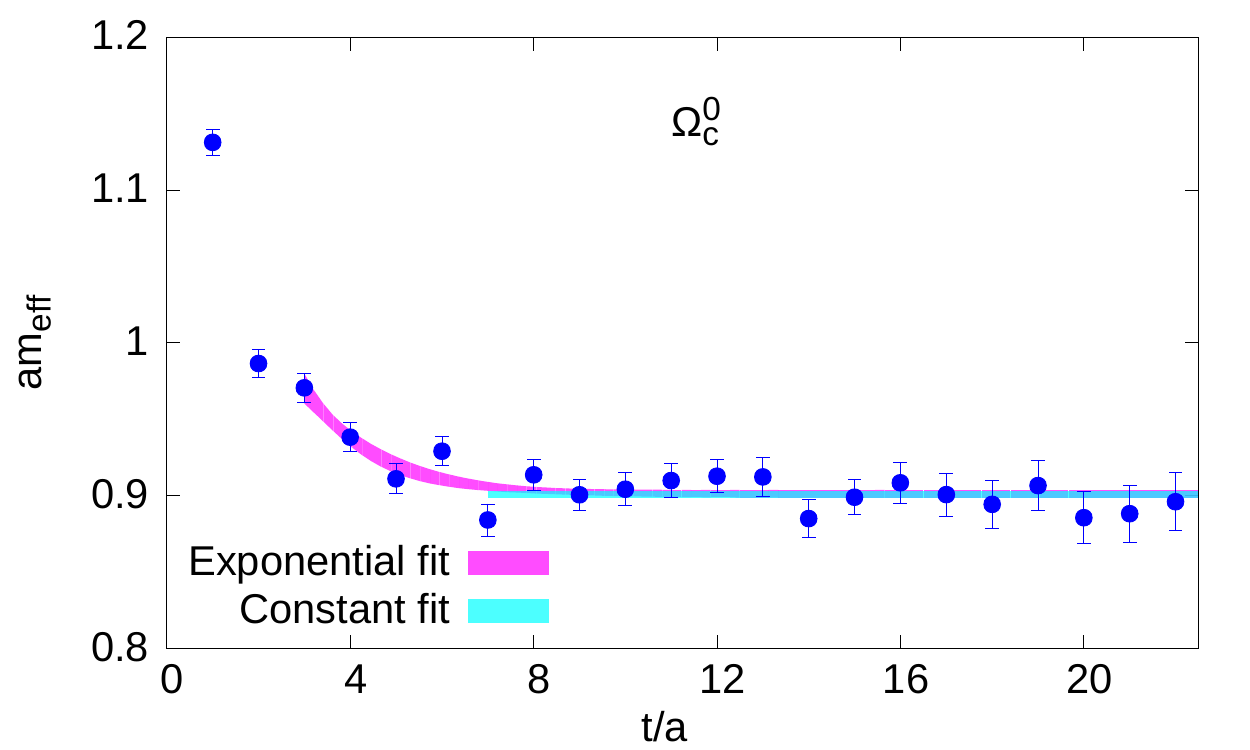}}
\end{minipage}
\caption{Representative effective mass plots for $\Xi^0$ (left) and $\Omega_c^0$ (right) at $\beta=2.10$, $a\mu_l=0.0015$. Both the constant and the exponential fits are displayed. }
\label{Fig:repr_effmass}
\end{figure}

The ground state mass of a given  hadron can be extracted by examining
the effective mass  defined by
\be \label{eq:eff_mass}
am_{\rm eff}^X (t) = \log\left(\frac{C_X(t)}{C_X(t+1)} \right) = am_X + \log \left( \frac{1+\sum_{i=1}^{\infty} c_i e^{-\Delta_i t}}{1+\sum_{i=1}^{\infty} c_i e^{-\Delta_i (t+1)}} \right) \tendto_{t\rightarrow \infty} am_X
\ee
where $\Delta_i = m_i - m_X$ is the mass difference of the excited state $i$ with respect to the ground mass $m_X$. All results in this work have been extracted from correlators where Gaussian smearing is applied both at the source and sink. In general, effective masses of  correlators of any  interpolating fields are expected to have the same value in the large time limit, but applying smearing on the interpolating fields suppresses excited states, therefore yielding a plateau region at earlier source-sink time separations and better accuracy in the extraction of the mass. Our fitting procedure to extract $m_X$ is as follows: The sum over excited states in the effective mass given in \eq{eq:eff_mass} is truncated, keeping only the first excited state, 
\be \label{eq:eff_mass_trunc}
am_{\rm eff}^X (t) \approx am_X + \log \left( \frac{1 + c_1 e^{-\Delta_1 t}}{1+ c_1 e^{-\Delta_1 (t+1)}} \right)\;.
\ee
The upper fitting time slice boundary is kept fixed, while allowing the
lower fitting time  to be  two or three time slices away from $t_{\rm source}$. We then fit the effective mass to the form given in \eq{eq:eff_mass_trunc}. This exponential fit yields an estimate for $c_1$ and $\Delta_1$ as well as for the ground state mass, which we denote by $m_X^{(E)}$. Then, we perform a constant fit to the effective mass increasing  the initial fitting time $t_1$. We denote the value extracted by $m_X^{(C)}(t_1)$. The final value of the mass is selected such that the ratio
\be \label{eq:crit_1}
\frac{| am_X^{(C)}(t_1) -  am_X^{(E)} |}{am_X^{\rm mean}}\quad,\; am_X^{\rm mean} = \frac{am_X^{(C)}(t_1) +  am_X^{(E)}}{2}
\ee
becomes less than 50\%  the statistical error on $m_X^{(C)}(t_1)$. This criterion is, in most cases, in agreement with $\chi^2/{\rm d.o.f.}$ becoming less than unity. In the cases in which this criterion is not satisfied a careful examination of the effective mass is made to ensure that the fit range is in the plateau region.
We show representative results of these fits to the effective mass of the baryons $\Xi^0$ and $\Omega_c^0$ in~\fig{Fig:repr_effmass}. The error bands on the constant and exponential fits are obtained using jackknife analysis. As can be seen the exponential and constant fits yield consistent results in the large time limit.


\subsection{Interpolating fields}

\begin{figure}[h]\vspace*{-0.2cm}
\center
\begin{minipage}{8cm}
\center
{\includegraphics[width=0.7\textwidth]{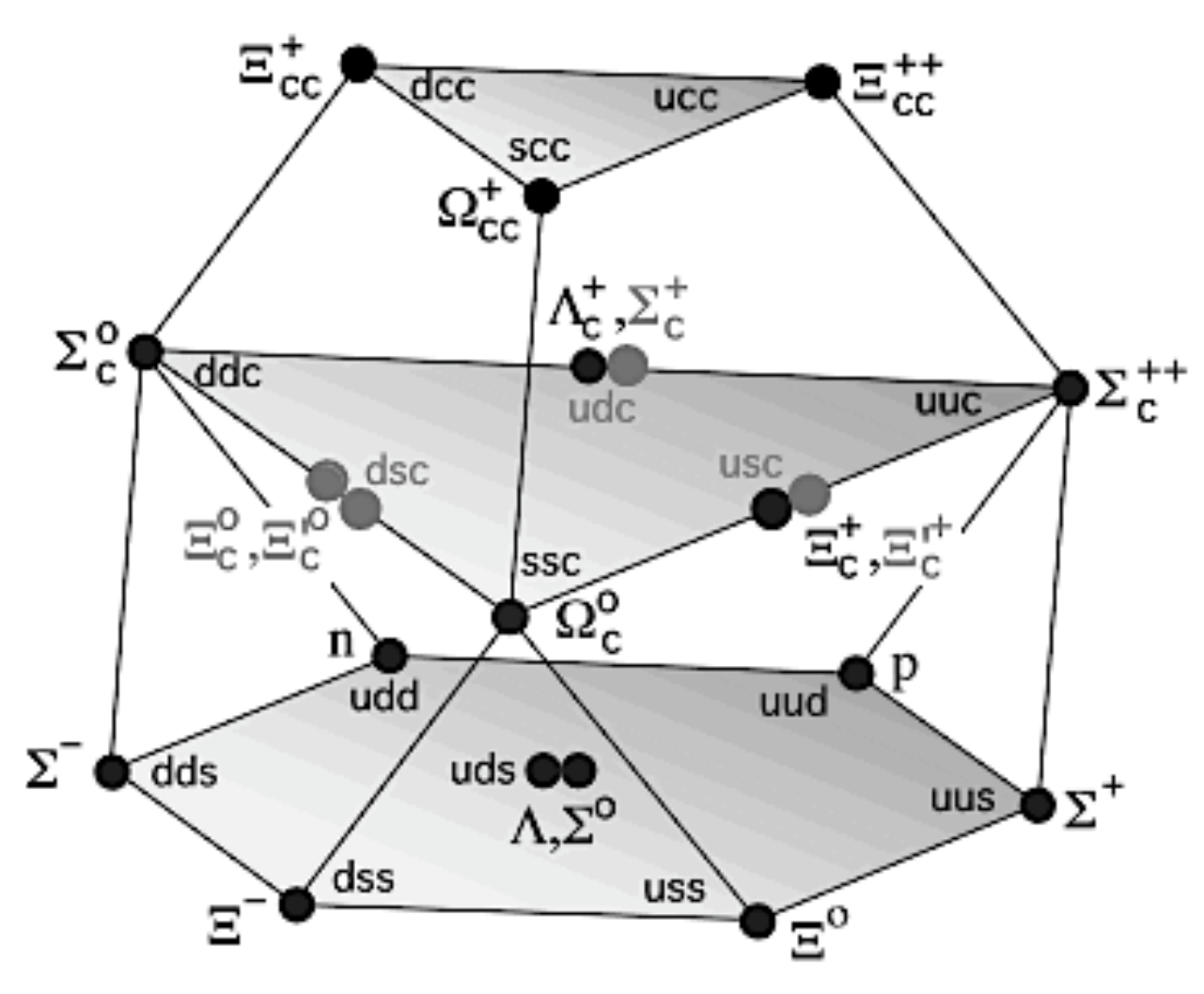}}
\caption{The $20^\prime$-plet of spin-1/2 baryons classified according to their charm content. The lowest level represents the $c=0$  SU(3) octet.}
\label{Fig:spin12_baryons}
\end{minipage}\hfill
\begin{minipage}{8cm}
\center
{\includegraphics[width=0.8\textwidth]{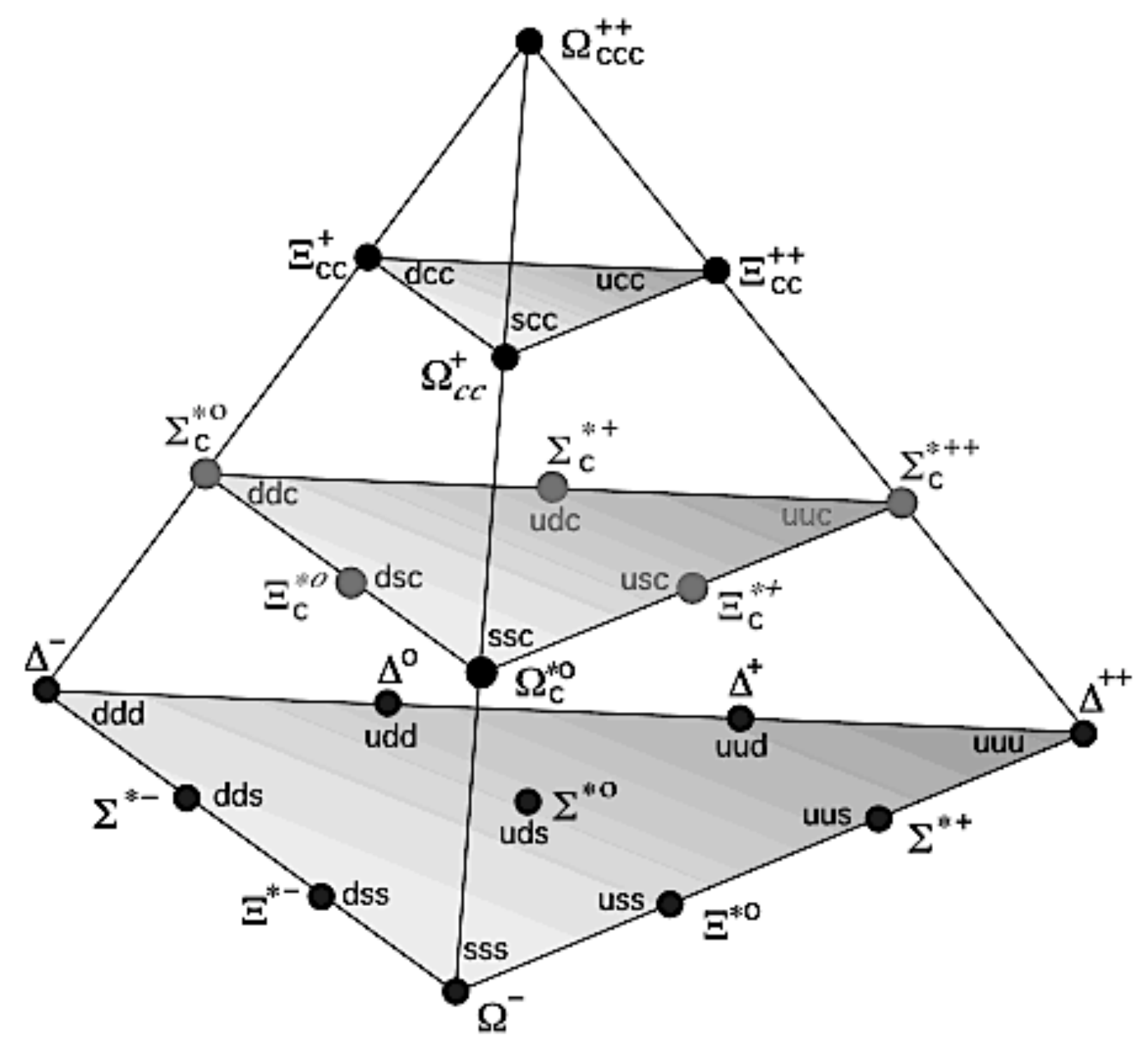}}
\caption{The 20-plet of spin-3/2 baryons classified according to their charm content. The lowest level represents the $c=0$ decuplet sub-group.}
\label{Fig:spin32_baryons}
\end{minipage}
\end{figure}

The baryon states are created from the vacuum with the use of interpolating fields that are constructed such that they have the quantum numbers of the baryon of interest and reduce to the quark model wave functions in the non-relativistic limit. We have a four-dimensional flavour space and therefore we consider SU(3) sub-groups to visualise baryons under SU(4) symmetry. The baryon states split into a $20^\prime$-plet of spin-1/2 states and a 20-plet of spin-3/2 states. There also exists a $\bar{4}$-plet, which is not considered in this work. Light, strange and charmed baryons can be classified according to their transformation properties under flavour SU(3)  and their charm content. This is shown schematically in \fig{Fig:spin12_baryons} and \fig{Fig:spin32_baryons}. The spin-1/2 $20^\prime$-plet decomposes into three horizontal levels. The first level  is the standard octet of the SU(3) symmetry that has no charm quarks, 
the $c=1$ is the second level that splits into two SU(3) multiplets, a
  {\bf 6} containing the $\Sigma_c$  and a $\bar{\bf 3}$ containing the $\Lambda_c$ and the $\Xi_c$  and the $c=2$ is a {\bf 3} multiplet of  SU(3) that forms the  top level. In a similar way, the 20-plet of spin-3/2 baryons contains the standard $c=0$ decuplet at the lowest level, the $c=1$ level {\bf 6} multiplet of  SU(3), the $c=2$  {\bf 3} multiplet  and a $c=3$ singlet at the top of the pyramid. The interpolating fields for these baryons, displayed \fig{Fig:spin12_baryons} and \fig{Fig:spin32_baryons}, are collected in the Tables~\ref{Table:spin12_intfields} and \ref{Table:spin32_intfields} of Appendix~\ref{App:int_fields}~\cite{Ioffe:1981kw,Leinweber:1990dv,Leinweber:1992hy}.

In other recent works where baryon properties are studied, e.g. in Ref~\cite{Durr:2012dw}, different interpolating fields to those we provide in Tables~\ref{Table:spin12_intfields} and \ref{Table:spin32_intfields} were used. These different interpolating fields are tabulated in \tbl{Table:alt_intfields} of Appendix \ref{App:int_fields}. In what follows we will compare the effective masses using the two different sets that have the same quantum numbers but different structure.

As local interpolating fields are not optimal for suppressing excited state contributions, we apply Gaussian smearing to each quark field $q({\bf x},t)$ \cite{Gusken:1989qx,Alexandrou:1992ti}. The smeared quark field is given by $q^{\rm smear}({\bf x},t) = \sum_y F({\bf x,y};U(t))q({\bf y},t)$, where we have used the gauge invariant smearing function
\be \label{eq:smear_func}
F({\bf x,y};U(t)) = \left(1+\alpha H\right)^n({\bf x,y};U(t)),
\ee
constructed from the hopping matrix understood as a matrix in coordinate, color and spin space,
\be \label{eq:hopping_matrix}
H({\bf x,y};U(t))=\sum_{i=1}^3 \left(U_i({\bf x},t)\delta_{{\bf x,y}-a\hat{i}}+U^\dag_i({\bf x}-a\hat{i},t)\delta_{{\bf x,y}+a\hat{i}} \right).
\ee
In addition, we apply APE smearing to the spatial links that enter the hopping matrix. The parameters $\alpha$ and $n$ of the Gaussian and APE smearing at each value of $\beta$ are collected in \tbl{Table:smear_params}.
\begin{table}[h]
\begin{center}
\renewcommand{\arraystretch}{1.6}
\renewcommand{\tabcolsep}{5.5pt}
\begin{tabular}{c|c||c|c||c|c}
  & \multirow{2}{*}{$a\mu_l\;, L/a$} 				& \multicolumn{2}{c||}{APE} & \multicolumn{2}{c}{Gaussian}   \\
\cline{3-6}
  &     &  	  n        &        $\alpha$        &           n           &           $\alpha$          \\
  \hline
\multirow{3}{*}{$\beta=1.90$}  &  0.0030, 32   &    20   &   0.5  &   50   &     4.0      \\
											  &  0.0040, 32   &    20   &   0.5  &   50   &     4.0      \\
											  &  0.0050, 32   &    20   &   0.5  &   50   &     4.0      \\
\hline\hline	
\multirow{4}{*}{$\beta=1.95$}  &  0.0025, 32   &    20   &   0.5  &   50   &     4.0      \\
											  &  0.0035, 32   &    20   &   0.5  &   50   &     4.0      \\
											  &  0.0055, 32   &    20   &   0.5  &   50   &     4.0      \\
											  &  0.0075, 32   &    20   &   0.5  &   50   &     4.0      \\
\hline\hline
\multirow{3}{*}{$\beta=2.10$}  &  0.0015, 48   &    50   &   0.5  &   110   &     4.0      \\
											  &  0.0020, 48   &    20   &   0.5  &   50   &     4.0      \\
											  &  0.0030, 48   &    20   &   0.5  &   50   &     4.0      \\
\end{tabular}
\end{center}
\caption{Smearing parameters for the ensembles at $\beta=1.90$, $\beta=1.95$ and $\beta=2.10$.}
\label{Table:smear_params}
\end{table}

The interpolating fields for the spin-3/2 baryons defined in \tbl{Table:spin32_intfields} have an overlap with spin-1/2 states. These overlaps can be removed with the incorporation of a spin-3/2 projector in the definitions of the interpolating fields
\be 
\mathcal{J}_{X_{3/2}}^\mu = P^{\mu\nu}_{3/2} \mathcal{J}_{\nu X} \, .
\ee
For non-zero momentum, $P^{\mu\nu}_{3/2}$ is defined by~\cite{Benmerrouche:1989uc}
\be \label{eq:proj32}
P^{\mu\nu}_{3/2} = \delta^{\mu\nu} - \reci{3}\gamma^\mu \gamma^\nu - \reci{3p^2}\left(\not{p}\gamma^\mu p^\nu + p^\mu \gamma^\nu \not{p}  \right) \, .
\ee
In correspondence, the spin-1/2 component $\mathcal{J}_{X_{1/2}}^\mu$ can be obtained by acting with the spin-1/2 projector $P^{\mu\nu}_{1/2}=\delta^{\mu\nu} - P^{\mu\nu}_{3/2}$ on $\mathcal{J}^\mu_X$. Elements with Lorentz indices $\mu,\nu=0$ will not contribute. In this work we study the mass spectrum of the baryons in the rest frame taking $\vec{p}=\vec{0}$. Since in that case the last term of \eq{eq:proj32} will contain $\delta_{0\mu}$, it will vanish. When the spin-3/2 and spin-1/2 projectors are applied to the interpolating field operators, the resulting two-point correlators for the spin-3/2 baryons acquire the form
\bea \label{eq:correlators_32_12}
C_{\frac{3}{2}} (t) &=& \frac{1}{3}\tr [C(t)] + \frac{1}{6} \sum_{i\ne j}^3 \gamma_i \gamma_j C_{ij}(t)\;, \nonumber\\
C_{\frac{1}{2}} (t) &=& \frac{1}{3}\tr [C(t)] - \frac{1}{3} \sum_{i\ne j}^3 \gamma_i \gamma_j C_{ij}(t)\;,
\eea 
where $\tr[C] = \sum_i C_{ii}$. When no projector is taken into account, the resulting two-point correlator would be $C = \frac{1}{3}\tr[C]$.

\begin{figure}[h]\vspace*{-0.2cm}
\center
\begin{minipage}{8cm}
\center
{\includegraphics[width=\textwidth]{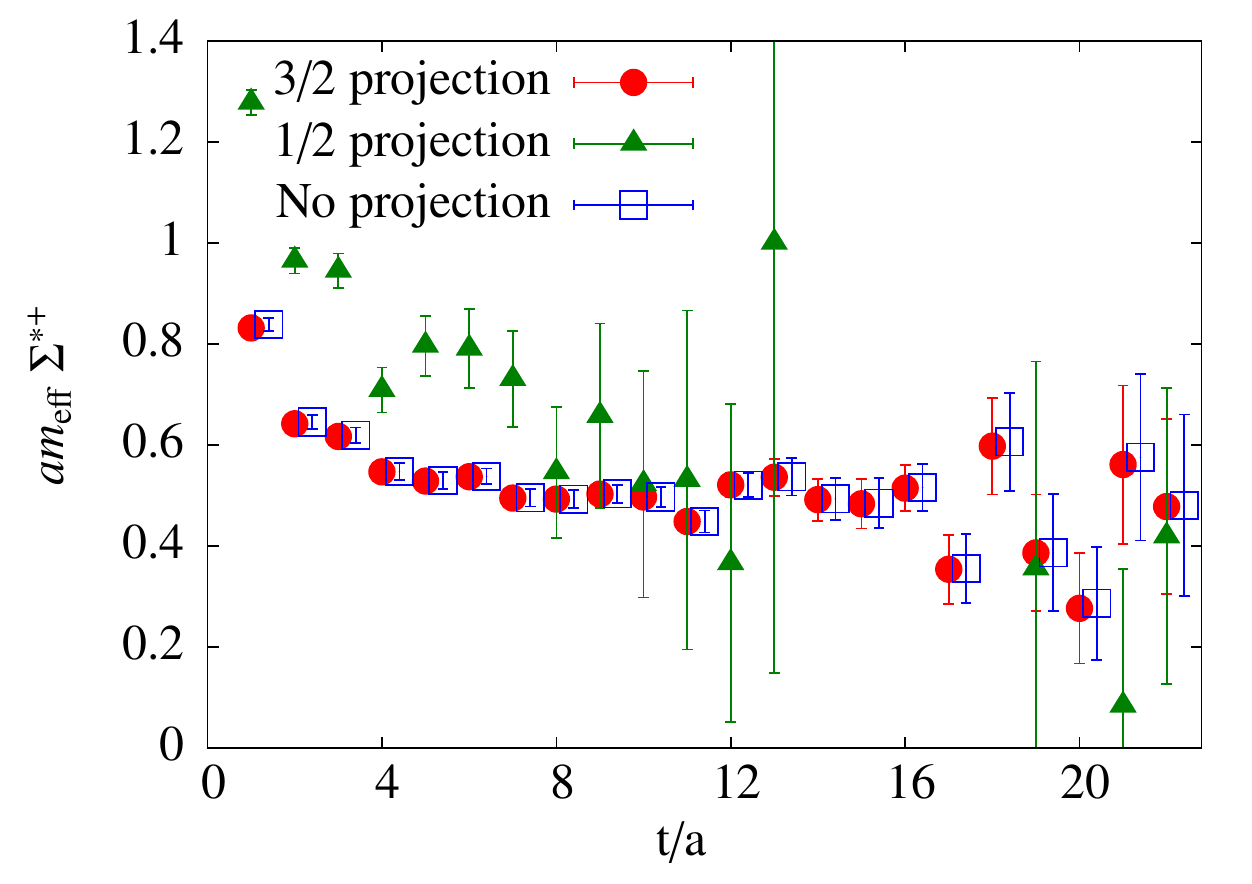}}
\end{minipage}\hfill
\begin{minipage}{8cm}
\center
\vspace{-0.05cm}
{\includegraphics[width=\textwidth]{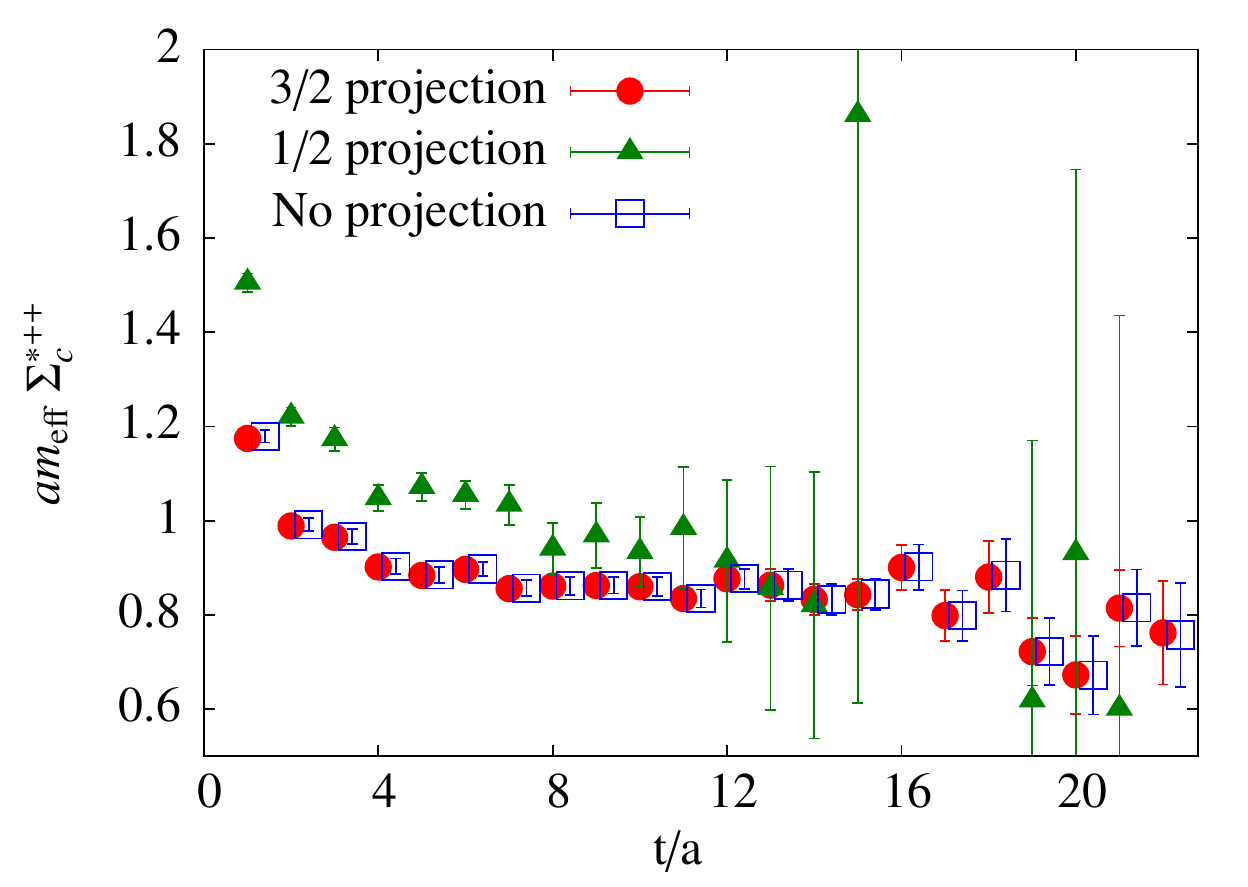}}
\end{minipage}
\caption{Comparison of effective masses extracted using  $\mathcal{J}_{\Sigma^{*+}}$ at $\beta=2.10$, $a\mu_l=0.0015$ (left) and using  $\mathcal{J}_{\Sigma_c^{*++}}$ at $\beta=1.95$, $a\mu_l=0.0055$ (right)  obtained with the spin-3/2 projection (red filled circles), spin-1/2 projection (green triangles) and without projection (blue open squares, shifted to the right for clarity).}
\label{Fig:sigmas_proj}
\end{figure}

\begin{figure}[h]\vspace*{-0.2cm}
\center
\begin{minipage}{8cm}
\center
{\includegraphics[width=\textwidth]{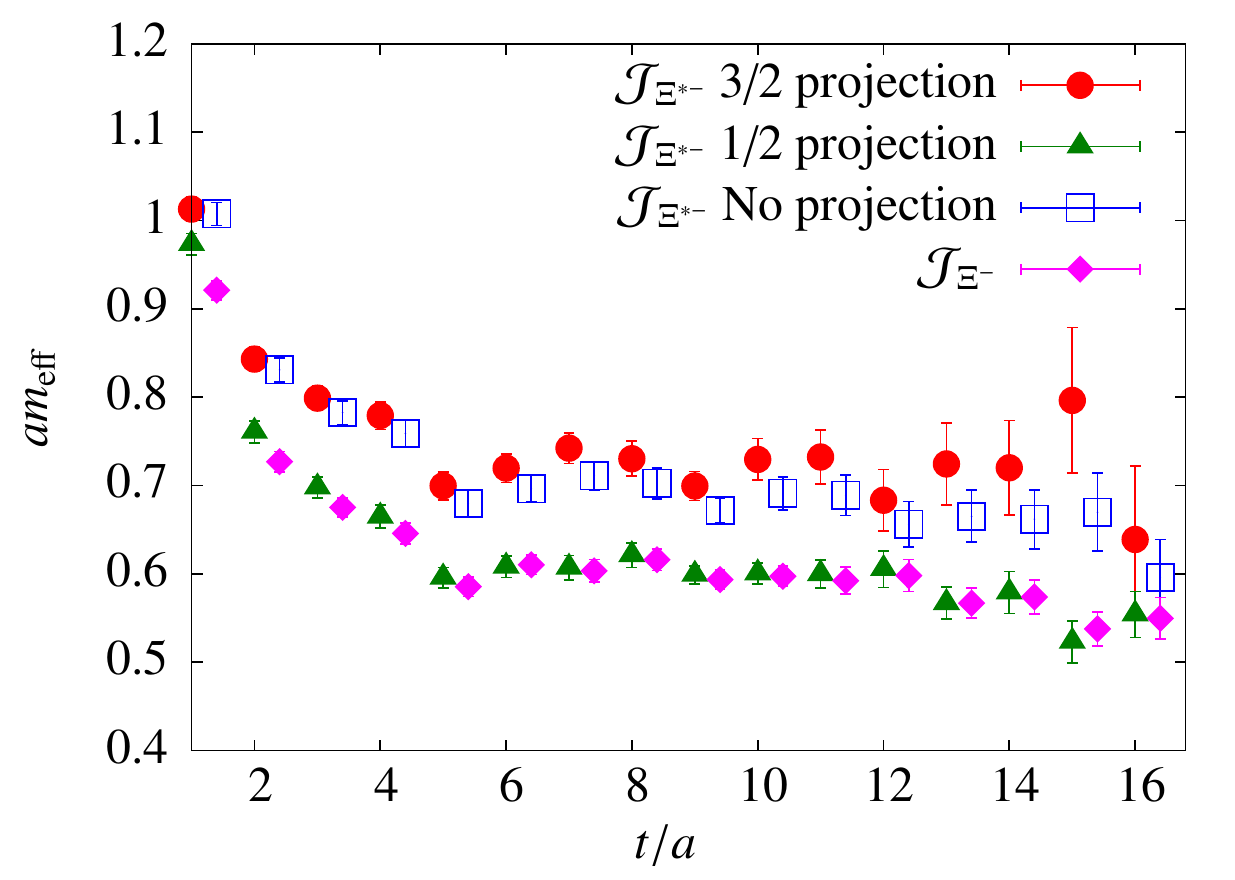}}
\caption{Comparison of effective masses extracted using for $\mathcal{J}_{\Xi^{*-}}$ at $\beta=1.95$, $a\mu_l=0.0025$ obtained with the spin-3/2 projection (red filled circles), without projection (blue open squares, shifted to the right for clarity) and with spin-1/2 projection (green triangles). Also plotted is the effective mass using $\mathcal{J}_{\Xi^-}$ (magenta diamonds). }
\label{Fig:xistarminus_proj}
\end{minipage}\hfill
\begin{minipage}{8cm}
\center
\vspace{-0.55cm}
{\includegraphics[width=\textwidth]{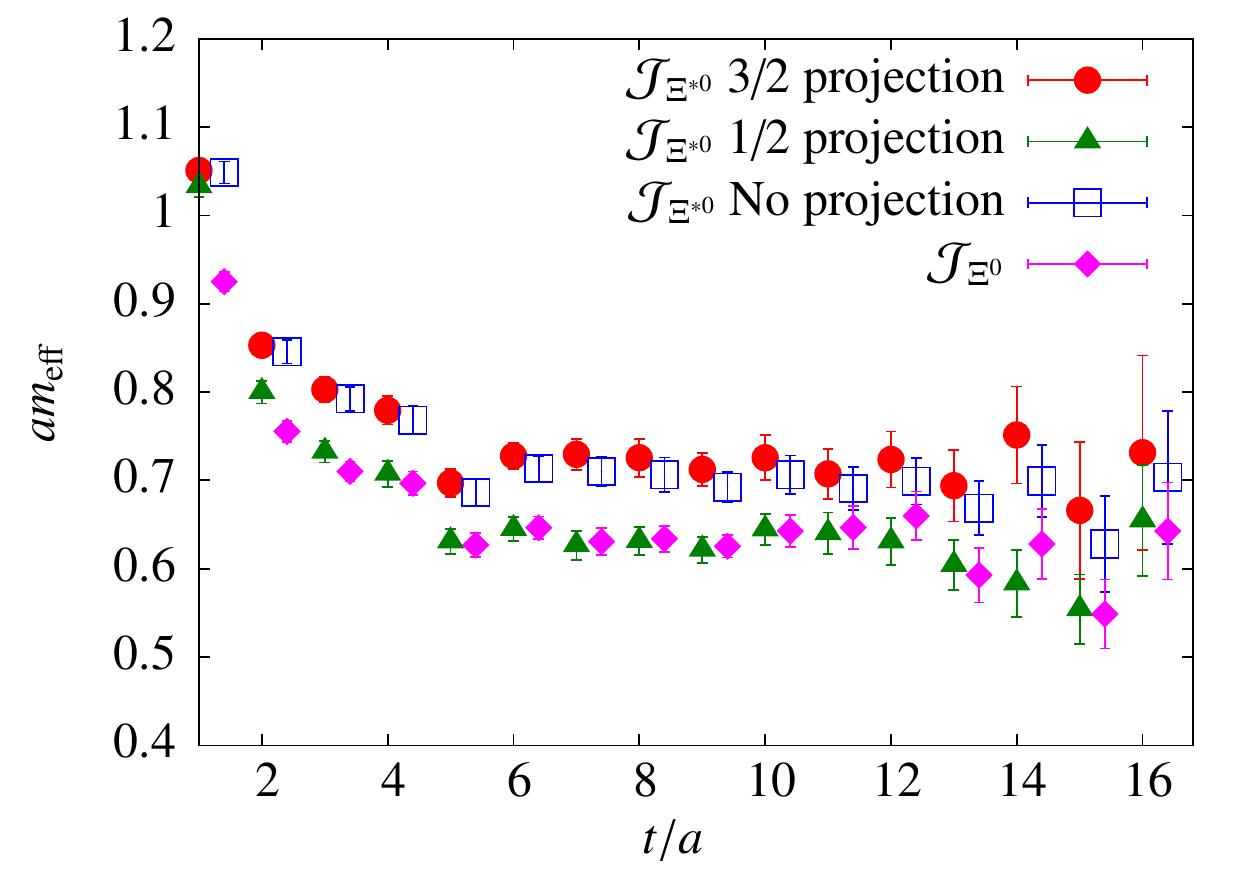}}
\caption{Comparison of effective masses for $\Xi^{*0}$ at $\beta=1.95$, $a\mu_l=0.0025$ obtained with the spin-3/2 projection, without projection and with spin-1/2 projection. Also plotted is the effective mass of $\Xi^0$. The notation is as in \fig{Fig:xistarminus_proj}.}
\label{Fig:xistarzero_proj}
\end{minipage}
\end{figure}

We have carried out an analysis to examine the results of the effective masses extracted from correlation functions with and without the spin-3/2 projection, as well as with the spin-1/2 projector using 100 gauge configurations, a number sufficiently large for the purpose of this comparison. 
In our comparison we also consider correlation functions obtained using the alternative interpolating fields given in \tbl{Table:alt_intfields}. To distinguish these two sets
we denote the interpolating fields of Tables~\ref{Table:spin12_intfields}  and \ref{Table:spin32_intfields} by $\mathcal{J}_B$ and those in \tbl{Table:alt_intfields} by $\tilde{\mathcal{J}}_B$. The left panel of \fig{Fig:sigmas_proj} compares effective masses extracted from correlators with $\mathcal{J}_{\Sigma^{*+}}$ at $\beta=2.10$, $a\mu_l=0.0015$. As can be seen, the results for the effective masses when applying the 3/2-projector and without any projection are perfectly consistent even at short source-sink time separations yielding the mass of  $\Sigma^{*+}$. On the other hand, the effective mass obtained using the spin-1/2 projected interpolating field is much more noisy and yields a higher value of the mass. The latter property suggests that  the 1/2-projected interpolating field $\mathcal{J}_{\Sigma^{*}}$ yields  an excited spin-1/2 state of the $\Sigma^*$  at least at small time slices. The large errors associated with the correlator with the spin-1/2 projector suggest that the overlap with this state is weak.  Another example is shown in the right panel of \fig{Fig:sigmas_proj}, where results are displayed for the correlator using $\mathcal{J}_{\Sigma_c^{*++}}$ at $\beta=1.95$, $a\mu_l=0.0055$. A similar behavior to ours for the  $\Sigma_c^{*++}$ was found in Ref.~\cite{Zanotti:2003fx} where the same spin projections are implemented. However, there are cases where the spin-3/2 projection is required. One example is the $\Xi^{*-}$ baryon, shown in \fig{Fig:xistarminus_proj}, where  the effective mass  when no projection is applied is persistently lower  than when using the spin-3/2 projector. It is also apparent from \fig{Fig:xistarminus_proj} that the spin-1/2 projected interpolating field $\mathcal{J}_{\Xi^{*-}}$ yields an effective mass, which is consistent with the corresponding results using the spin-1/2 interpolating field $\mathcal{J}_{\Xi^{-}}$ and thus the mass of  $\Xi^-$. A similar case to this is the $\Xi^{*0}$, as can be seen from \fig{Fig:xistarzero_proj}. Therefore, it is crucial in order to obtain the correct spin-3/2 mass to project out the lower-lying spin-1/2 state.
\begin{figure}[h]\vspace*{-0.2cm}
\center
\begin{minipage}{8cm}
\center
{\includegraphics[width=\textwidth]{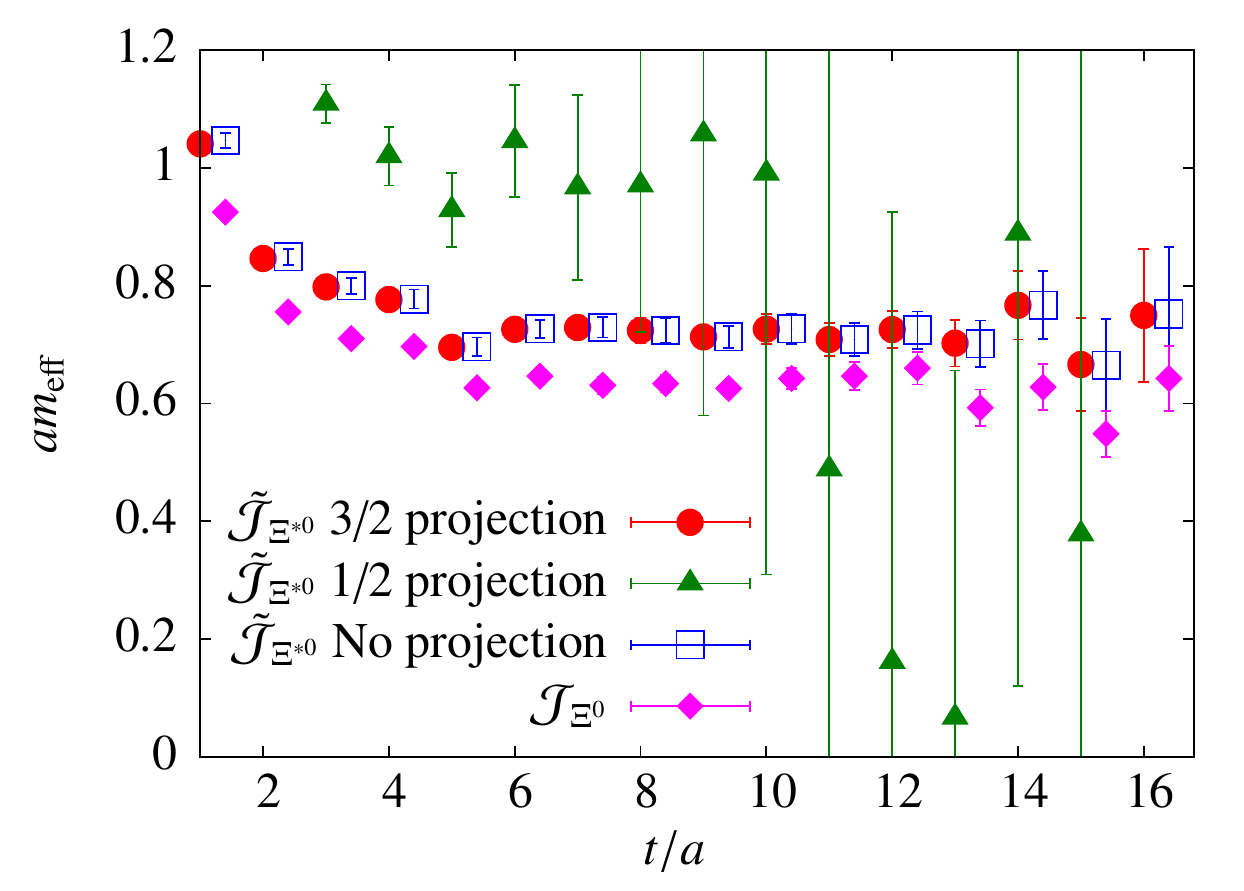}}
\caption{Effective masses obtained using $\tilde{\mathcal{J}}_{\Xi^{*0}}$ at $\beta=1.95$, $a\mu_l=0.0025$ with the spin-3/2 projection (red filled circles), without projection (blue open squares, shifted to the right for clarity) and with spin-1/2 projection (green triangles). Also plotted is the effective masses using $\tilde{\mathcal{J}}_{\Xi^0}$ (magenta diamonds). }
\label{Fig:xistarzero_extra_proj1}
\end{minipage}\hfill
\begin{minipage}{8cm}
\center
\vspace{-0.25cm}
{\includegraphics[width=\textwidth]{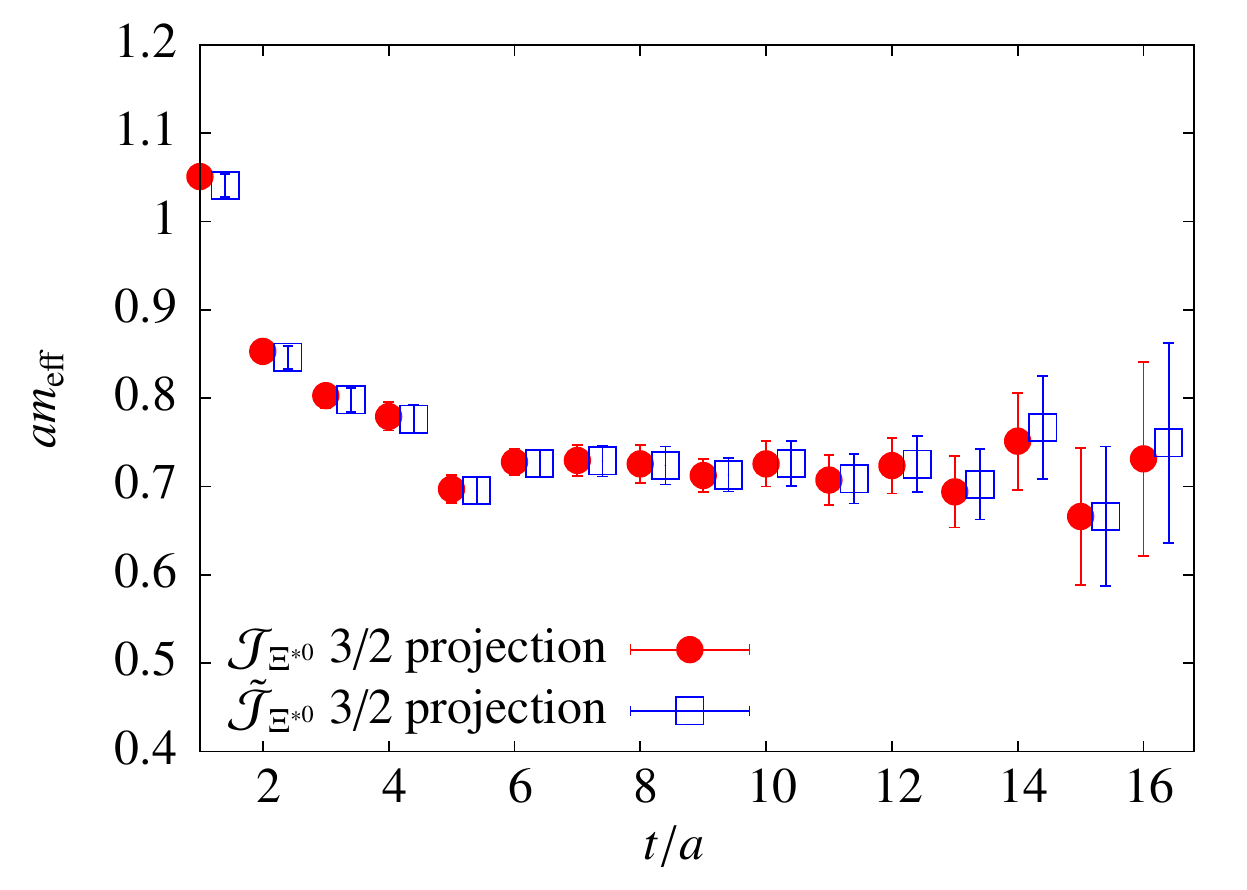}}
\caption{Comparison of effective masses for $\Xi^{*0}$ at $\beta=1.95$, $a\mu_l=0.0025$ obtained from $\mathcal{J}_{\Xi^{*0}}$ (red filled circles) and $\tilde{\mathcal{J}}_{\Xi^{*0}}$ (blue open squares, shifted to the right for clarity) using the spin-3/2 projection. Results from the two interpolating fields are fully consistent.}
\label{Fig:xistarzero_extra_proj2}
\end{minipage}
\end{figure}

\begin{figure}[h]\vspace*{-0.2cm}
\center
\begin{minipage}{8cm}
\center
\vspace{-1.1cm}
{\includegraphics[width=\textwidth]{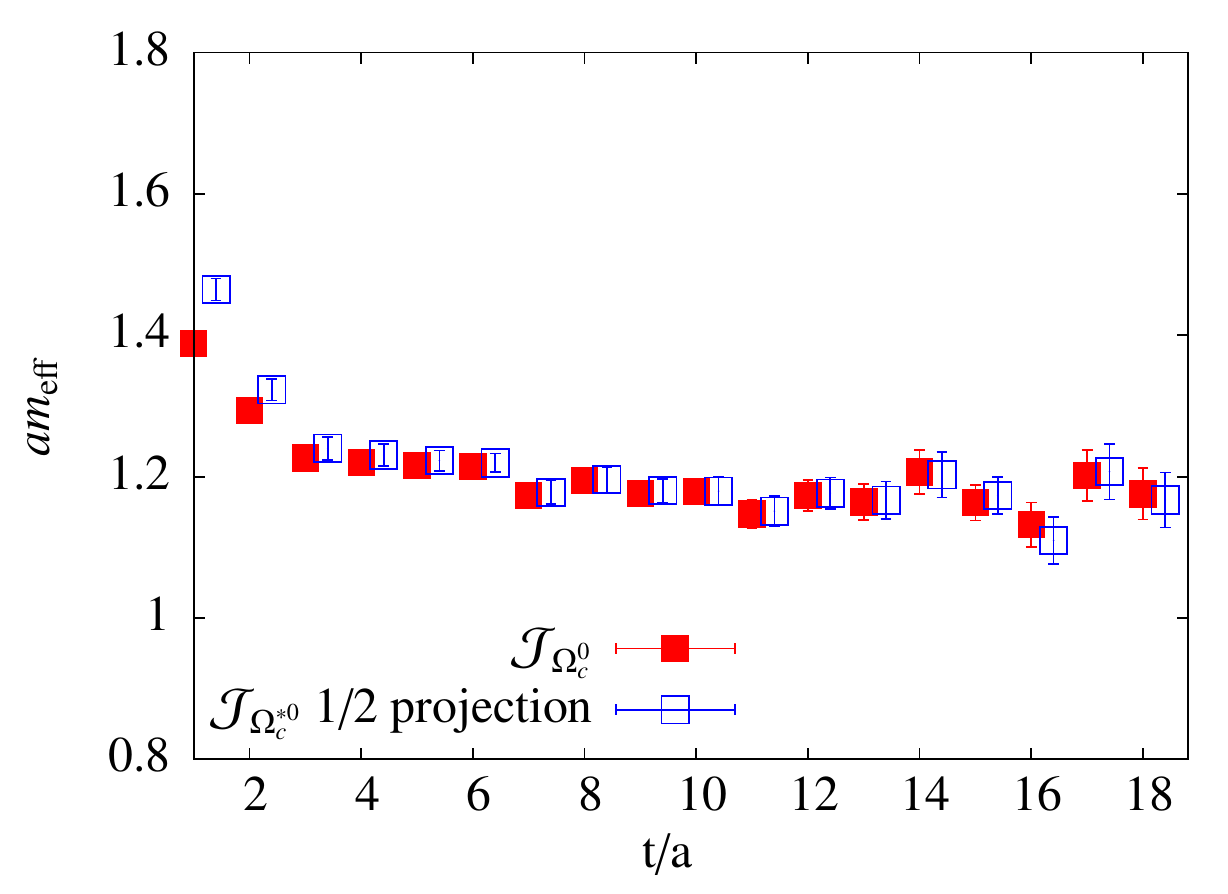}}
\caption{Effective mass results obtained for $\Omega_{c}^{0}$ (red filled squares) and from $\mathcal{J}_{\Omega_{c}^{*0}}$ using the spin-1/2 projection (blue open squares). The results are in agreement. }
\label{Fig:omegac_proj1}
\end{minipage}\hfill
\begin{minipage}{8cm}
\center
{\includegraphics[width=\textwidth]{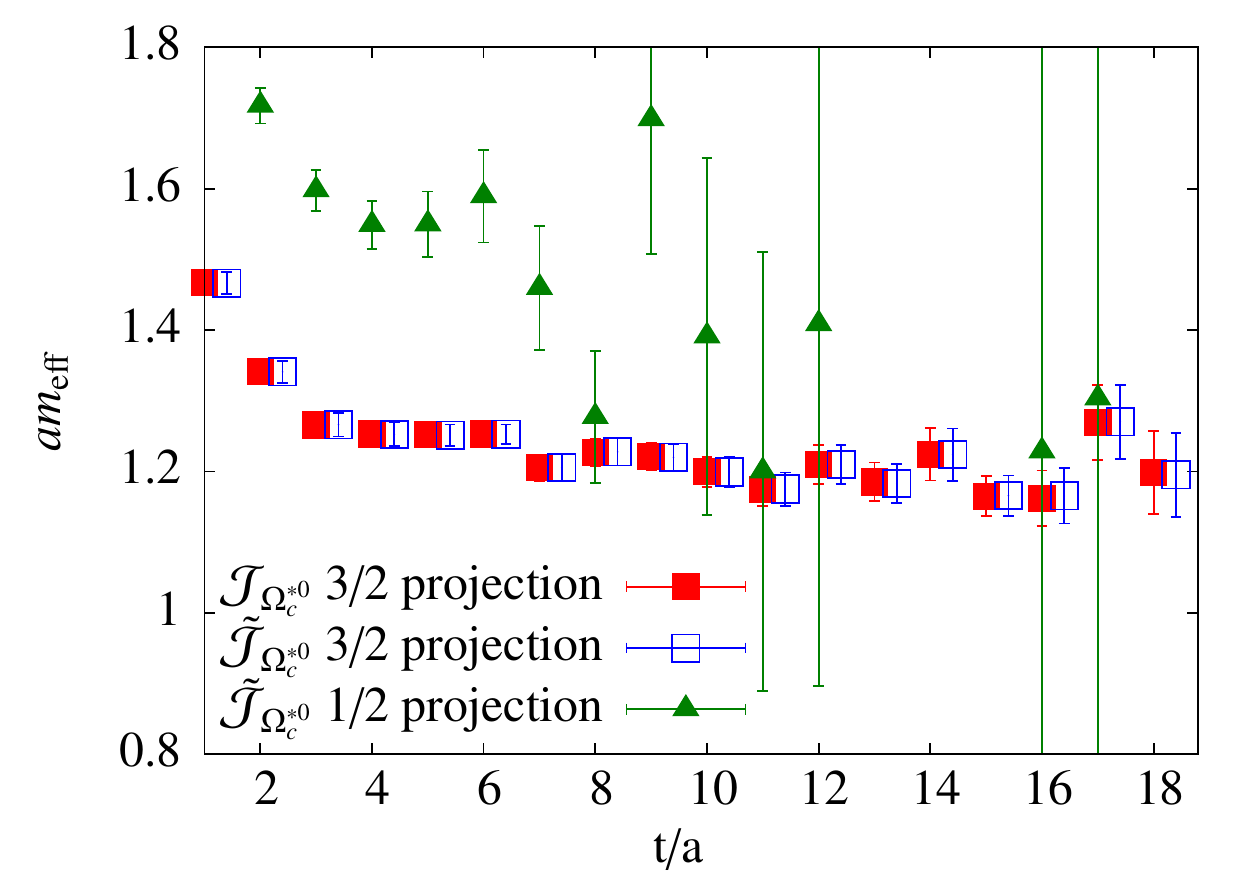}}
\caption{Effective mass results of $\Omega_{c}^{*0}$ obtained from the spin-3/2 projections of $\mathcal{J}_{\Omega_{c}^{*0}}$ (red filled squares) and $\tilde{\mathcal{J}}_{\Omega_{c}^{*0}}$ (blue open squares) as well as from the spin-1/2 projection of $\tilde{\mathcal{J}}_{\Omega_{c}^{*0}}$ (green triangles).  More details are given in the text.}
\label{Fig:omegac_proj2}. 
\end{minipage}
\end{figure}

In order to further examine the properties of the interpolating fields, we also include effective mass results from the alternative  set of interpolating fields. We plot effective mass results obtained from $\tilde{\mathcal{J}}_{\Xi^{*0}}$ as well as the effective mass of the spin-1/2 $\Xi^0$ at $\beta=1.95$, $a\mu_l=0.0025$ in \fig{Fig:xistarzero_extra_proj1}, in correspondance with \fig{Fig:xistarzero_proj}. As shown, the results from using spin-3/2 projection and when applying no projection on $\tilde{\mathcal{J}}_{\Xi^{*0}}$ are now consistent. In contrast with $\mathcal{J}_{\Xi^{*0}}$, the spin-1/2 projection of $\tilde{\mathcal{J}}_{\Xi^{*0}}$  yields an excited spin-1/2 state of $\Xi^{*0}$. However, as can be seen from \fig{Fig:xistarzero_extra_proj2}, the spin-3/2 projections of the two interpolating fields for $\Xi^{*0}$ yield fully consistent results, as expected. Similar behavior is observed in the other baryon states as well. We demonstrate this by showing results for $\Omega_{c}^{*0}$ at $\beta=1.95$, $a\mu_l=0.0075$ in Figs. \ref{Fig:omegac_proj1} and \ref{Fig:omegac_proj2}.

The main conclusion of this analysis is that the set of spin-3/2 $\tilde{\mathcal{J}}$ interpolating fields   do not need any spin-3/2 projection, whereas
the $\mathcal{J}$ in general do.   After spin-3/2 projection they both   give consistent results for the mass of the spin-3/2 state they represent, as expected.  Therefore from now on 
we use only results from spin-3/2 projected interpolating fields and limit
ourselves to the interpolating fields $\mathcal{J}$ listed in Tables~\ref{Table:spin12_intfields} and \ref{Table:spin32_intfields}.


\subsection{Determination of the lattice spacing}

Since in this work the observables discussed are the masses of baryons, the physical nucleon mass is the most appropriate quantity to set the scale. In order
to  determine the values of the lattice spacings as accurate as possible
 we have carried out a high statistics analysis of the nucleon masses for a total of 17 $N_f=2+1+1$ gauge ensembles at $\beta=1.90$, $\beta=1.95$ and $\beta=2.10$ on a range of pion masses and volumes.  We average over the masses of the proton and neutron to  further gain on statistics. The resulting nucleon masses for each of the gauge ensembles are collected in \tbl{Table:nucleon_masses}.
\begin{table}[h]
\begin{center}
\renewcommand{\arraystretch}{1.4}
\renewcommand{\tabcolsep}{5.8pt}
\begin{tabular}{c|c|c|c|c|c|c}
\hline\hline
 Volume & Statistics & $a\mu_l$ & $am_\pi$ & $m_\pi$ (GeV) & $am_N$ & $m_N$ (GeV) \\
\hline\hline
\multicolumn{7}{c}{$\beta=1.90$} \\
\hline
\multirow{3}{*}{$32^3\times64$} & 740    &  0.0030   & 0.1240 &  0.2607  &  0.5239(87)  & 1.1020(183)  \\
 & 1556   &  0.0040   & 0.1414 &  0.2975  &  0.5192(112) & 1.0921(235)  \\
 & 387    &  0.0050   & 0.1580 &  0.3323  &  0.5422(62)  & 1.1407(130)  \\
\hline
\multirow{3}{*}{$24^3\times48$} & 2092   &  0.0400   & 0.1449 &  0.3049  &  0.5414(84)  & 1.1389(176)  \\
 & 1916   &  0.0060   & 0.1728 &  0.3634  &  0.5722(48)  & 1.2036(101)  \\
 & 1796   &  0.0080   & 0.1988 &  0.4181  &  0.5898(50)  & 1.2407(104)  \\
 & 2004   &  0.0100   & 0.2229 &  0.4690  &  0.6206(43)  & 1.3056(90)   \\
\hline
$20^3\times 48$ & 617    &  0.0040   & 0.1493 &  0.3140  &  0.5499(195) & 1.1568(410)  \\
\hline\hline
\multicolumn{7}{c}{$\beta=1.95$}\\
\hline
\multirow{4}{*}{$32^3\times64$} & 2892   &  0.0025   & 0.1068 &  0.2558  &  0.4470(59)  & 1.0706(141)  \\
 & 4204   &  0.0035   & 0.1260 &  0.3018  &  0.4784(48)  & 1.1458(114)  \\
 & 18576  &  0.0055   & 0.1552 &  0.3716  &  0.5031(16)  & 1.2049(39)   \\
 & 2084   &  0.0075   & 0.1802 &  0.4316  &  0.5330(42)  & 1.2764(100)  \\
\hline
$24^3\times 48$  & 937    &  0.0085   & 0.1940 &  0.4645  &  0.5416(50)  & 1.2970(121)  \\
\hline\hline
\multicolumn{7}{c}{$\beta=2.10$}\\
\hline
\multirow{3}{*}{$48^3\times96$} & 2424   &  0.0015   & 0.0698 &  0.2128  &  0.3380(41)  & 1.0310(125)  \\
 & 744    &  0.0020   & 0.0805 &  0.2455  &  0.3514(70)  & 1.0721(215)  \\
 & 226    &  0.0030   & 0.0978 &  0.2984  &  0.3618(68)  & 1.1038(208)  \\
\hline
 $32^3\times 64$ & 1905   &  0.0045   & 0.1209 &  0.3687  &  0.3944(26)  & 1.2032(79)   \\
\hline\hline
\end{tabular}
\caption{Values of the nucleon masses with the associated statistical error.}
\label{Table:nucleon_masses}
\end{center}
\vspace*{-.0cm}
\end{table} 

\begin{figure}[!ht]\vspace*{-0.2cm}
\center
\begin{minipage}{8.5cm}
{\includegraphics[width=\textwidth]{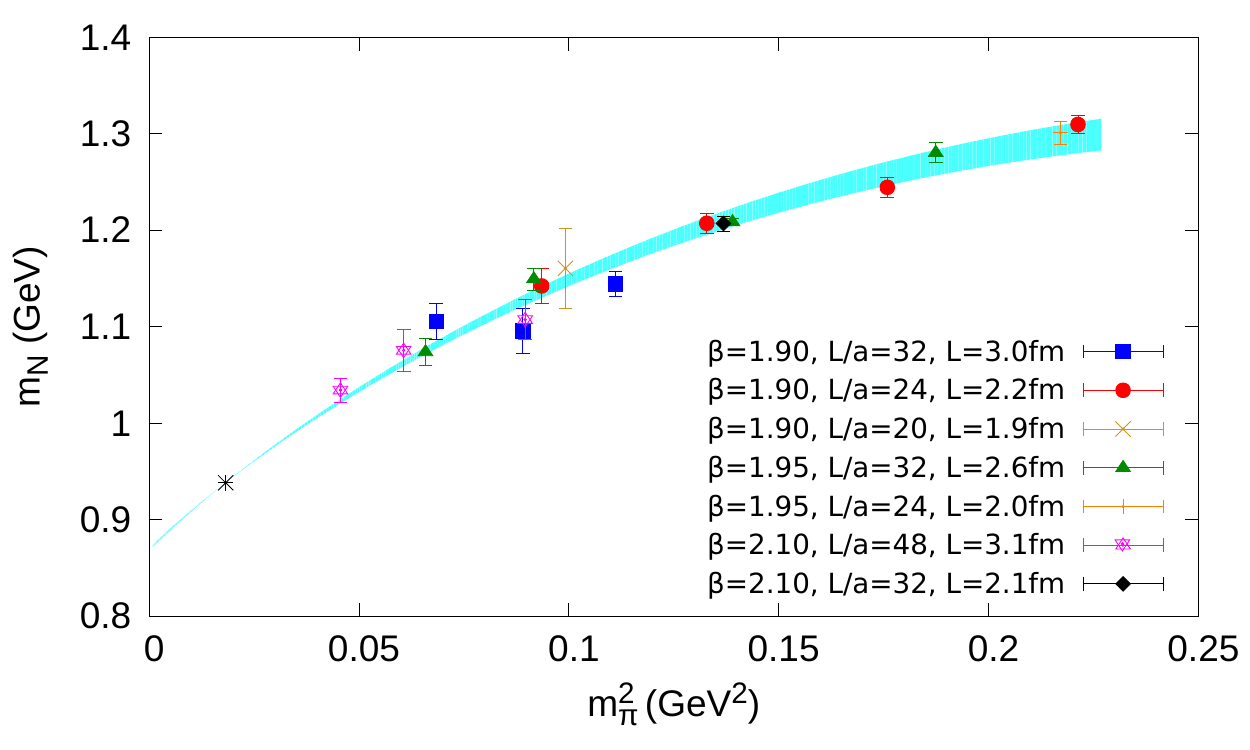}}
\end{minipage}\hfill
\begin{minipage}{8.5cm}
{\includegraphics[width=\textwidth]{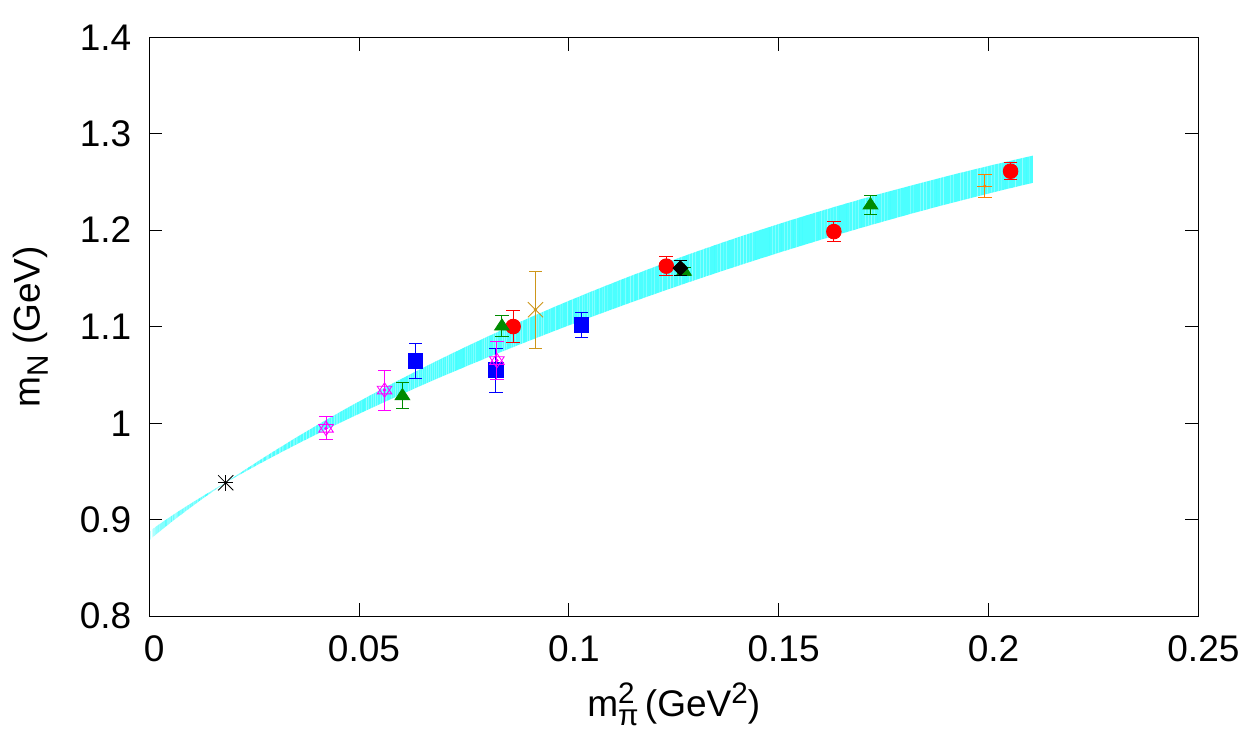}}
\end{minipage}
\caption{Nucleon masses at the three values of the lattice spacing. On the left panel the solid band represents a fit to the lowest order $\mathcal{O}(p^3)$ expansion from HB$\chi$PT. The band on the right panel is a fit to $\mathcal{O}(p^4)$ with explicit $\Delta$ degrees of freedom in the so called small scale expansion (SSE). The physical nucleon mass is denoted with the asterisk.}
\label{Fig:nucleon_mass}
\end{figure}

The nucleon masses as function of $m_\pi^2$ are presented in \fig{Fig:nucleon_mass}. As can be seen, cut-off effects are negligible, therefore we can use continuum chiral perturbation theory to extrapolate to the physical pion mass using all the lattice results. To this end we consider SU(2) chiral perturbation theory ($\chi$PT) \cite{Tiburzi:2008bk} and the well-established $\mathcal{O}(p^3)$ result of the nucleon mass dependence on the pion mass, given by
\be \label{eq:nucleon_p3}
m_N = m_N^{(0)} - 4c_1m_\pi^2 - \frac{3g_A^2}{32\pi f_\pi^2} m_\pi^3
\ee
where $m_N^0$ is the nucleon mass at the chiral limit and together with $c_1$ are treated as fit parameters. This lowest order result for the nucleon in HB$\chi$PT, first derived in Ref.~\cite{Gasser:1987rb}, and describes well lattice data~\cite{Alexandrou:2008tn,Alexandrou:2009qu}. Since this result is well established as the leading contribution irrespective of the various approaches to compute higher orders such as in HB$\chi$PT with dimensional and infra-red regularization with and without the $\Delta$ degree of freedom explicitly included, we will use it to fix the lattice spacing from the nucleon mass. The lattice spacings $a_{\beta=1.90}$, $a_{\beta=1.95}$ and $a_{\beta=2.10}$ are considered as additional independent fit parameters in a combined fit of our data at $\beta=1.90$, $\beta=1.95$ and $\beta=2.10$.  We constrain our fit so that the fitted curve passes through the physical point by fixing the value of $c_1$. The physical values of $f_\pi$ and $g_A$ are used in the fits, namely $f_\pi=0.092419(7)(25)$ GeV and $g_A=1.2695(29)$, which is common practice in chiral fits to lattice data on the nucleon mass~\cite{Bernard:2005fy,Pascalutsa:2005nd,Procura:2006bj}. The left panel of \fig{Fig:nucleon_mass} shows the fit to the $\mathcal{O}(p^3)$ result of \eq{eq:nucleon_p3} on the nucleon mass. The error band and the errors on the fit parameters are obtained from super-jackknife analysis~\cite{Bratt:2010jn}. As can be seen, the $\mathcal{O}(p^3)$ result provides a very good fit to our lattice data, which in fact confirms that cut-off and finite volume effects are small for the $\beta$-values used. In addition, our lattice results exhibit a curvature which supports the presence of the $m_\pi^3$-term.

In order to estimate the systematic error due to the chiral extrapolation we also perform a fit using heavy baryon chiral perturbation theory (HB$\chi$PT) to $\mathcal{O}(p^4)$ in the so-called small scale expansion (SSE) \cite{Procura:2006bj}. This form includes explicit $\Delta$ degrees of freedom by introducing as an additional parameter the $\Delta$-nucleon mass splitting, $\Delta\equiv m_\Delta - m_N$, taking $\mathcal{O}(\Delta /m_N) \sim \mathcal{O} (m_\pi /m_N)$. In SSE the nucleon mass is given by
\bea \label{eq:nucleon_p4}
	m_N &=& m_N^0-4c_1 m_\pi^2-\frac{3g_A^2}{32\pi f_\pi^2} m_\pi^3 -4E_1(\lambda)m_\pi^4 - \frac{3\left(g_A^2+3c_A^2 \right)}{64\pi^2f_\pi^2 m_N^0}m_\pi^4 - \frac{\left(3g_A^2+10c_A^2 \right)}{32\pi^2f_\pi^2 m_N^0}m_\pi^4 \log\left(\frac{m_\pi}{\lambda}\right) \nonumber\\	 
	&-& \frac{c_A^2}{3\pi^2f_\pi^2} \left(1+\frac{\Delta}{2m_N^0}\right)\left[\frac{\Delta}{4}m_\pi^2 + \left(\Delta^3-\frac{3}{2}m_\pi^2\Delta\right) \log\left(\frac{m_\pi}{2\Delta}\right) + \left(\Delta^2-m_\pi^2\right) R\left(m_\pi\right) \right]
\eea   
where $R\left(m_\pi\right)= - \sqrt{m_\pi^2 - \Delta^2}\cos^{-1}\left(\frac{\Delta}{m_\pi}\right)$ for $m_\pi > \Delta$ and $R\left(m_\pi\right)= \sqrt{\Delta^2 - m_\pi^2} \log\left( \frac{\Delta}{m_\pi}+\sqrt{\frac{\Delta^2}{m_\pi^2}-1} \right)$ for $m_\pi < \Delta$. We take the cut-off scale $\lambda=1$ GeV, $c_1=1.127$~\cite{Procura:2006bj} and treat the counter-term $E_1$ as an additional fit parameter. As in the $\mathcal{O}(p^3)$ case we use the physical values of $g_A$ and $f_\pi$. The corresponding plot is shown on the right panel of \fig{Fig:nucleon_mass}. The error band as well as the errors on the fit parameters are obtained using super-jackknife analysis. One can see that this formulation provides a good description of the lattice data as well and yields values of the lattice spacings and $m_N^0$ which are consistent with those obtained in $\mathcal{O}(p^3)$ of HB$\chi$PT. We take the difference between the results of the $\mathcal{O}(p^3)$ and $\mathcal{O}(p^4)$ fits as an estimate of the uncertainty due to the chiral extrapolation. This is found to be about three times the statistical error. The final values of the lattice spacing are shown in \eq{eq:lat_spacings}. The first parenthesis is the statistical error and the systematic error is given is the second parenthesis. The rest of the fit parameters for the two expansions and the $\chi^2$/d.o.f. are given in \tbl{Table:nucleon_fitparams}.
\bea \label{eq:lat_spacings}
a_{\beta=1.90} &=& 0.0936(13)(35)\; \rm{fm}\;, \nonumber\\
a_{\beta=1.95} &=& 0.0823(10)(35)\; \rm{fm}\;, \nonumber\\
a_{\beta=2.10} &=& 0.0646(7)(25) \;  \rm{fm}\;.
\eea
\begin{table}[h]
\begin{center}
\renewcommand{\arraystretch}{1.2}
\renewcommand{\tabcolsep}{7.5pt}
\begin{tabular}{l|ccccc}
\hline\hline
 & $m_N^0$        &  $-4c_1 (\rm{GeV}^{-1})$ & $E_1(\lambda)$  (GeV$^{-3}$) &  $\sigma_{\pi N}$ (MeV)    &  $\chi^2/{\rm d.o.f}$   \\				
\hline
$\mathcal{O}(p^3)$ HB$\chi$PT  &  0.8667(15)  &   4.5735        &                            &  64.9(1.5)   &  1.5779  \\
$\mathcal{O}(p^4)$ SSE               &   0.8813(47) &   3.7282        &  -2.5858(2480)   &  45.3(4.3)   &  1.0880 \\
\hline\hline
\end{tabular}
\end{center}
\caption{Fit parameters $m_N^0$ in GeV and $E_1(\lambda)$ in GeV$^{-3}$ from $\mathcal{O}(p^3)$ HB$\chi$PT and $\mathcal{O}(p^4)$ SSE, as well as the fixed value of $-4c_1$. Also included is the value of the  $\sigma$-term for each fit. }
\label{Table:nucleon_fitparams}
\end{table}

In order to better assess discretization effects we perform a fit to $\mathcal{O}(p^3)$ at each of the $\beta$ values separately. The values we find are $a_{\beta=1.90} = 0.0923(20)\; \rm{fm}$, $a_{\beta=1.95} = 0.0821(16)\; \rm{fm}$ and $a_{\beta=2.10} = 0.0657(12) \; \rm{fm}$. These values are fully consistent with those obtained in \eq{eq:lat_spacings} from the combined fit, indicating that discretization effects are small, thus confirming a posteriori the validity of the assumption that cut-off effects are small for the nucleon mass. A different
way of demonstraing this is to include a quadratic term $da^2$ to Eqns.~\ref{eq:nucleon_p3} and~\ref{eq:nucleon_p4}, treating $d$ as an additional fit parameter. Performing the fits with the $da^2$ term gives a value
of $d=0.017(17)$~GeV$^{3}$ i.e. consistent with zero. The same is true for the $\Delta$ mass
 confirming that cut-off effects are negligible in the light quark sector.

We will use the values given in \eq{eq:lat_spacings} to convert to physical units all the quantities studied in this work. We note that when performing these fits only statistical errors are taken into account and systematic errors due to the choice of the plateau are not included.  The lattice spacings for these $\beta$ values were also calculated from a pion decay constant analysis using NLO SU(2) chiral perturbation theory for the extrapolations~\cite{Baron:2011sf}. In that preliminary analysis only a subset of the ensembles used here was included, yielding values of the lattice spacings that are smaller compared to the values we extract using the nucleon mass in this work. Specifically, the lattice spacings at $\beta = 1.90\;, 1.95\;$ and $2.10$ were found to be $a_{f_\pi} = 0.0863(4)\;, 0.0779(4)$ and $0.607(2)$ respectively, where $a_{f_\pi}$ denotes the lattice spacing determined using the pion decay constant. This implies that the values of the pion masses in physical units we quote in this paper are equivalently smaller than those obtained using $f_\pi$ to convert to physical units. A comprehensive study of the different lattice spacing determinations is on-going.

\begin{figure}[!ht]\vspace*{-0.2cm}\hspace*{-4.5cm}
\center
{\includegraphics[width=0.7\textwidth]{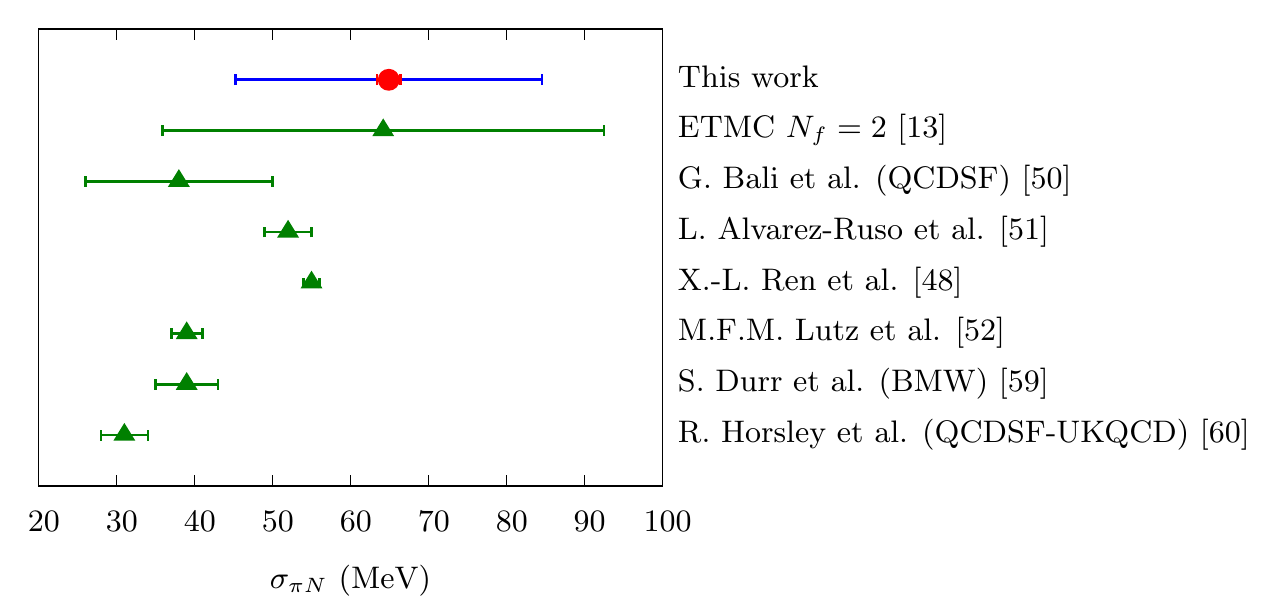}}
\caption{Comparison of lattice results for  $\sigma_{\pi N}$ in MeV, extracted from the $\mathcal{O}(p^3)$ analysis of this work with the results from other lattice calculations.
 Our result shows the statistical error in red and a systematic error  in blue  taken as  the difference between the value obtained using the $\mathcal{O}(p^3)$ and $\mathcal{O}(p^4)$ expressions (Eqns. (\ref{eq:nucleon_p3}) and (\ref{eq:nucleon_p4}) respectively) providing an estimate of the uncertainty due to the chiral extrapolation. }
\label{Fig:nucleon_sterm_compare}
\end{figure}

Having determined the parameters of the chiral fit we can compute the nucleon $\sigma_{\pi N}$-term  by evaluating $m_\pi^2 \partial m_N/\partial m_\pi^2$ where we have taken the leading order relation $m_\pi^2 \sim \mu_l$.  Using \eq{eq:nucleon_p3}  we find $\sigma_{\pi N} = 64.9\pm1.5$ MeV. This value is fully consistent with previous values extracted using this lowest order fit by ETMC on $N_f=2$ quark flavor ensembles~\cite{Alexandrou:2008tn,Alexandrou:2009qu}. 
Performing the same calculation using the $\mathcal{O}(p^4)$ expression we obtain a lower value of $\sigma_{\pi N} = 45.3\pm4.3$ MeV showing the sensitivity to the chiral extrapolation.
It is worth mentioning that such a difference in the determination of
the $\sigma_{\pi N}$-term is known in the literature. For example,
 a latest $\pi N$ scattering study~\cite{Alarcon:2011zs}, reporting a value $\sigma_{\pi N} = 59\pm7$ MeV, while  higher values were also obtained using the Feynman-Hellmann theorem to analyse lattice QCD data yielding $\sigma_{\pi N}=55\pm1$ MeV~\cite{Ren:2014vea}.
Lower values are associated with the well-known result of $\sigma_{\pi N}=45\pm 8$ MeV  extracted from an earlier chiral perturbation analysis of experimental scattering data~\cite{Gasser:1990ce}, as well as, with the values extracted in other lattice QCD calculations, such as the analysis of the QCDSF collaboration~\cite{Bali:2011ks}, where a value $\sigma_{\pi N} = 38\pm 12$ MeV is obtained and of Ref. \cite{Alvarez-Ruso:2013fza} where a value of $\sigma_{\pi N} = 52\pm 3\pm8$ is extracted from a flavour SU(2) extrapolation of a large set of lattice data on the nucleon mass.  A very recent result is obtained using the relativistic chiral Lagrangian from Ref.~\cite{Lutz:2014oxa}, suggests a rather smaller value of $\sigma_{\pi N}=39+2-1$ MeV. We summarize lattice  results on $\sigma_{\pi N}$  in \fig{Fig:nucleon_sterm_compare} we we show our ${\cal O}(p^3)$ value. We take difference between the value extracted from the $\mathcal{O}(p^4)$ expression of \eq{eq:nucleon_p4} and the  ${\cal O}(p^3)$ value as
an estimate for the error arising from chiral extrapolation. As can be seen from the values in Table~\ref{Table:nucleon_fitparams} the chiral extrapolation error is large
showing the sensitivity on the chiral 
extrapolation, which explains the large error shown on our $\sigma_{\pi N}$ results.
 It is apparent that, despite the long efforts, the precise determination of the nucleon $\sigma$-terms is still an open issue and direct techniques as those described in for example Ref.~\cite{Abdel-Rehim:2013wlz} are welcome.
%

\subsection{Tuning of the bare strange and charm quark masses}

 A tuning of the bare strange and charm quark masses is performed 
 using the physical mass of the $\Omega^-$ and the $\Lambda_c^+$ baryons respectively.
 For the tuning we calculate the $\Omega^-$ and $\Lambda_c^+$ masses
at a given value of the renormalized strange and charm quark mass  for all $\beta$ values. For this we need the renormalization constants $Z_P$ for the three $\beta$ values. These were computed in Ref.~\cite{Carrasco:2014cwa} and  we quote, for the convenience of the reader, the values computed in the $\overline{\rm MS}$ scheme at 2~GeV:
\be\label{eq:Zps}
 Z_P^{\beta=1.90}=0.529(7),\; Z_P^{\beta=1.95}=0.509(4),\; Z_P^{\beta=2.10}=0.516(2).
 \ee
For the $\Omega^-$ we use the leading one-loop result from SU(2) $\chi$PT, given by
\be \label{eq:omega_tuning}
m_{\Omega} = m_{\Omega}^{(0)} - 4c_{\Omega}^{(1)}m_\pi^2\;,
\ee
where the mass $m_{\Omega}^{(0)}$ and $c_{\Omega}^{(1)}$ are treated as fit parameters. For the $\Lambda_c^+$ baryon, we use the result motivated by SU(2) HB$\chi$PT to leading one-loop order given by
\be \label{eq:lambdac_tuning}
m_{\Lambda_c} = m_{\Lambda_c}^{(0)} + c_1 m_\pi^2 + c_2 m_\pi^3 \;,
\ee
where $m_{\Lambda_c}^{(0)}$ and the coefficients $c_i$ are treated as fit parameters. We include cut-off effects, by adding a quadratic term $da^2$ to the  Eqns. (\ref{eq:omega_tuning}) and (\ref{eq:lambdac_tuning}), where $d$ is treated as an additional fit parameter. The fit then yields the result at the physical point in the continuum limit. We use the lattice spacings given in Eq.~(\ref{eq:lat_spacings}) extracted from the nucleon mass to convert the $\Omega^-$ and $\Lambda_c$ masses to physical units.

\begin{figure}[!ht]\vspace*{-0.2cm}
\center
\begin{minipage}{8.5cm}
{\includegraphics[width=0.9\textwidth]{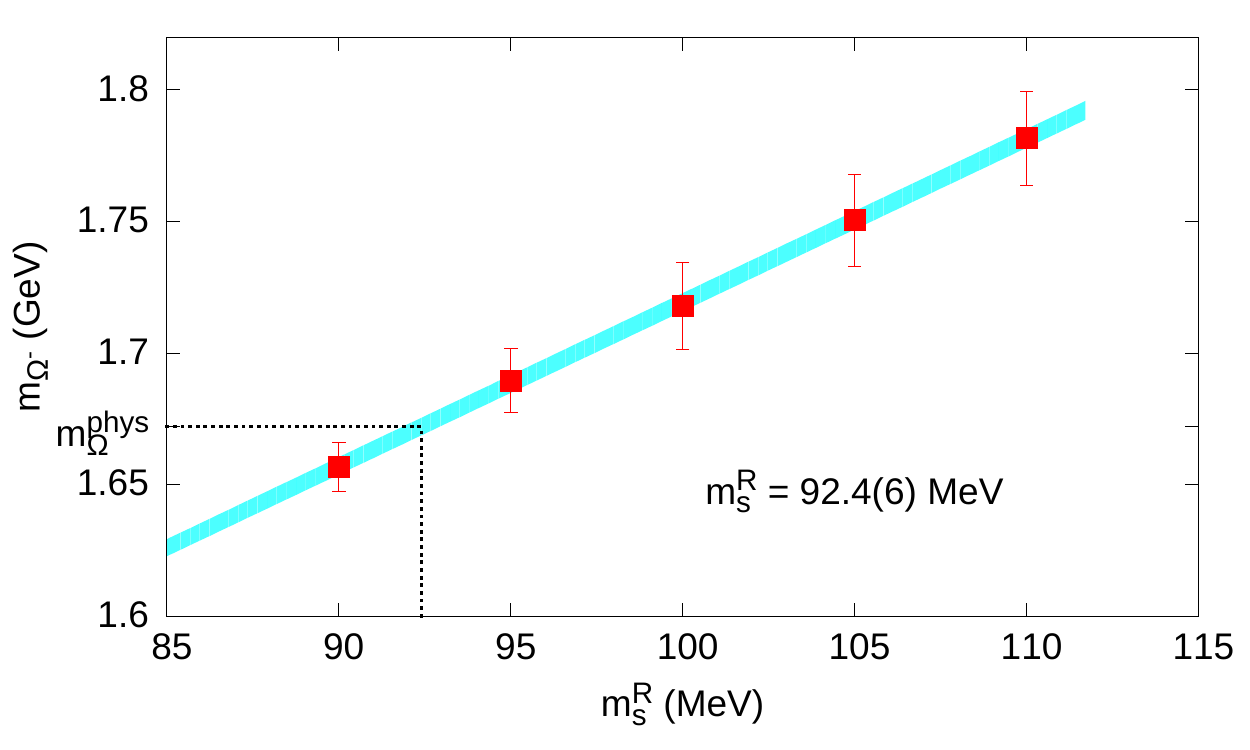}}
\end{minipage}\hfill
\begin{minipage}{8.5cm}
{\includegraphics[width=0.9\textwidth]{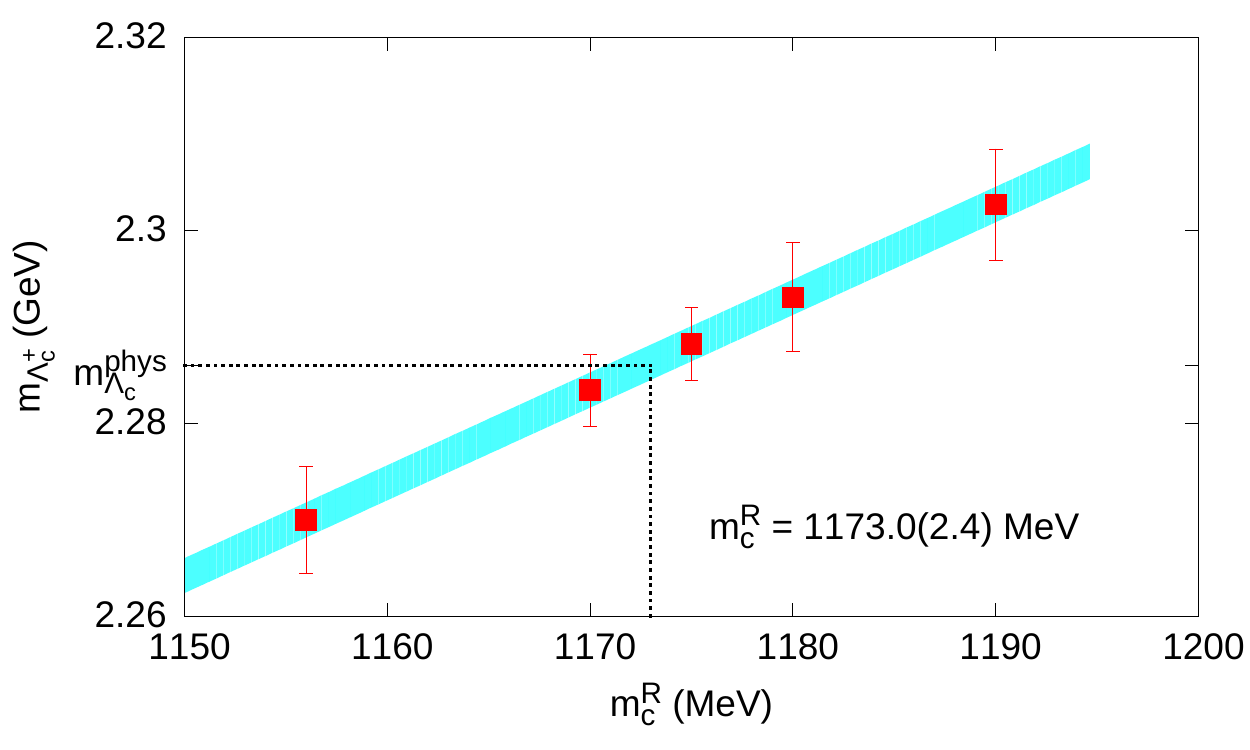}}
\end{minipage}
\caption{Tuning of the renormalized strange and charm quark masses with the experimental values of the $\Omega$ (left) and $\Lambda_c^+$ (right) masses respectively.}
\label{Fig:omega_lambda_ms_mc}
\end{figure}

\begin{figure}[!ht]\vspace*{-0.2cm}
\center
\begin{minipage}{8cm}
{\includegraphics[width=\textwidth]{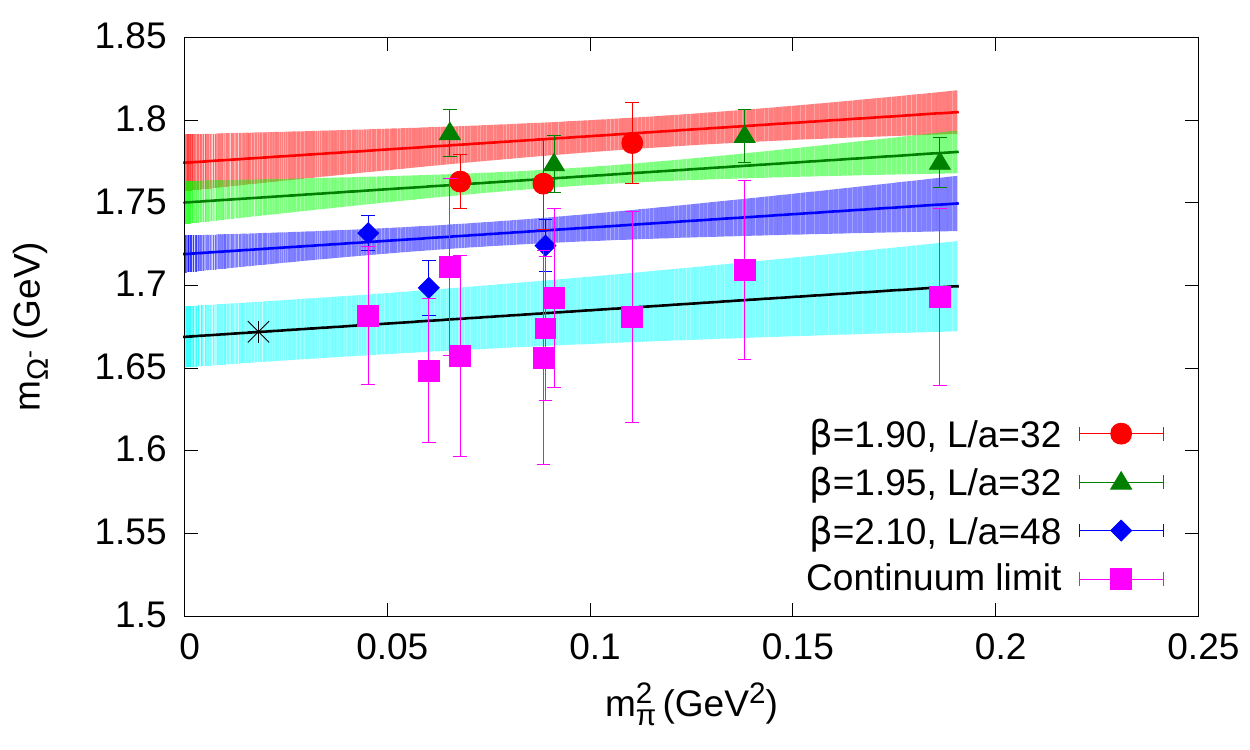}}
\end{minipage}\hfill
\begin{minipage}{8cm}
{\includegraphics[width=\textwidth]{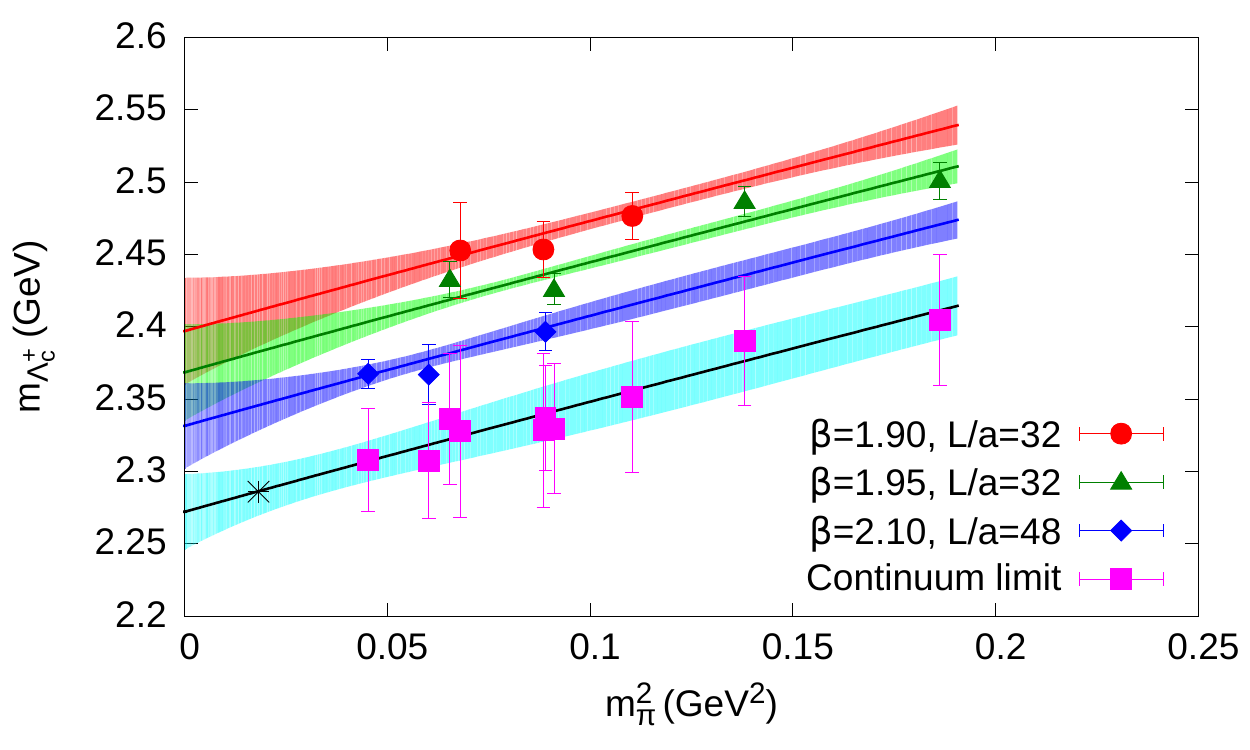}}
\end{minipage}
\caption{Chiral extrapolations of the lattice data for $\Omega^-$ (left) and $\Lambda_c$ (right) at the fixed values of the renormalized strange and charm quark masses of \eq{eq:strange_charm_values} respectively. In these figures, the lattice data for each $\beta$ value as well as the continuum extrapolated values are plotted. The physical masses of $\Omega^-$ and $\Lambda_c$ are reproduced at the continuum limit and at the physical pion mass.}
\label{Fig:chiral_omega_lambdac}
\end{figure}

In order to perform the tuning we use several values of the strange and charm quark masses for the gauge ensembles considered in this work, as listed in \tbl{Table:quark_masses_tuning}. 
\begin{table}[h]
\begin{center}
\renewcommand{\arraystretch}{1.2}
\renewcommand{\tabcolsep}{7.5pt}
\begin{tabular}{c|c|c|c||c|c}
\multicolumn{2}{c|}{Ensemble}                                                    &$am_s$   & $m_s^R$ (GeV) & $am_c$   & $m_c^R$ (GeV)  \\
  \hline\hline
\multirow{6}{*}{$\beta=1.90$} & \multirow{2}{*}{$a\mu_l=0.0030, L/a=32$}      &  0.0229  & 0.0904   & 0.2968 & 1.1737   \\
											 & 																	  & 0.0234   & 0.0924   & 0.2999 & 1.1860   \\
\cline{2-6}
											 & \multirow{3}{*}{$a\mu_l=0.0040, L/a=32$}      &  0.0232  & 0.0917   & \multirow{3}{*}{\begin{tabular}{cc}0.2851 \\ 0.2999\end{tabular}}  & 			                             																																 \multirow{3}{*}{\begin{tabular}{cc}1.1272 \\ 1.1860\end{tabular}}   \\
                     							 & 				        						 						 &  0.0234  & 0.0924   &   &      \\
											 &																		 &	0.0264   & 0.1043   &   &	   \\
\cline{2-6}
											& \multirow{2}{*}{$a\mu_l=0.0050, L/a=32$}      & \multirow{2}{*}{0.0234}  & \multirow{2}{*}{0.0924}   & 0.2943 & 1.1637   \\
											&																		&				 						&				      				   & 0.2999  & 1.1860    \\
\hline\hline
\multirow{15}{*}{$\beta=1.95$} & \multirow{4}{*}{$a\mu_l=0.0025, L/a=32$}   &  0.0182   & 0.0862 & 0.2350 & 1.1122 \\
                                                  &            														&  0.0192   & 0.0909 & 0.2506  & 1.1860 \\
                                                  &           															&  0.0195	 &  0.0924 & 0.2550 & 1.2069 \\
                                                  &																	&  0.0200   & 0.0947 & 0.2694 & 1.2752 \\
\cline{2-6}
											   & \multirow{4}{*}{$a\mu_l=0.0035, L/a=32$}    &  \multirow{4}{*}{\begin{tabular}{cc}0.0187 \\ 0.0195 \\ 0.0200\end{tabular}}   & \multirow{4}{*}{\begin{tabular}{cc}0.0883 \\ 0.0924 \\ 0.0970\end{tabular}}  & 0.2250 & 1.0649  \\
                                                 &      															     &    &  & 0.2450 & 1.1596  \\
                                                 &															         &    &  & 0.2506 & 1.1860  \\
										       &																	&     &  & 0.2580 & 1.2210  \\	
\cline{2-6}
											  & \multirow{4}{*}{$a\mu_l=0.0055, L/a=32$}    &  \multirow{4}{*}{\begin{tabular}{cc}0.0186 \\ 0.0195 \\ 0.0200\end{tabular}}   & \multirow{4}{*}{\begin{tabular}{cc}0.0879 \\ 0.0924 \\ 0.0970\end{tabular}} & 0.2350 & 1.1122  \\
                                                 & 													                 &    &  & 0.2506 & 1.1860  \\
			                                    & 																    &    &  & 0.2570 & 1.2164  \\
			                                    & 																    &    &  & 0.2715 & 1.2848  \\
\cline{2-6}
											  & \multirow{3}{*}{$a\mu_l=0.0075, L/a=32$}    &  \multirow{3}{*}{\begin{tabular}{cc}0.0195 \\ 0.0200\end{tabular}}   & \multirow{3}{*}{\begin{tabular}{cc}0.0924 \\ 0.0970\end{tabular}} & 0.2240 & 1.0602   \\
                                                 &														             &   &  & 0.2440 & 1.1548   \\
											  &																	& 	 &	 & 0.2506 & 1.1860   \\	
\hline\hline
\multirow{10}{*}{$\beta=2.10$} & \multirow{4}{*}{$a\mu_l=0.0015, L/a=48$}   &  0.0155   & 0.0919 & 0.1850 & 1.0959 \\
                                                &															             &  0.0156	  & 0.0924 & 0.2000 & 1.1847   \\
                                                &																         &  0.0162   & 0.0959 & 0.2002 & 1.1860   \\
											 &																		 &  0.0169   & 0.1002 & 0.2195 & 1.3002   \\
\cline{2-6}
											 & \multirow{3}{*}{$a\mu_l=0.0020, L/a=48$}      &  0.0156	  & 0.0924 & 0.1900 & 1.1255   \\
                                                &   													                  &  0.0158   & 0.0936 & 0.2002 & 1.1860   \\
                                                &   													                  &  0.0165   & 0.0977 & 0.2150 & 1.2736   \\
\cline{2-6}
											 & \multirow{3}{*}{$a\mu_l=0.0030, L/a=48$}     &  \multirow{3}{*}{\begin{tabular}{cc}0.0156 \\ 0.0163\end{tabular}}   & \multirow{3}{*}{\begin{tabular}{cc}0.0924 \\ 0.0965\end{tabular}} & 0.1800 & 1.0662   \\
                                                & 													                 &   &  & 0.2002 & 1.1860   \\
                                                & 													                 &   &  & 0.2080 & 1.2321   \\
\hline\hline
\end{tabular}
\caption{The values of the strange and charm quark masses for each ensemble used for the tuning.}
\label{Table:quark_masses_tuning}
\end{center}
\end{table}
Our strategy  is to interpolate the $\Omega^-$ and $\Lambda_c^+$ masses to a given  value of the renormalized strange and charm quark mass, respectively, and then extrapolate to the physical point using Eqns. (\ref{eq:omega_tuning}) and (\ref{eq:lambdac_tuning}) to compare with the experimental values.
The value of the renormalized quark mass is then changed iteratively until the extrapolated continuum values agree with the experimental ones.
This determines the tuned values of $m_s^R$ and $m_c^R$ that reproduce the physical masses of $\Omega^-$ and $\Lambda_c^+$, respectively. In \fig{Fig:omega_lambda_ms_mc} we show representative plots from the determination of $m_S^R$ and $m_c^R$. We obtain the following values in $\overline{\rm MS}$ at 2~GeV:
\bea \label{eq:strange_charm_values}
m_s^R &=& 92.4(6)(2.0) \rm MeV \nonumber\\
m_c^R &=& 1173.0(2.4)(17.0) \rm MeV\;. 
\eea
The error in the first parenthesis  is the statistical error on the fit parameters  and in the second parenthesis is the error associated with the tuning estimated 
by allowing the renormalized mass to vary within the statistical errors  of the $\Omega^-$ and $\Lambda_c^+$ mass at the physical point. The latter  systematic uncertainty due to the tuning will be included in the final errors we quote for the baryon masses. 
 In Ref.~\cite{Carrasco:2014cwa} the mass of the kaon and D-meson were used
to tune the strange and charm quark masses, obtaining $m_s^R=99.6(4.1)$~MeV and  $m_c^R=1176(36)$~MeV in $\overline{\rm MS}$ at 2~GeV, respectively, both in agreement with our values.
The corresponding plots of the chiral extrapolations for $\Omega^-$ ($\Lambda_c^+$) at the fixed value of the strange (charm) quark mass after correcting for cut-off effects are shown in \fig{Fig:chiral_omega_lambdac}, where  indeed all data fall on the same curve and the physical masses of the $\Omega^-$ and $\Lambda_c^+$ baryons are reproduced. The fit parameters $m_\Omega^{(0)}$, $c_{\Omega}^{(1)}$ and $c_i$ are collected in \tbl{Table:tuning_fitparams}. The results in lattice units and the continuum extrapolated values in physical units for $\Omega^-$ and $\Lambda_c^+$ are listed in \tbl{Table:omega_lambdac_masses}.

%
\begin{table}[h]
\begin{center}
\renewcommand{\arraystretch}{1.25}
\renewcommand{\tabcolsep}{5pt}
\begin{tabular}{l|ll|ll}
\hline\hline
$\; a\mu_l$ & $\quad\; am_\Omega$  & $\; m_\Omega$ (GeV) & $\quad\; am_{\Lambda_c^+}$  & $\; m_{\Lambda_c^+}$ (GeV)  \\
\hline\hline
\multicolumn{5}{c}{$\qquad\;\beta=1.90$} \\
\hline
0.0030 & 0.8380(77)   & 1.6575(609)   &  1.1651(157) &  2.3223(729)  \\
0.0040 & 0.8374(131)  & 1.6562(648)   &  1.1714(92)  &  2.3356(678)  \\
0.0050 & 0.8491(118)  & 1.6808(637)   &  1.1816(78)  &  2.3571(670)  \\
\hline\hline
\multicolumn{5}{c}{$\qquad\;\beta=1.95$}\\
\hline
0.0025 & 0.7484(60)   & 1.7111(535)   &  1.0236(52)  &  2.3523(584)  \\
0.0035 & 0.7406(72)   & 1.6924(544)   &  1.0261(45)  &  2.3581(581)  \\
0.0055 & 0.7477(67)   & 1.7093(540)   &  1.0434(43)  &  2.3997(580)  \\
0.0075 & 0.7409(62)   & 1.6931(536)   &  1.0468(53)  &  2.4077(585)  \\
\hline\hline
\multicolumn{5}{c}{$\qquad\;\beta=2.10$}\\
\hline
0.0015 & 0.5676(34)   & 1.6816(418)   &  0.7817(33)  &  2.3234(459)  \\
0.0020 & 0.5568(54)   & 1.6484(437)   &  0.7796(68)  &  2.3171(494)  \\
0.0030 & 0.5651(51)   & 1.6740(434)   &  0.7883(43)  &  2.3438(467)  \\
\hline\hline
\end{tabular}
\caption{Masses of the $\Omega$ and $\Lambda_c^+$ baryons in lattice and physical units with the associated statistical error. The values in physical units are continuum extrapolated.}
\label{Table:omega_lambdac_masses}
\end{center}
\vspace*{-.0cm}
\end{table}

Given the fact that we have performed a  high statistics run (see \tbl{Table:params}) using  $m_c^R=1186$ MeV, which was our first estimate for $m_c^R$ and since
this value is consistent with the final tuned value given in \eq{eq:strange_charm_values} we will use the high statistics results to obtain the values
of the charmed baryon masses at the physical point. We have checked  that interpolation of our lattice data for the charm baryons at the tuned value of $m_c^R=1173(2.4)$ yield masses at the physical point which are totally consistent with the ones obtained at $m_c^R=1186(2.4)$, albeit with larger  errors  due to the interpolation of the lattice results. Thus, we avoid interpolation and use  the  results obtained directly  at $m_c^R=1186$~MeV in what follows.
\begin{table}[h]
\begin{center}
\renewcommand{\arraystretch}{1.2}
\renewcommand{\tabcolsep}{7.5pt}
\begin{tabular}{l|c}
\hline\hline
\multicolumn{2}{c}{$\Omega^-$ (1.672)}\\
\hline
  $m_\Omega^{(0)}$ (GeV) 				&   1.669(19)      \\
\hline
$-4c_{\Omega}^{(1)}$ (GeV$^{-1}$)  &   0.161(124)     \\
\hline
$d$  (GeV$^3$)      &   0.466(123)   \\
\hline
$\chi^2$/d.o.f.            &   2.24  \\
\hline
$m$ (GeV)  &   1.672(18) \\
\hline\hline
\multicolumn{2}{c}{$\Lambda_c^+$ (2.286)} \\
\hline
$m_{\Lambda_c}^{(0)}$ (GeV)		               				&   2.272(26)       \\
\hline
$c_1$ (GeV$^{-1}$)					  		&   0.799(935)     \\
\hline
$c_2$ (GeV$^{-2}$)		 			  		&   -0.118(1.834) \\
\hline
$d$ (GeV$^3$) & 0.553(104) \\
\hline
$\chi^2$/d.o.f.              &   1.33 \\
\hline
$m$ (GeV)   &   2.286(17) \\
\hline\hline
\end{tabular}
\end{center}
\caption{Fit parameters and physical point values determined from the chiral fits to the $\Omega^-$ and $\Lambda_c^+$ using Eqns. (\ref{eq:omega_tuning}) and (\ref{eq:lambdac_tuning}) respectively.}
\label{Table:tuning_fitparams}
\end{table}
%
%


\section{Lattice Results}

Lattice results are obtained for three lattice spacings allowing to assess cut-off effects. We start by addressing any possible isospin breaking effects on the baryon masses. 

\subsection{Isospin symmetry breaking}

The twisted mass action breaks isospin explicitly to $\mathcal{O}(a^2)$ and the size of the $\mathcal{O}(a^2)$ terms determines how large this breaking is. 
Any isospin splitting should vanish  in the continuum limit. In general, isospin symmetry breaking manifests itself as a mass splitting among baryons belonging to the same multiplets. We note that there is still a symmetry when interchanging a u- with a d-quark, which means for example that the proton and the neutron are still degenerate as are  the $\Delta^{++}$ and the $\Delta^{-}$ as well as the $\Delta^+$ and $\Delta^0$. However, mass splitting could be seen between the $\Delta^{++}$ and the $\Delta^+$. Also, isospin breaking effects maybe present in the hyperons and charmed baryons in particular given that we consider only the  $s^+$ and $c^+$, as explained in section II.A.

\begin{figure}[!ht]\vspace*{-0.2cm}
\center
{\includegraphics[width=0.45\textwidth]{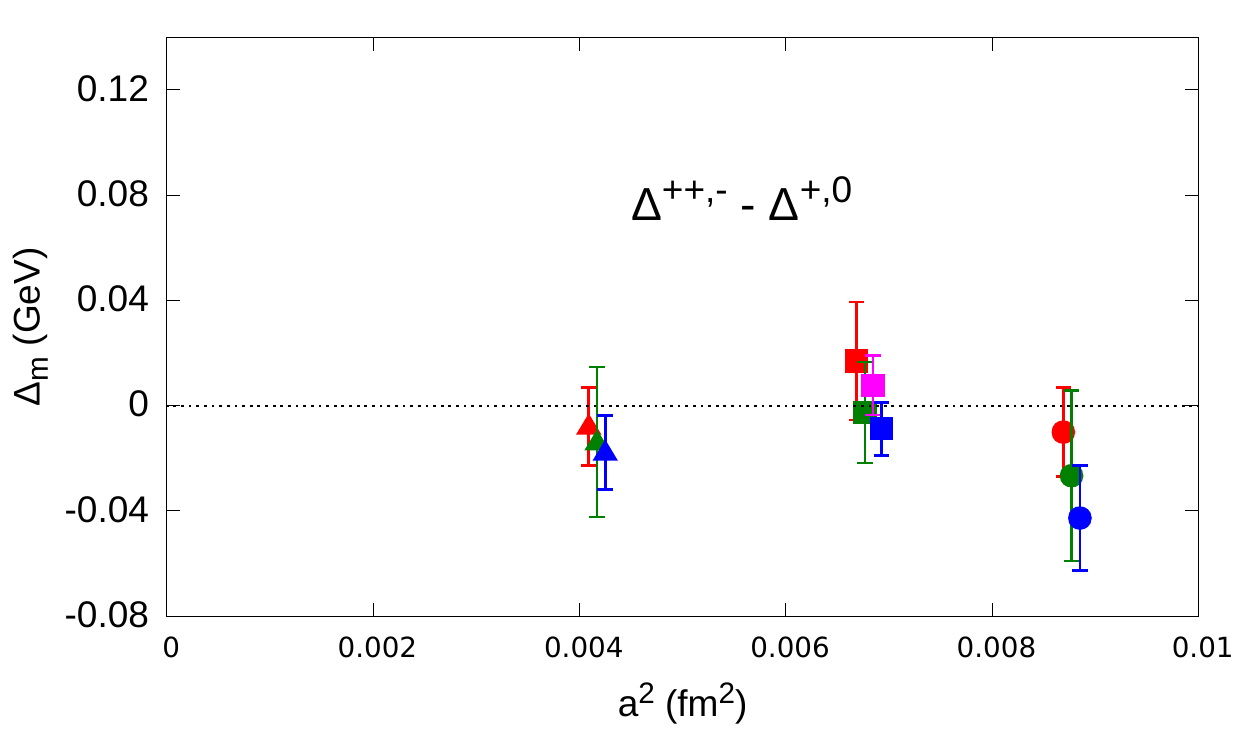}}
\caption{Mass differences for the $\Delta$ baryons for our three lattice spacings (circles for $\beta=1.90$, squares for $\beta=1.95$ and triangles for $\beta=2.10$) examined and for all pion masses. Symbols for each lattice spacing have been shifted to the left and right for clarity. Red symbols represent the lightest pion mass and blue symbols the heaviest pion mass for each lattice spacing. For $\beta=1.95$, the green symbol is the second lightest pion mass and the magenta symbol is the second heaviest pion mass.}
\label{Fig:delta_massdiff}
\end{figure}

\begin{figure}[!ht]\vspace*{-0.2cm}
\center
\begin{minipage}{8cm}
{\includegraphics[width=0.8\textwidth]{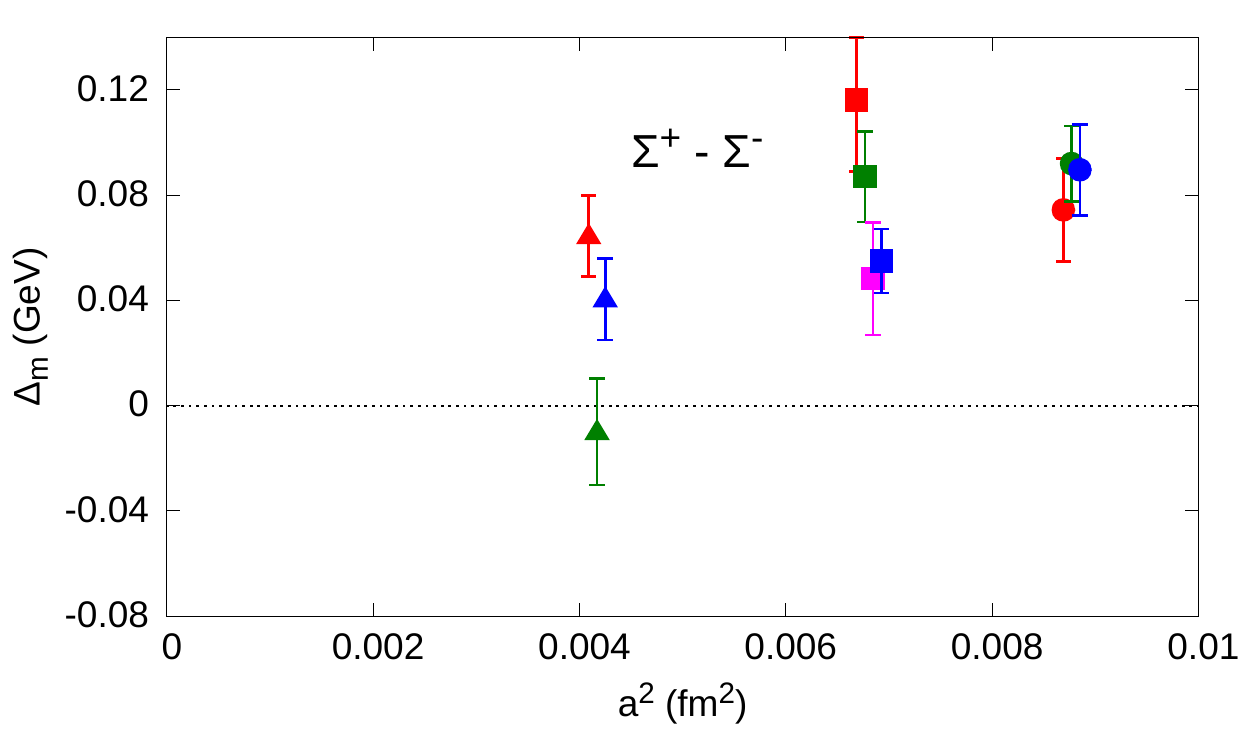}}
\end{minipage}\hfill
\begin{minipage}{8cm}
{\includegraphics[width=0.8\textwidth]{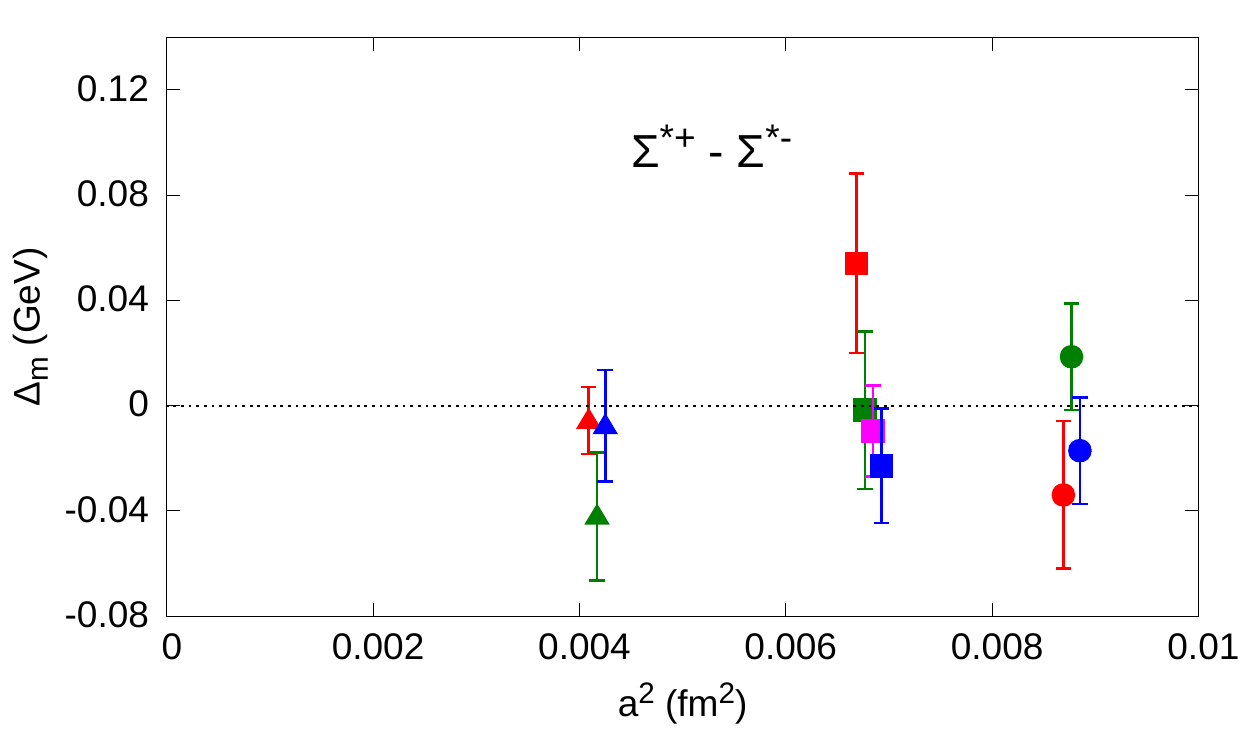}}
\end{minipage}
\begin{minipage}{8cm}
{\includegraphics[width=0.8\textwidth]{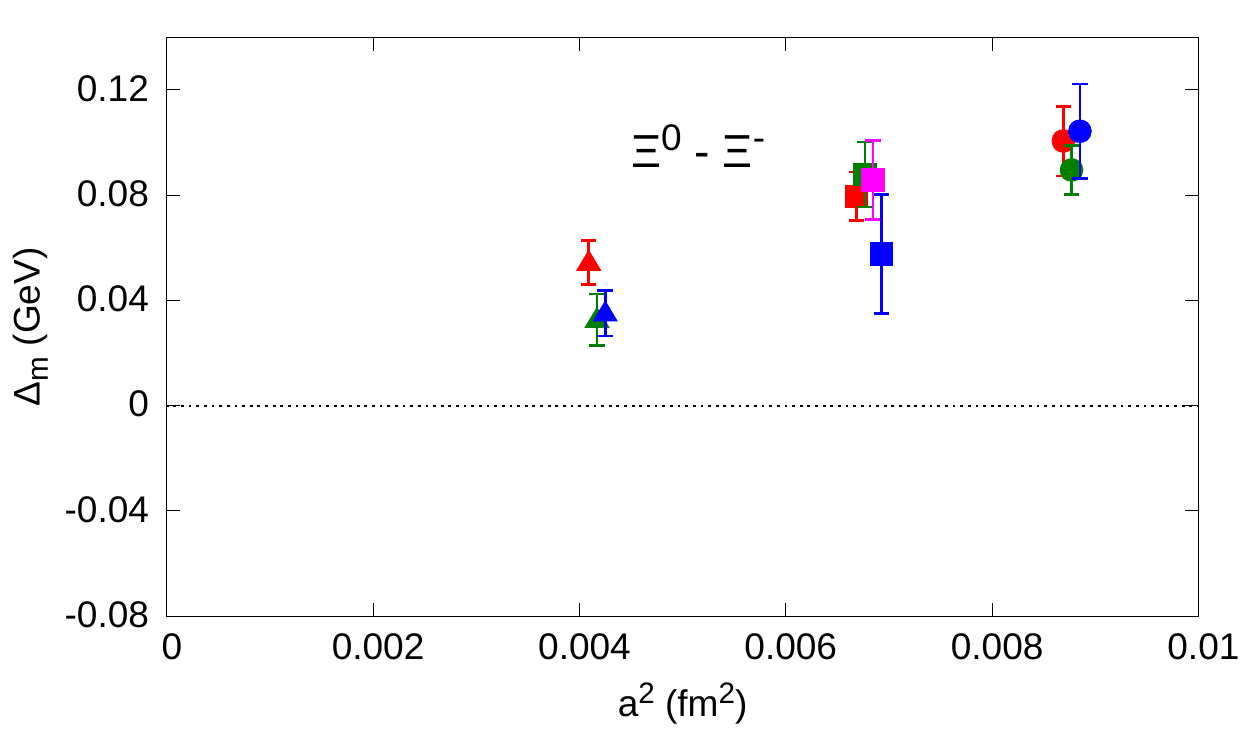}}
\end{minipage}\hfill
\begin{minipage}{8cm}
{\includegraphics[width=0.8\textwidth]{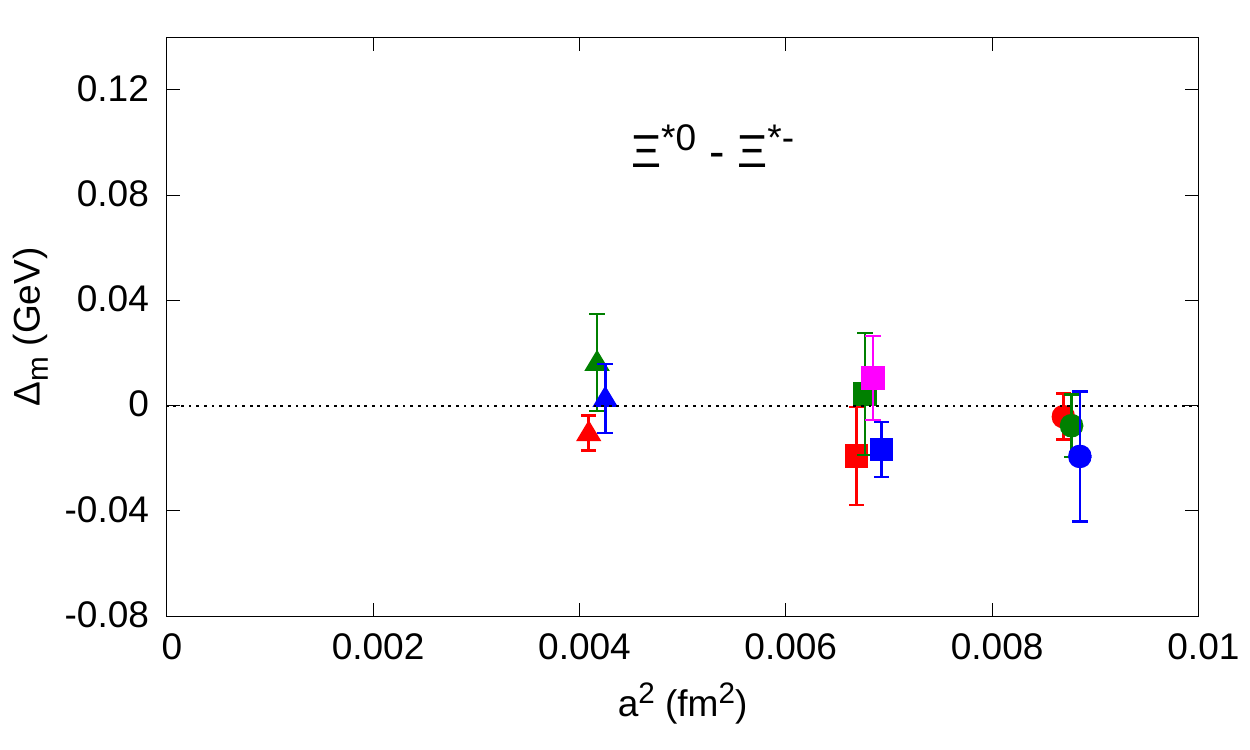}}
\end{minipage}
\caption{Mass differences for the octet (left) and decuplet (right) hyperons for our three lattice spacings examined. Small non-zero mass differences are observed for the octet hyperons. The symbol notation is as in \fig{Fig:delta_massdiff}.}
\label{Fig:strange_massdiff}
\end{figure}

We begin this analysis by plotting the mass difference as a function of $a^2$ for the $\Delta$ baryons. We average over $\Delta^{++}$ and $\Delta^-$ as well as over $\Delta^+$ and $\Delta^0$ and take the difference between the two averages. The corresponding plot is shown in \fig{Fig:delta_massdiff}, where as one can see, the mass difference is consistent with zero, indicating that isospin breaking effects are small for the $\Delta$ baryons at the $\beta$ values analysed. We also examine the mass difference of the strange baryons in \fig{Fig:strange_massdiff}. We observe that the mass difference between the $\Sigma^+$ and $\Sigma^-$ and between the $\Xi^0$ and $\Xi^-$ are indeed decreasing linearly with $a^2$ being almost zero at our smallest lattice spacing. For the strange spin-3/2 baryons the results are fully consistent with zero at all lattice spacings. 

\begin{figure}[!ht]\vspace*{-0.2cm}
\center
\begin{minipage}{8cm}
{\includegraphics[width=0.8\textwidth]{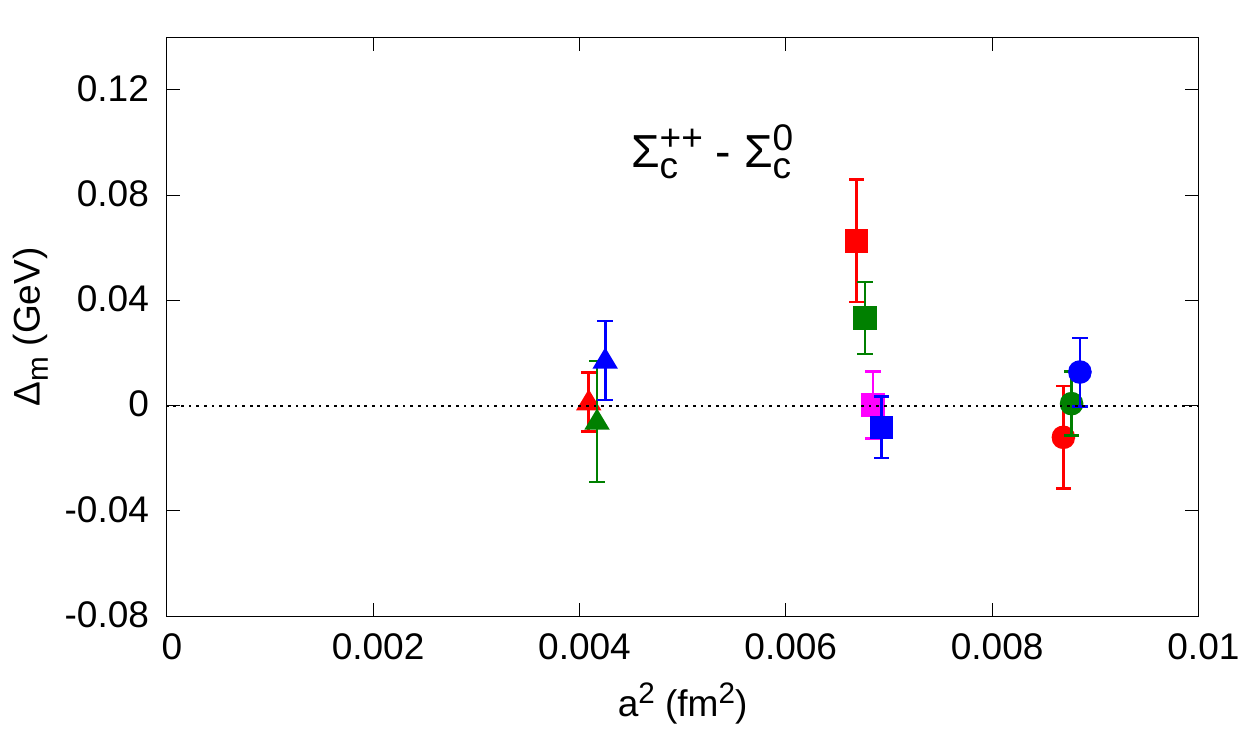}}
\end{minipage}\hfill
\begin{minipage}{8cm}
{\includegraphics[width=0.8\textwidth]{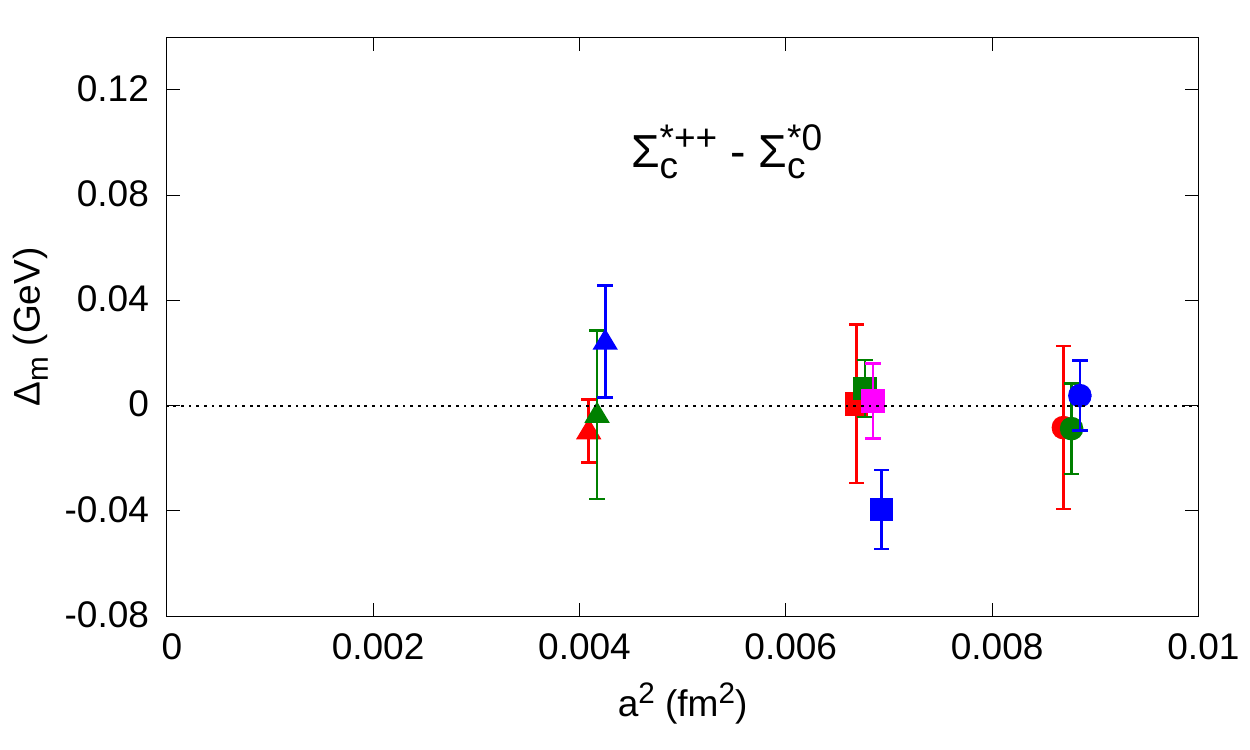}}
\end{minipage}
\begin{minipage}{8cm}
{\includegraphics[width=0.8\textwidth]{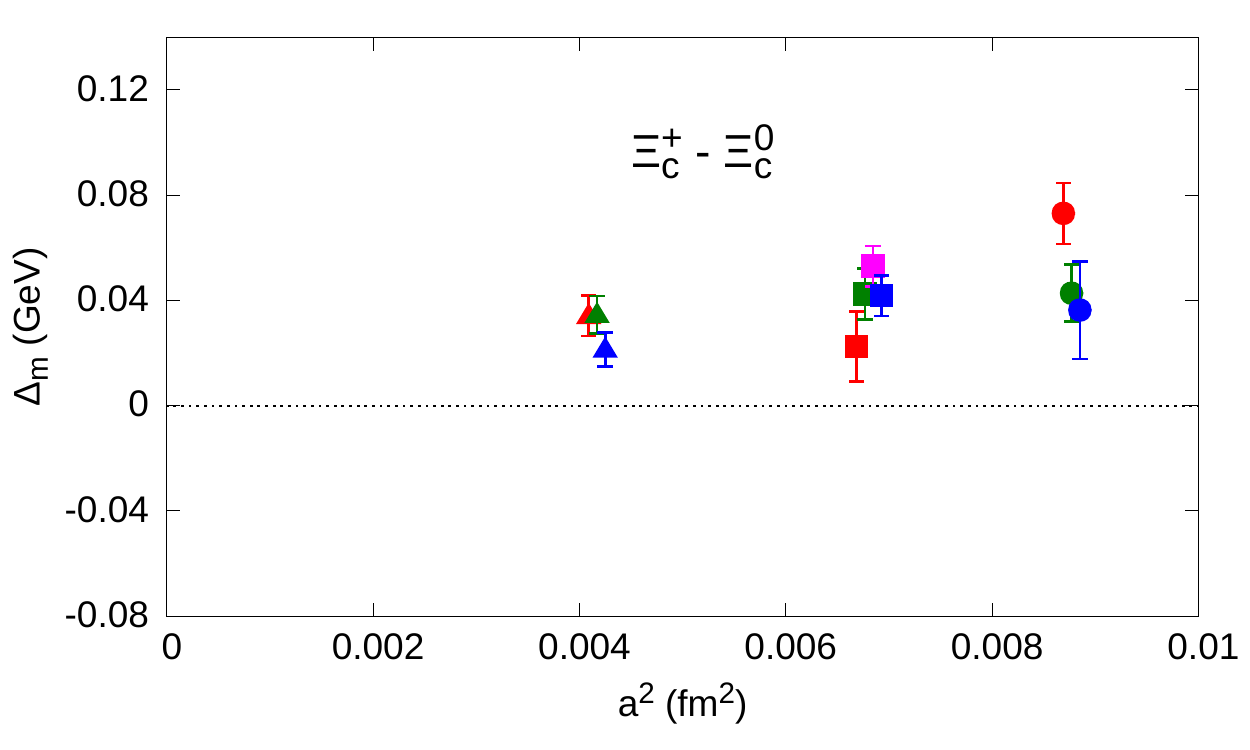}}
\end{minipage}\hfill
\begin{minipage}{8cm}
{\includegraphics[width=0.8\textwidth]{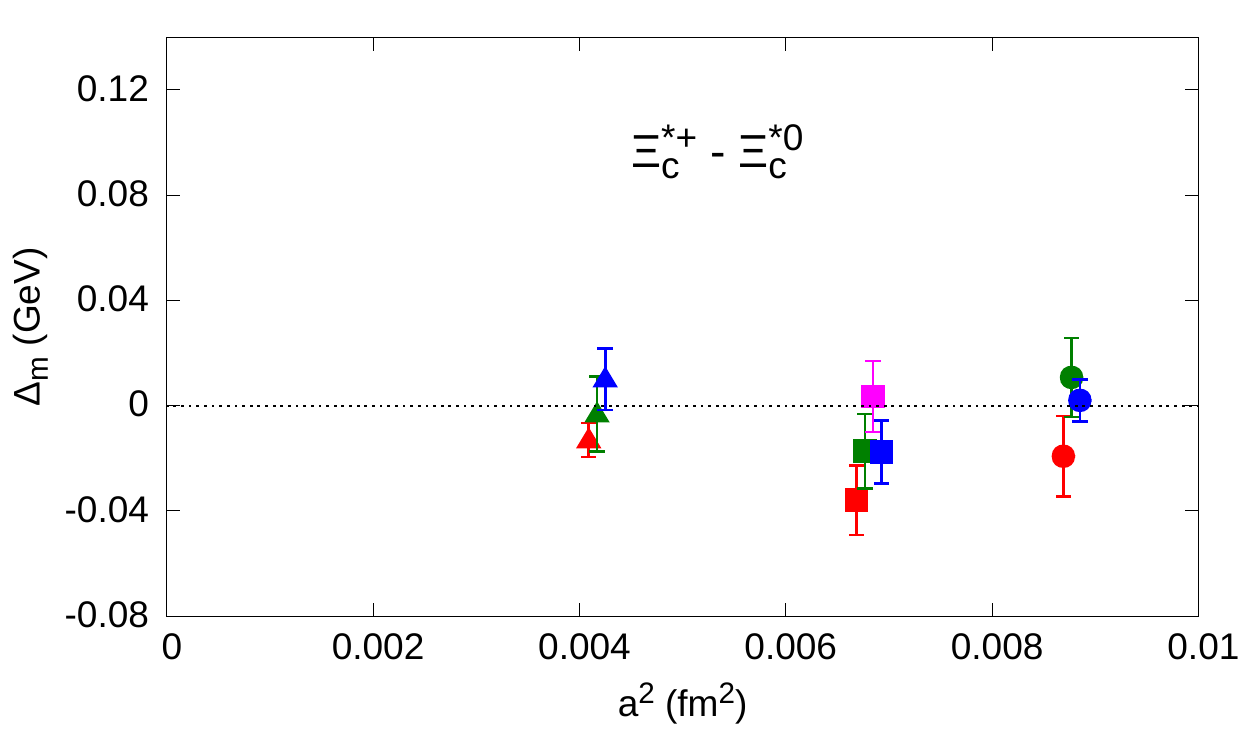}}
\end{minipage}
\begin{minipage}{8cm}
{\includegraphics[width=0.8\textwidth]{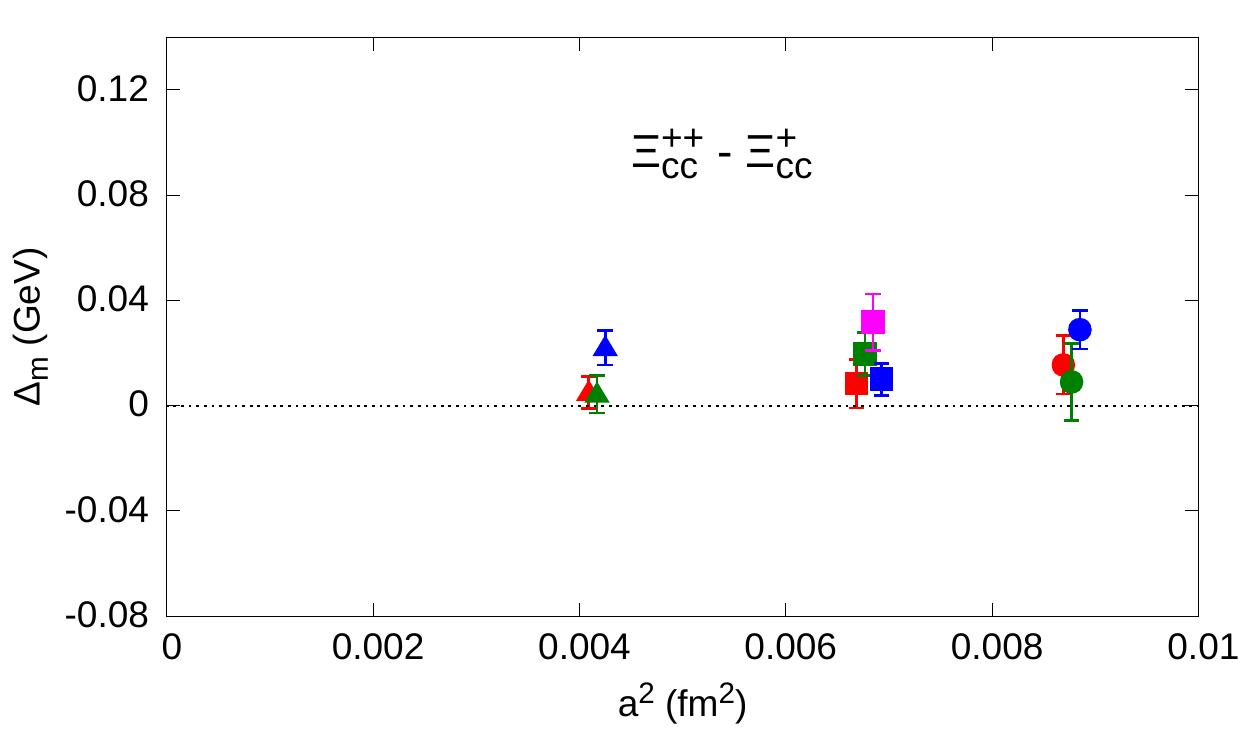}}
\end{minipage}\hfill
\begin{minipage}{8cm}
{\includegraphics[width=0.8\textwidth]{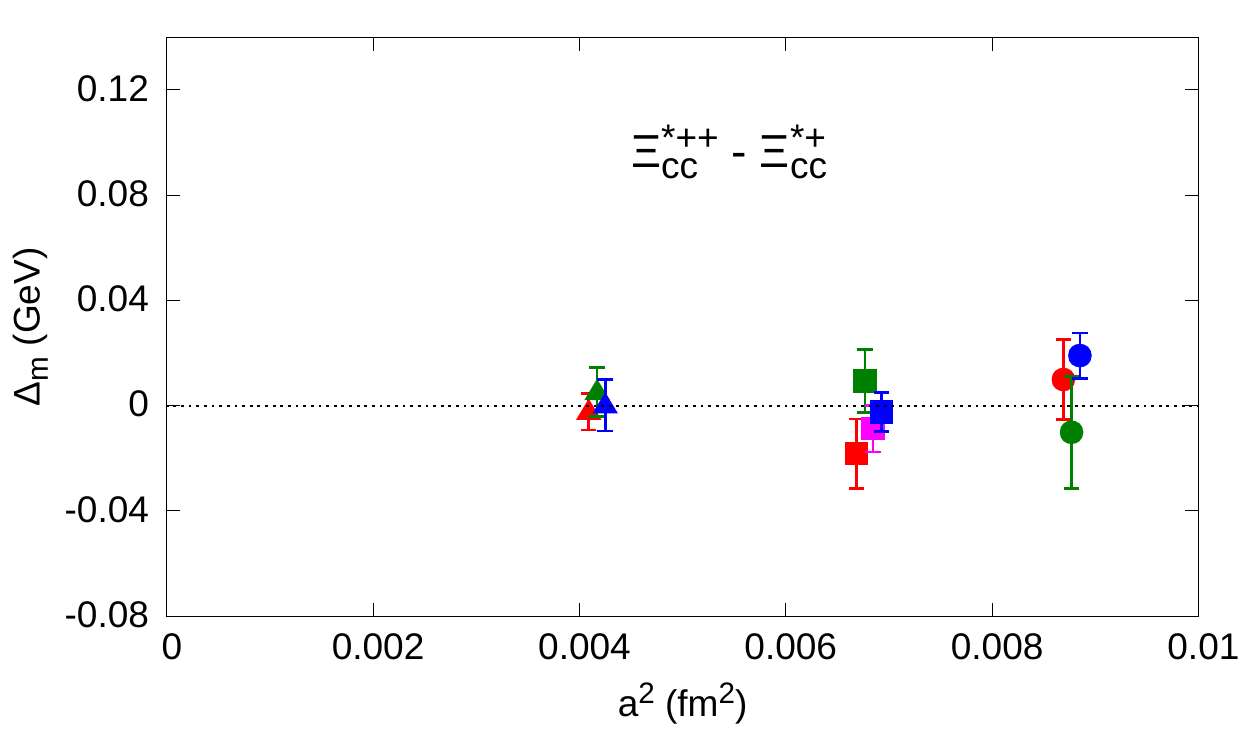}}
\end{minipage}
\begin{minipage}{8cm}
\center
{\includegraphics[width=0.8\textwidth]{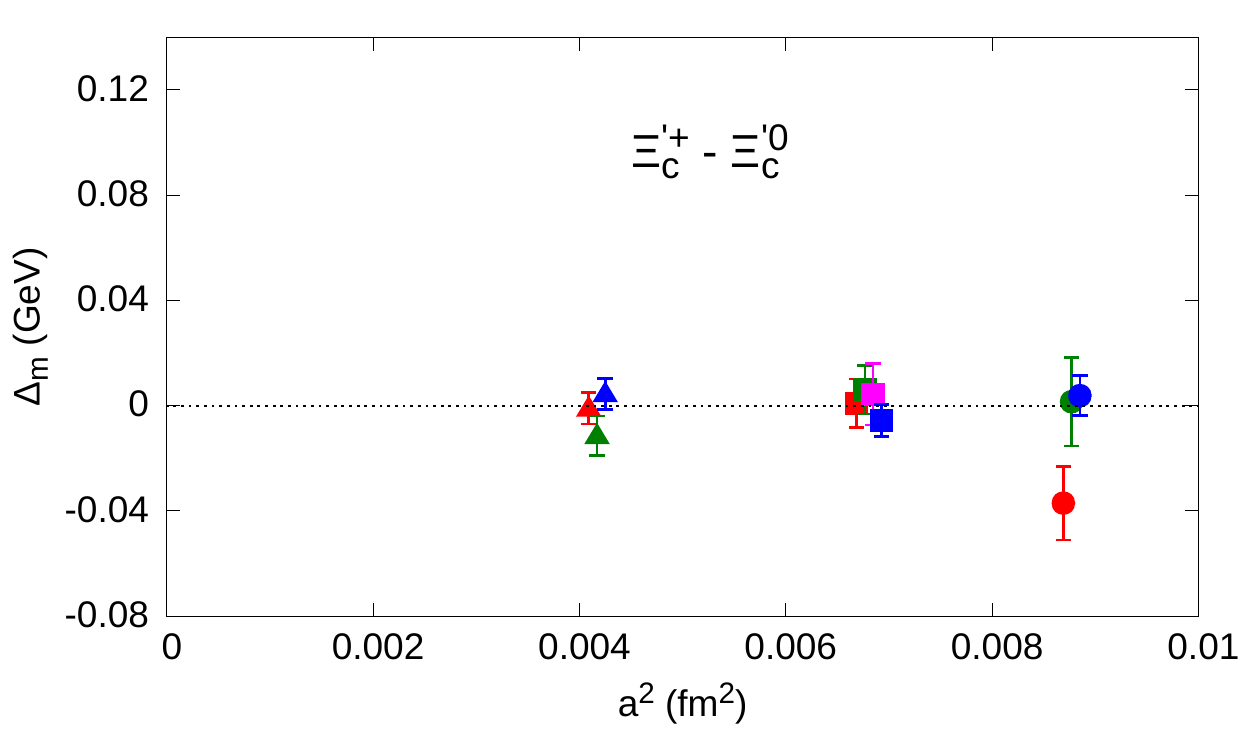}}
\end{minipage}
\caption{Mass differences between the charm baryons belonging to the same isospin multiplets for the three lattice spacings. Small non-zero differences which are reduced as the lattice spacing gets smaller are seen between the $\Xi_c$ states. The notation is the same as that in \fig{Fig:delta_massdiff}.}
\label{Fig:charm_massdiff}
\end{figure}

We continue our analysis by studying the isospin breaking for the charm baryons. We show in \fig{Fig:charm_massdiff}  the mass difference between the  $\Sigma_c$, $\Xi_c$ and $\Xi_{cc}$ multiplets at the three lattice spacings for all pion masses considered in this work. As in the strange sector, non-zero values are obtained at the largest lattice spacing, which do not exceed 3\% the average mass of these baryons. As expected, the mass splitting  vanishes as the continuum limit is approached. In the same figure we also show the mass difference between $\Xi_c^{\prime +}$ and $\Xi_c^{\prime 0}$, which is consistent with zero indicating that isospin breaking effects are small at all values of the lattice spacing.
As in the case of the strange decuplet, the isospin splitting for the charmed spin-3/2 baryons is consistent with zero.

Having several pion masses at a given lattice spacing one can ask how the isospin mass splitting depends on the pion mass.
As shown in Figs. \ref{Fig:strange_massdiff} and \ref{Fig:charm_massdiff}, the baryon mass differences are independent of the light quark mass to the present 
accuracy of our results.

%


\section{Chiral and continuum extrapolation}

In order to extrapolate our lattice results to the physical pion mass
we allow for cut-off effects by including a term quadratic in the lattice spacing and then apply continuum chiral perturbation theory at our results. 

For the strange baryon sector we consider SU(2) heavy baryon chiral perturbation theory (HB$\chi$PT). The same expressions were used in other twisted mass fermion studies \cite{Alexandrou:2008tn,Alexandrou:2009qu,Alexandrou:2012xk} and were found to describe lattice data satisfactory. The leading one-loop results for the octet and decuplet baryons \cite{Nagels:1979xh,Nagels:1978sc} are given by
\bea \label{eq:octet_expressions_lo}
	m_\Lambda (m_\pi) &=& m_\Lambda^{(0)} - 4c_\Lambda^{(1)}m_\pi^2 - \frac{g_{\Lambda\Sigma}^2}{16\pi f_\pi^2} m_\pi^3 \nonumber\\
	m_\Sigma (m_\pi)    &=& m_\Sigma^{(0)}   - 4c_\Sigma^{(1)}m_\pi^2    - \frac{2g_{\Sigma\Sigma}^2+g_{\Lambda\Sigma}^2/3}{16\pi f_\pi^2} m_\pi^3 \nonumber\\
	m_\Xi (m_\pi)          &=& m_\Xi^{(0)}          - 4c_\Xi^{(1)}m_\pi^2          - \frac{3g_{\Xi\Xi}^2}{16\pi f_\pi^2} m_\pi^3
\eea
for the octet baryons and
\bea \label{eq:decuplet_expressions_lo}
	m_\Delta (m_\pi)        &=& m_\Delta^{(0)}        - 4c_\Delta^{(1)}m_\pi^2         - \frac{25}{27}\frac{g_{\Delta\Delta}^2}{16\pi f_\pi^2} m_\pi^3 \nonumber\\
	m_{\Sigma^*} (m_\pi) &=& m_{\Sigma^*}^{(0)} - 4c_{\Sigma^*}^{(1)}m_\pi^2  - \frac{10}{9}  \frac{g_{\Sigma^*\Sigma^*}^2}{16\pi f_\pi^2} m_\pi^3 \nonumber\\
	m_{\Xi^*} (m_\pi)       &=& m_{\Xi^*}^{(0)}        - 4c_{\Xi^*}^{(1)}m_\pi^2         - \frac{5}{3}   \frac{g_{\Xi^*\Xi^*}^2}{16\pi f_\pi^2} m_\pi^3 \nonumber\\
	m_{\Omega} (m_\pi)       &=& m_{\Omega}^{(0)}        - 4c_{\Omega}^{(1)}m_\pi^2
\eea
for the decuplet baryons.
In addition we consider the next-to-leading order SU(2) $\chi$PT results \cite{Tiburzi:2008bk}. For completeness, we include the expressions in Appendix \ref{App:chiral_expressions_nlo}.

 We fix the nucleon axial charge $g_A$ and pion decay constant $f_\pi$ to their experimental values (we use the convention such that $f_\pi=0.092419(7)(25)$ GeV) as was done in the case of determining  the lattice spacings from fitting the nucleon mass. The remaining pion-baryon axial coupling constants are taken from the following SU(3) relations~\cite{Tiburzi:2008bk}:
\renewcommand{\arraystretch}{1.4}
\be\label{eq:gA_all}
\begin{array}{lllll}
{\rm Octet:}        &    g_A=D+F & g_{\Sigma\Sigma}=2F, & g_{\Xi\Xi} = D-F, & g_{\Lambda\Sigma}=2D \\
{\rm Decuplet:}   &	g_{\Delta\Delta}=\mathcal{H}, & g_{\Sigma^*\Sigma^*}=\frac{2}{3}\mathcal{H}, & g_{\Xi^*\Xi^*}=\frac{1}{3}\mathcal{H} &  \\
{\rm Transition:} & g_{\Delta N}=\mathcal{C},  & g_{\Sigma^*\Sigma}=\frac{1}{\sqrt{3}}\mathcal{C}, & g_{\Xi^*\Xi} = \frac{1}{\sqrt{3}}\mathcal{C}, & g_{\Lambda\Sigma^*} = -\frac{1}{\sqrt{2}}\mathcal{C} \\
\end{array}
\ee
In the octet case, once $g_A$ is fixed, the axial coupling constants depend on a single parameter $\alpha$ such that $\alpha=\frac{D}{D+F}$. Its value is poorly known. It can be taken either from the quark model ($\alpha = 3/5$), from the phenomenology of semi-leptonic decays or from hyperon-nucleon scattering. As in Ref.~\cite{Tiburzi:2008bk}, we take $\alpha=0.58$ or $2D=1.47$. The axial couplings in the decuplet case depend only on $\mathcal{H}$ for which we take the value $\mathcal{H}=2.2$, again from Ref.~\cite{Tiburzi:2008bk}. This value is close to the prediction by SU(6), namely $\mathcal{H}=\frac{9}{5}g_A=2.29$. The latter was used in a previous work \cite{Alexandrou:2008tn}, resulting in the same cubic term for the nucleon and $\Delta$. When fixing the octet-decuplet transition couplings we take $\mathcal{C}=1.48$ from Ref. \cite{Tiburzi:2005na}.
Having fixed the coupling constants this way, the LO, the one-loop as well as the NLO expressions are left with $m_X^{(0)}$ and $c_X^{(1)}$ as independent fit parameters. Unlike in Ref. \cite{Tiburzi:2008bk} where a universal mass parameter $m_X^{(0)}$ was used for all baryons with the same strangeness, in this work we treat all mass parameters $m_X^{(0)}$ independently.
The chiral extrapolation is applied to the  average over all states belonging to the same isospin multiplets, except for the charged states of the $\Sigma$, $\Xi$ and $\Xi_c$ where small non zero mass differences exist due to isospin breaking effects. For these particles we first extrapolate to the continuum limit 
to ensure that they are degenerate and then take the average of their continuum values.

\begin{figure}[!ht]\vspace*{-0.2cm}
\center
\begin{minipage}{8cm}
{\includegraphics[width=0.9\textwidth]{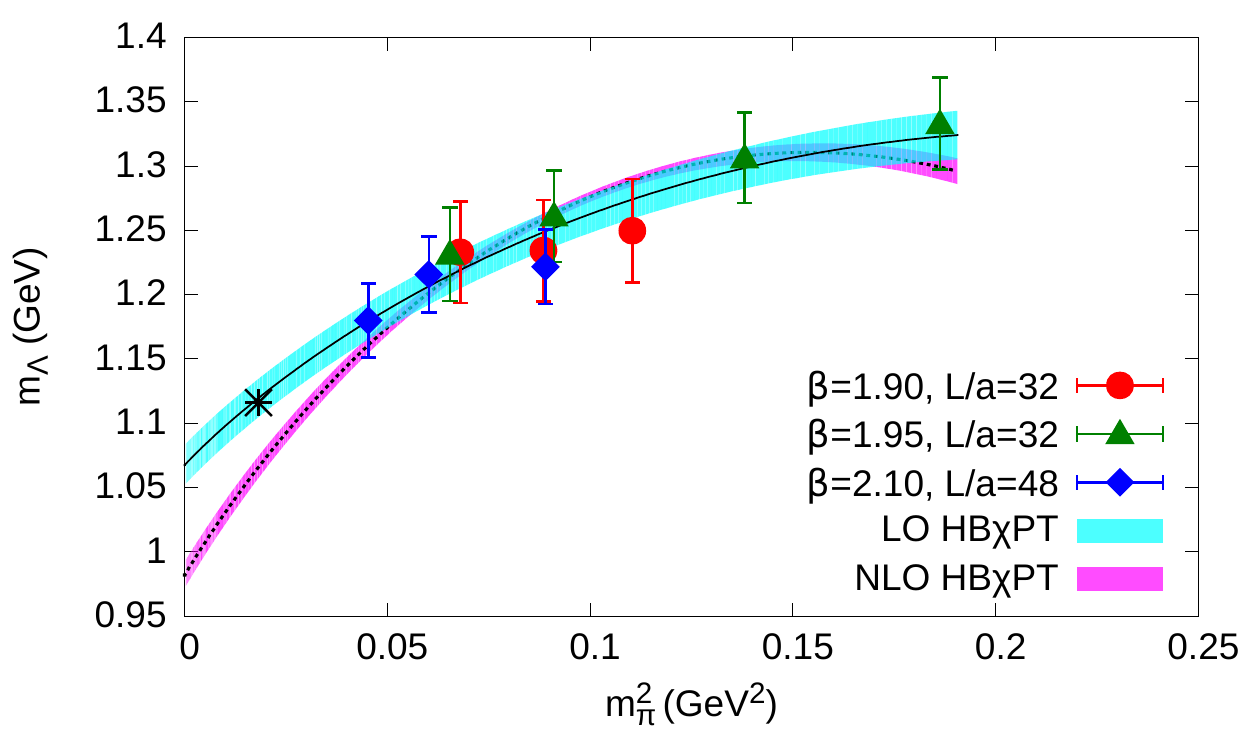}}
\end{minipage}\hfill
\begin{minipage}{8cm}
{\includegraphics[width=0.9\textwidth]{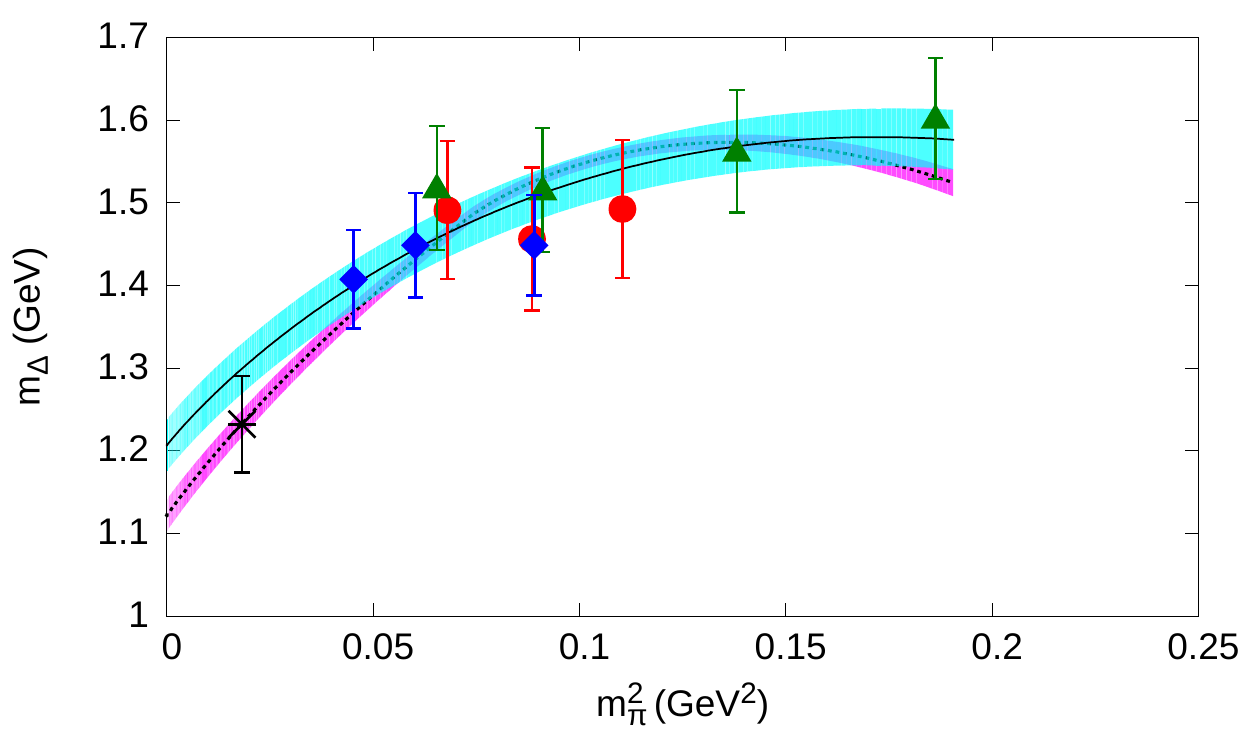}}
\end{minipage}
\begin{minipage}{8cm}
{\includegraphics[width=0.9\textwidth]{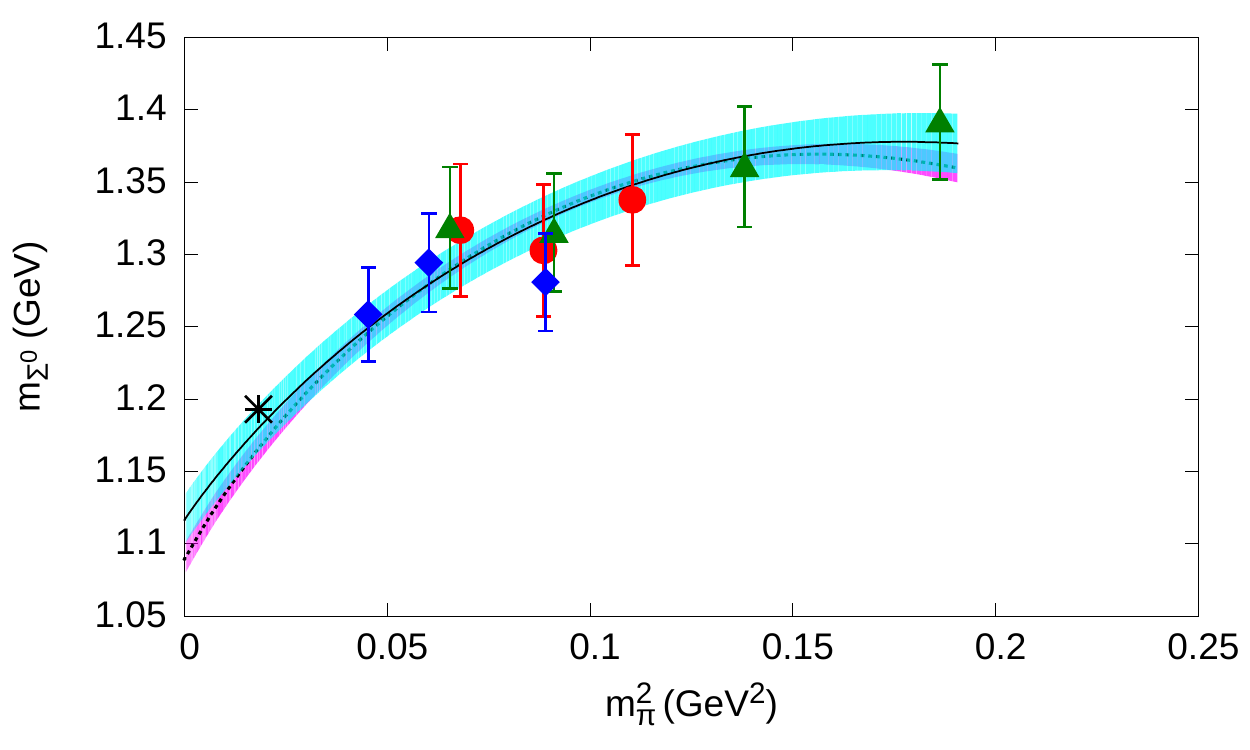}}
\end{minipage}\hfill
\begin{minipage}{8cm}
{\includegraphics[width=0.9\textwidth]{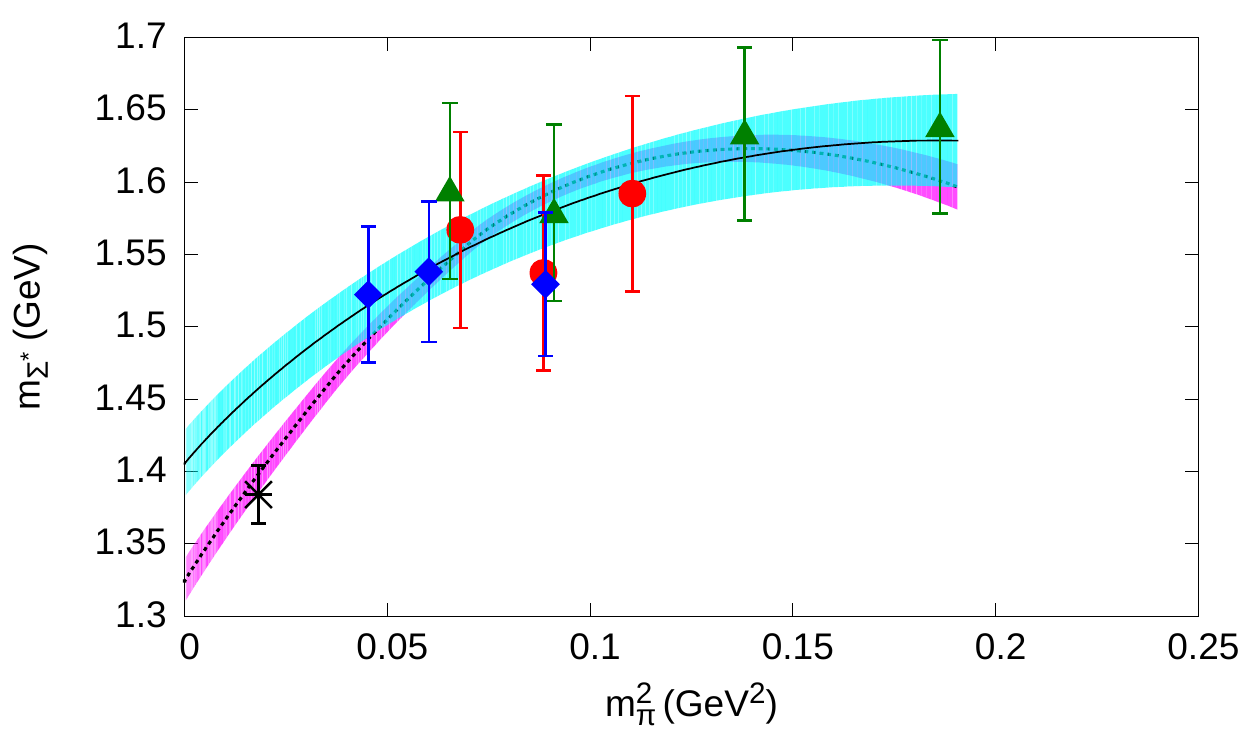}}
\end{minipage}
\begin{minipage}{8cm}
{\includegraphics[width=0.9\textwidth]{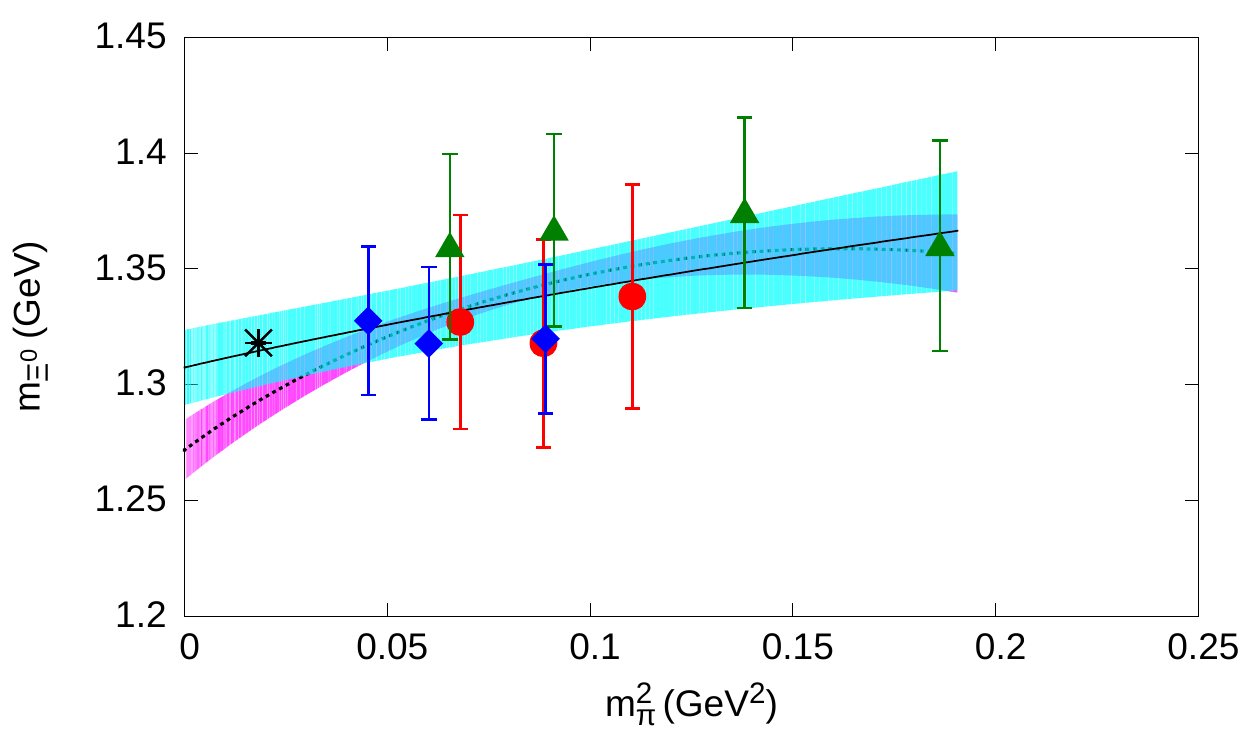}}
\end{minipage}\hfill
\begin{minipage}{8cm}
{\includegraphics[width=0.9\textwidth]{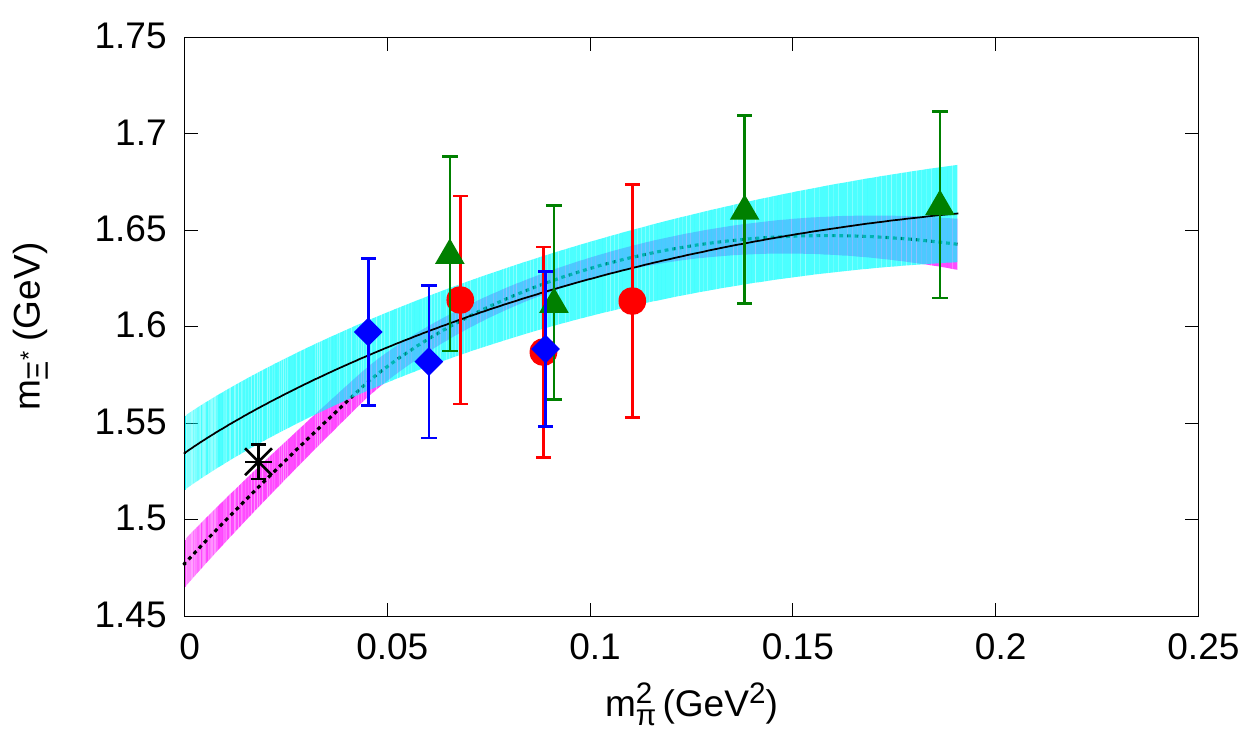}}
\end{minipage}
\caption{Chiral extrapolations of the octet (left) and decuplet (right) baryons in physical units, using the leading one-loop expressions of Eqns. (\ref{eq:octet_expressions_lo}) and (\ref{eq:decuplet_expressions_lo}) respectively as well as the NLO expressions of Eqns. (\ref{eq:octet_expressions_nlo}) and (\ref{eq:decuplet_expressions_nlo}). The lattice values  are continuum extrapolated.The  notation is given in the legend  in the top left plot. The experimental value is shown with the black asterisk.}
\label{Fig:octet_decuplet_chiral}
\end{figure}

We give the fit parameters extracted from fitting our lattice results for the octet and decuplet baryons to the leading one-loop order (Eqns. (\ref{eq:octet_expressions_lo}) and (\ref{eq:decuplet_expressions_lo})) and NLO (Eqns. (\ref{eq:octet_expressions_nlo}) and (\ref{eq:decuplet_expressions_nlo})) in \tbl{Table:fit_params_810}. We also show the baryon masses at the physical point obtained from the leading order fits in \tbl{Table:phys_masses}. The lattice results for the octet and decuplet baryons at the three $\beta$ values are collected in Appendix. \ref{App:numerical_results}. The deviation of the values obtained at the physical pion mass from the two fitting procedures provide an estimate of the systematic error due to the chiral extrapolation. This error on the  masses  is given in the second parenthesis  in \tbl{Table:phys_masses}. Since for the $\Omega$ the LO and NLO expressions have no difference, we do not quote a systematic error due to the chiral extrapolation. We show representative plots of the chiral fits for the octet and decuplet baryons in \fig{Fig:octet_decuplet_chiral}. Our results shown here are continuum extrapolated and thus the errors on the points are larger than those on the raw data. The error band for the leading one-loop order and NLO fits are constructed using the super-jackknife procedure~\cite{Bratt:2010jn}.
As can be seen, the data are well described by the LO fits and the physical masses of $\Lambda$, $\Sigma^0$ and $\Xi^0$ are reproduced. For the $\Delta$ and $\Xi^*$ the physical point is missed by about one standard deviation, while the results for $\Sigma^*$ extrapolate to a $5$\% higher value. The NLO fits also describe the lattice data satisfactory but in general extrapolate to a lower value at the physical point. Taking the difference between the value found using the LO and NLO expressions we estimate the systematic error due to the chiral extrapolation, and this yields agreement with the experimental values also in the cases of $\Delta$, $\Sigma^*$ and $\Xi^*$.

\begin{figure}[!ht]\vspace*{-0.2cm}
\center
\begin{minipage}{8cm}
{\includegraphics[width=0.9\textwidth]{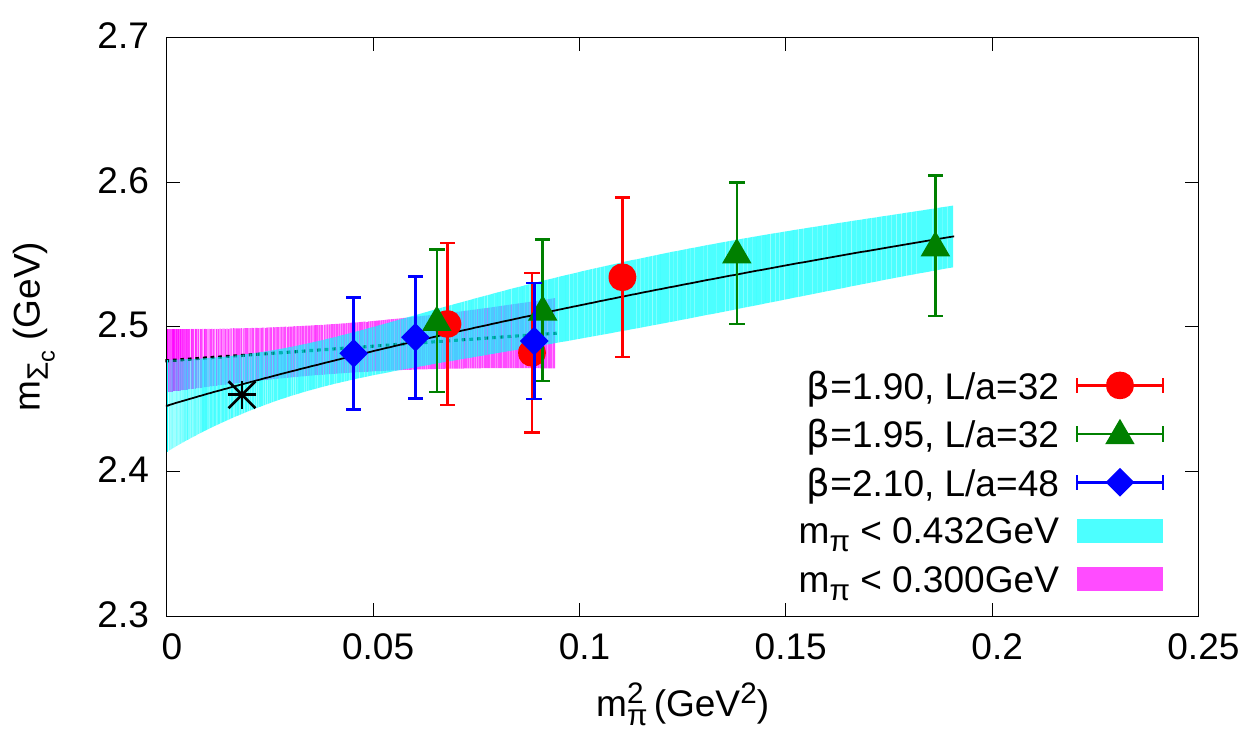}}
\end{minipage}\hfill
\begin{minipage}{8cm}
{\includegraphics[width=0.9\textwidth]{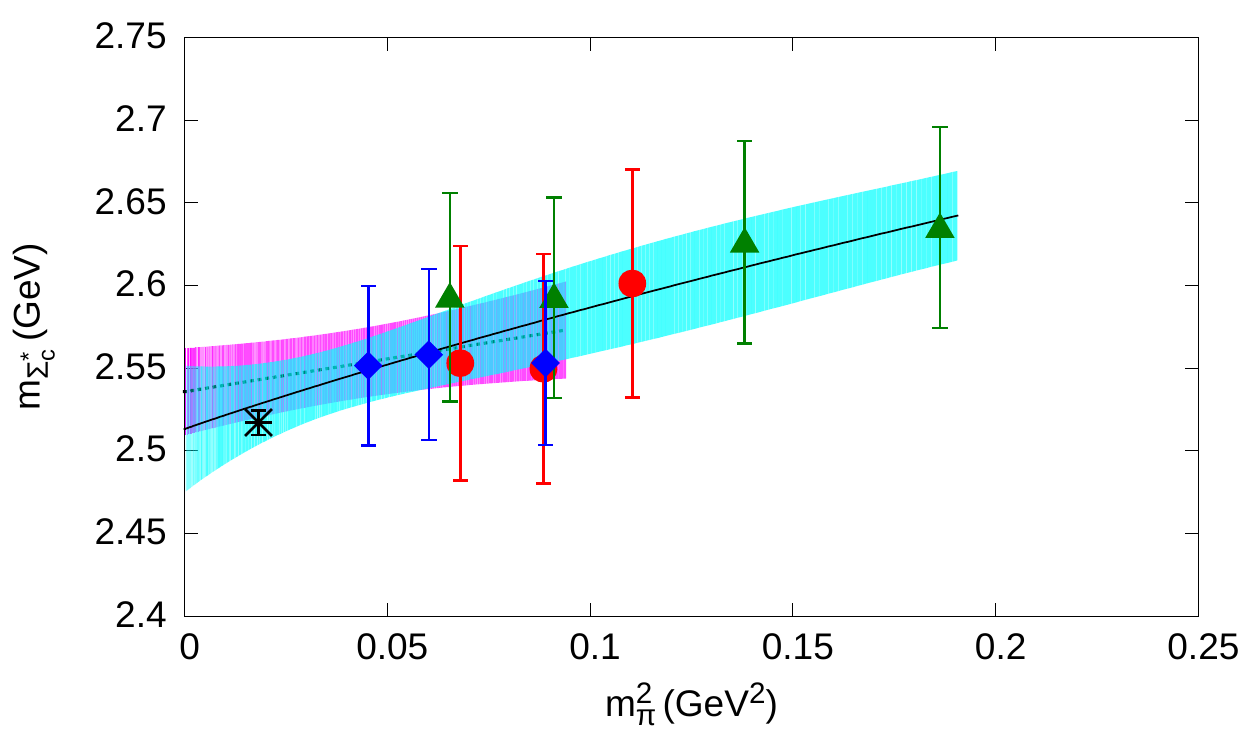}}
\end{minipage}
\begin{minipage}{8cm}
{\includegraphics[width=0.9\textwidth]{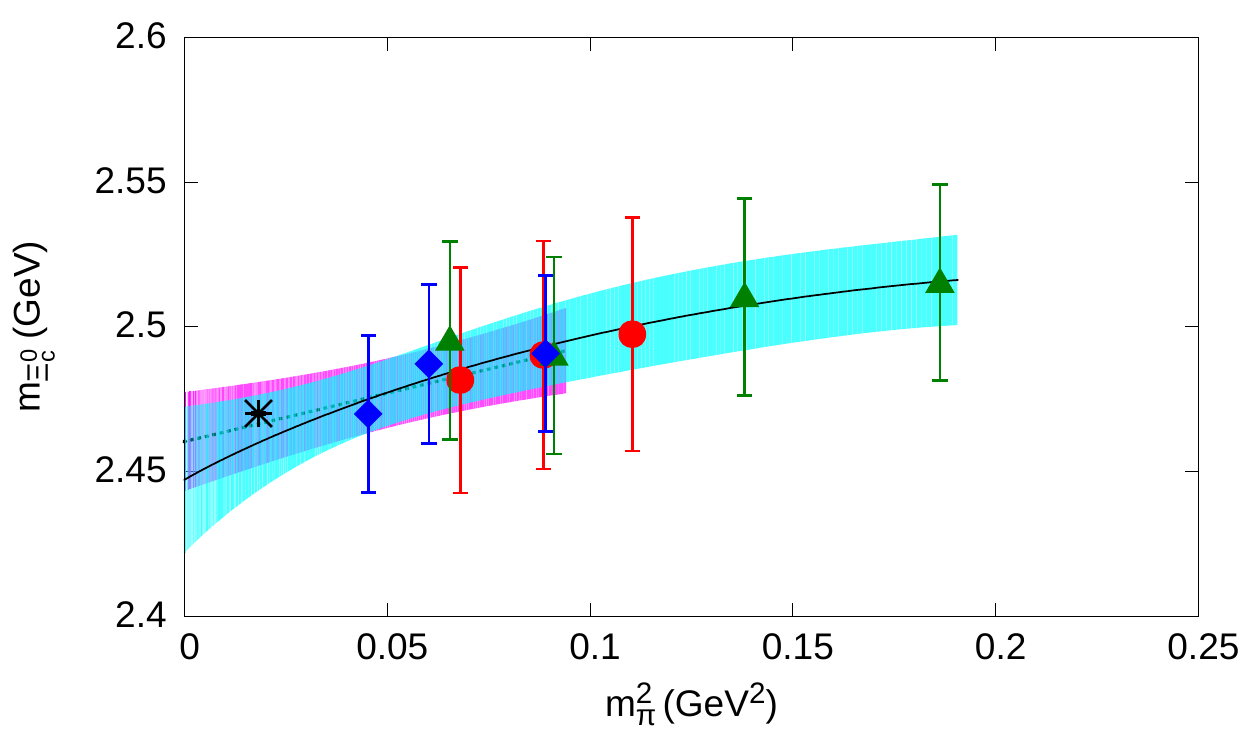}}
\end{minipage}\hfill
\begin{minipage}{8cm}
{\includegraphics[width=0.9\textwidth]{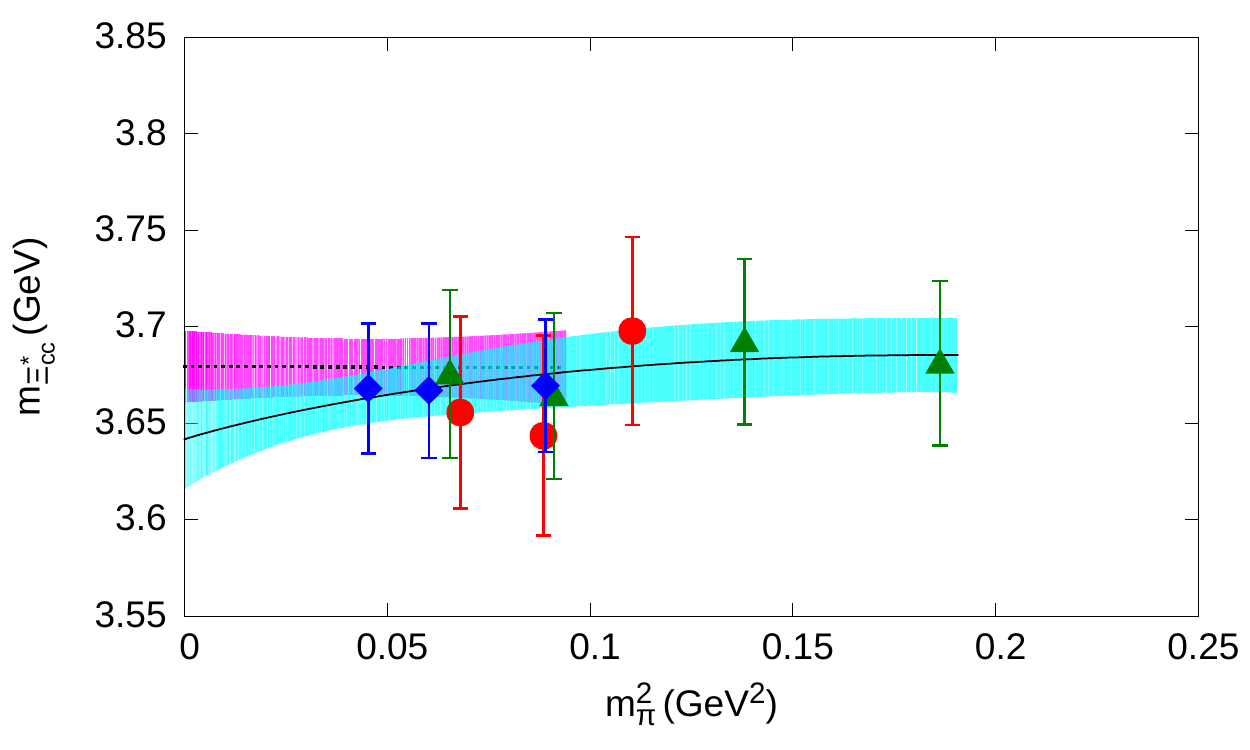}}
\end{minipage}
\begin{minipage}{8cm}
{\includegraphics[width=0.9\textwidth]{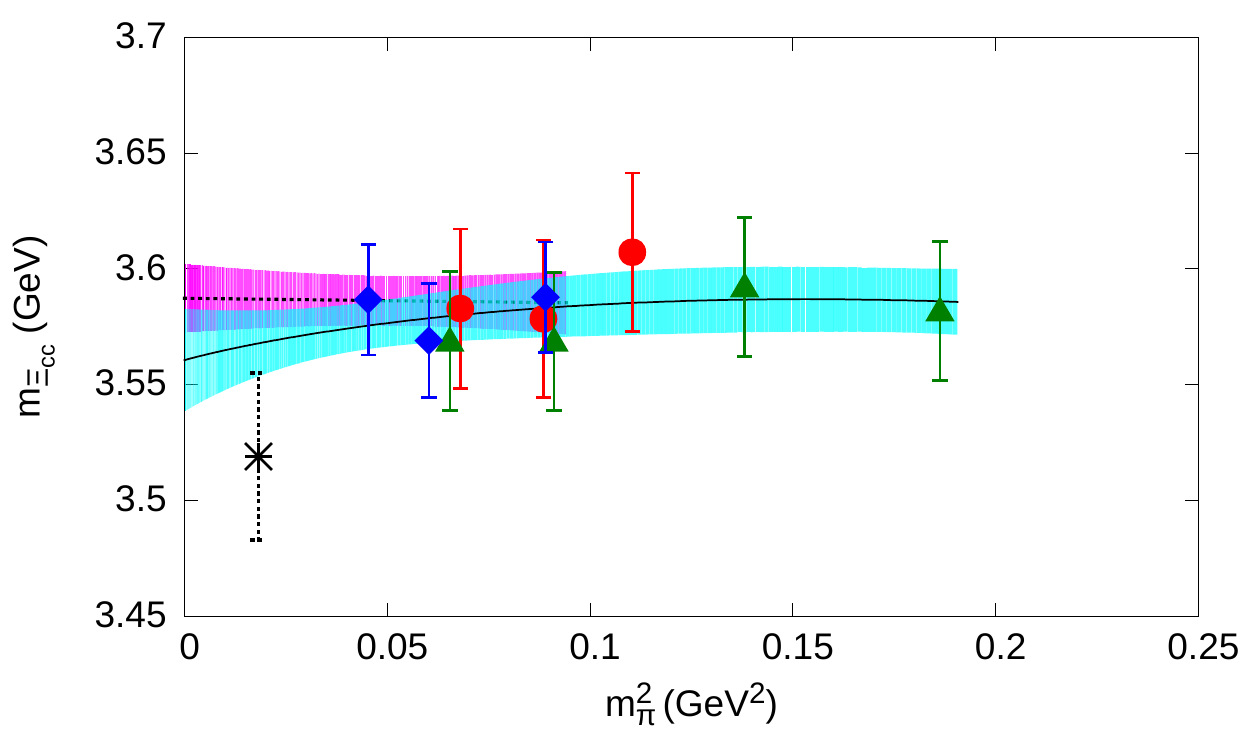}}
\end{minipage}\hfill
\begin{minipage}{8cm}
{\includegraphics[width=0.9\textwidth]{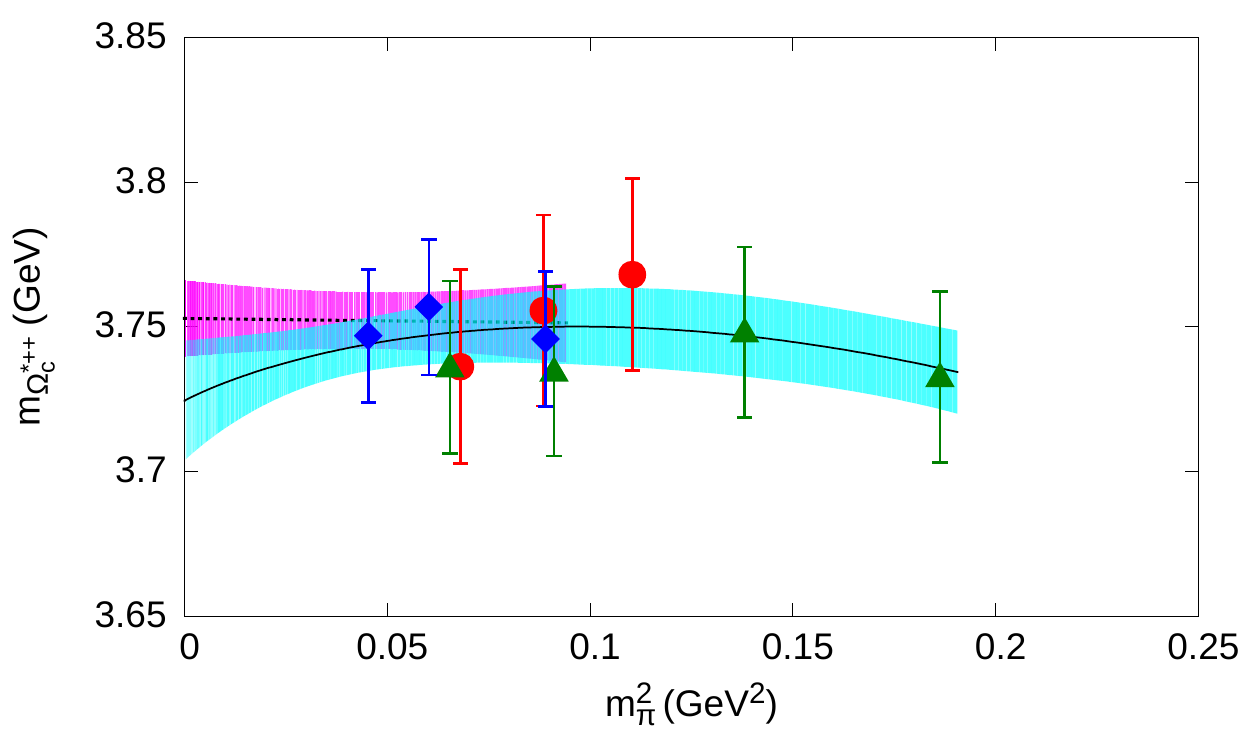}}
\end{minipage}
\caption{Representative chiral fits of the charm spin-1/2 (left) and spin-3/2 (right) baryon results in physical units, using the Ansatz of \eq{eq:charm_ansatz}. The lattice results  are the continuum extrapolated ones. The notation is shown in the legend of the  top left plot.}
\label{Fig:charm_chiral}
\end{figure}

For the charm baryons we use the Ansatz
\be \label{eq:charm_ansatz}
m_B = m_B^{(0)} + c_1 m_\pi^2 + c_2 m_\pi^3\;.
\ee
This expression is motivated by SU(2) HB$\chi$PT to leading one-loop order, where $m_B^{(0)}$ and $c_i$ are treated as independent fit parameters. As before, we add the term $da^2$ in the fits in order to simultaneously extrapolate to the continuum and we average over the states belonging to the same isospin multiplets. We show representative plots of the chiral fits for the charm baryons in \fig{Fig:charm_chiral}. The resulting fit parameters from the fits are listed in \tbl{Table:fit_params_charm}. The masses at the physical point are shown in \tbl{Table:phys_masses}. The lattice results for all charm baryons at the three $\beta$ values are collected in Appendix. \ref{App:numerical_results}. As can be seen from the chiral fits, setting $c_2=0$ in the Ansatz would lead to satisfactory fits as well. This is also reflected by the large uncertainties on this fit parameter, making it consistent with zero. As in the strange baryon sector, our continuum data are described well by \eq{eq:charm_ansatz}, yielding values at the physical point which in general  are consistent with experiment. For the $\Omega_c^0$ and $\Omega_c^{*0}$ the lattice data extrapolate to a lower value by one and two standard deviations respectively. In order to estimate a systematic error due to the chiral extrapolation in the charm sector, we perform the chiral fits using \eq{eq:charm_ansatz} with our lattice data only up to $m_\pi\sim 300$MeV and setting $c_2=0$. The deviation of the values obtained at the physical pion mass from fitting using the whole pion mass range and fitting up to $m_\pi\sim 300$MeV yields an estimation of the systematic error due to the chiral extrapolation.

\begin{figure}[!ht]\vspace*{-0.2cm}
\center
\begin{minipage}{8cm}
{\includegraphics[width=\textwidth]{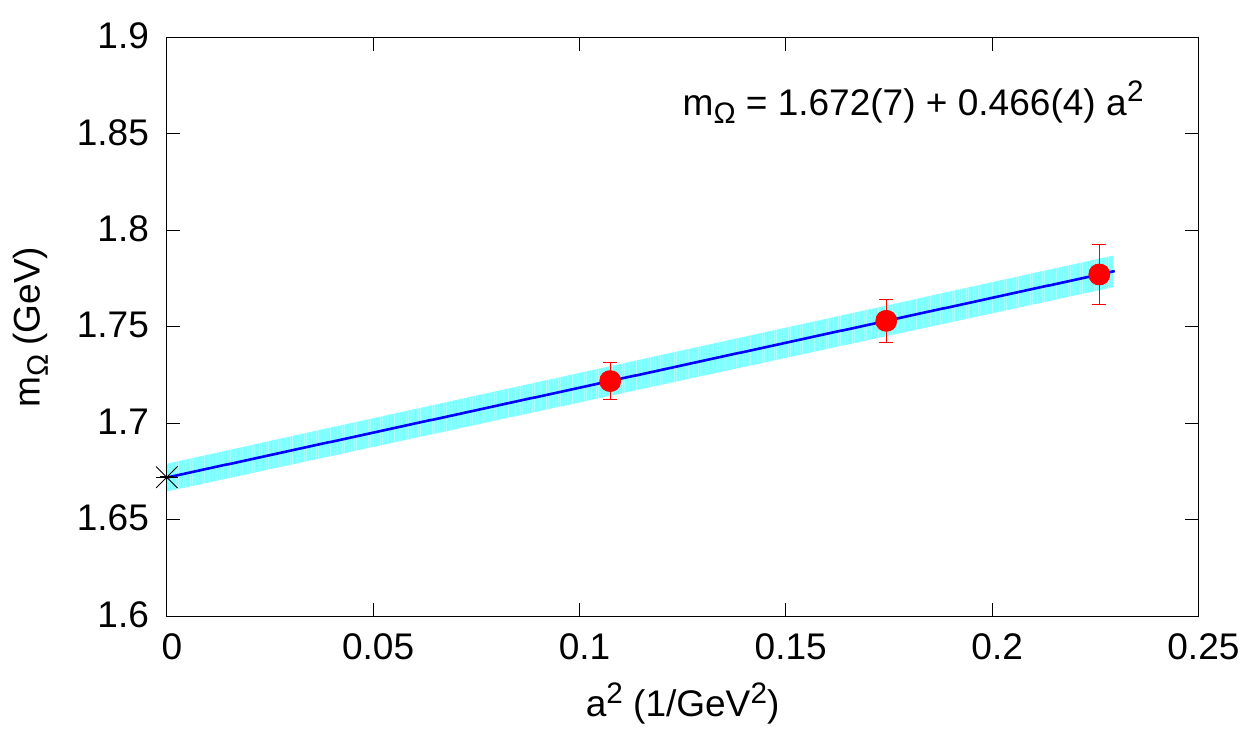}}
\end{minipage}\hfill
\begin{minipage}{8cm}
{\includegraphics[width=\textwidth]{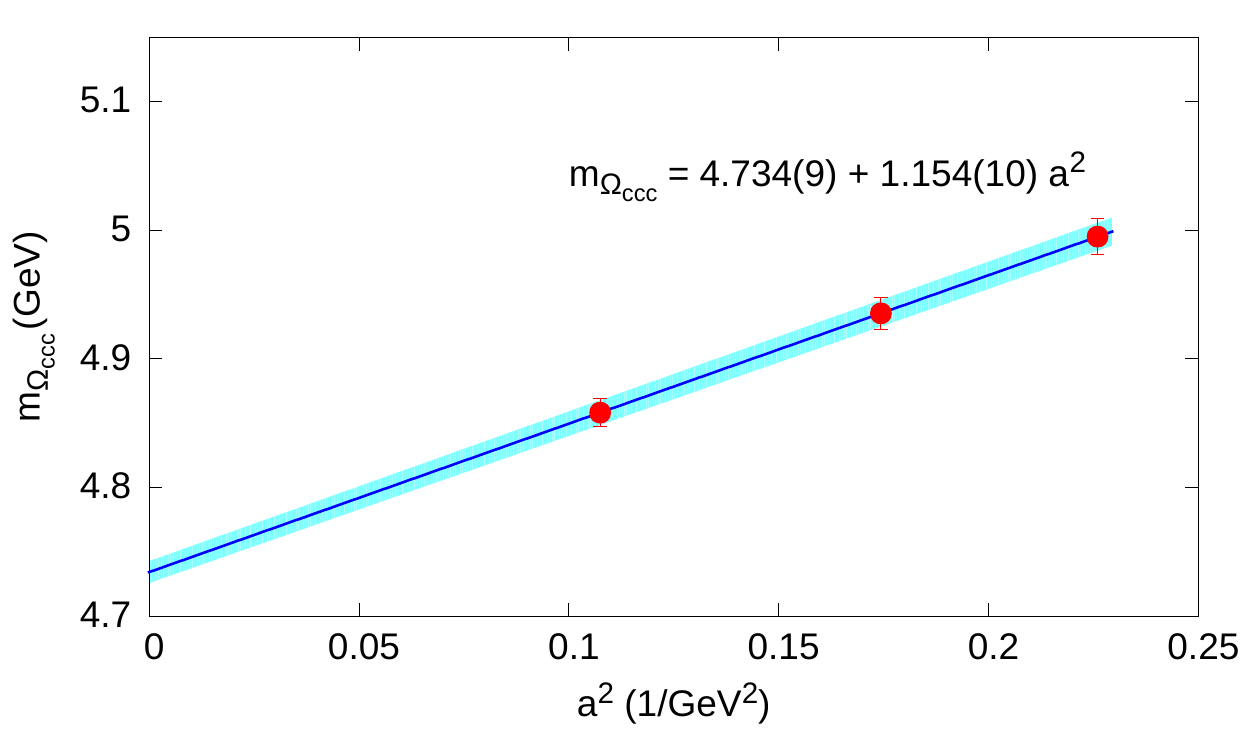}}
\end{minipage}
\label{fig:cutoff}
\caption{Dependence of the $\Omega^-$ (left) and $\Omega_{ccc}$ (right) mass on the lattice spacing.}
\end{figure}

The size of the cut-off effects in both the strange and charm quark  sectors
are small. This can be seen by the values of the fit parameter $d$, which are $\mathcal{O}(1)$, and thus the cut-off effects are indeed $\mathcal{O}(a^2)$.  As an example, we show in Fig.~\ref{fig:cutoff} the $a$-dependence of the mass of the  $\Omega^-$ and $\Omega_{ccc}$ for fixed quark masses. The correction at the largest value of $a$ is 6\% for the $\Omega^-$ and 5\% for the $\Omega_{ccc}$. 
In \tbl{Table:cut_off_parameter} we give the values of the parameter $d$ and the finite lattice spacing corrections in percentage of the mass at each $\beta$ value for the doubly and triply charmed baryon masses.

\begin{table}[h]
\begin{center}
\renewcommand{\arraystretch}{1.25}
\renewcommand{\tabcolsep}{7.5pt}
\begin{tabular}{l|c|ccc}
\hline
\multirow{2}{*}{Baryon}   &   \multirow{2}{*}{$d$ (GeV$^3$)}  & \multicolumn{3}{c}{\% correction} \\
\cline{3-5}
& &     $\beta=1.90$      &    $\beta=1.95$   &   $\beta=2.10$         \\
\hline
$\Xi_{cc}$              &  1.08   &  6.3    & 5.0    &  3.1    \\
$\Xi_{cc}^*$          &  1.01   &  5.9    & 4.6    &  2.9     \\
$\Omega_{cc}$      &  1.20   &  6.9    & 5.4    &  3.4     \\
$\Omega_{cc}^*$  &  1.10   &  6.2    & 4.9    &  3.0     \\
$\Omega_{ccc}$    &  1.15   &  5.1    & 4.1    &  2.6     \\
\end{tabular}
\end{center}
\caption{The value of the fit parameter $d$ and the finite lattice spacing 
correction as  percentage of the mass for  the doubly and triply charmed baryons .}
\label{Table:cut_off_parameter}
\end{table}

We also estimate a systematic uncertainty due to the tuning for all strange and charm baryons. This is done by evaluating the baryon masses when the strange and charm quark masses take the upper and lower bound allowed by the error in their tuned values (\eq{eq:strange_charm_values}). The deviation of the mass  extracted using $\chi$PT to leading order provides an estimate of the systematic error due to the tuning, given in the third parenthesis in \tbl{Table:phys_masses}. In the strange sector, the systematic error due to the tuning on the strange baryon masses gives an upper bound of the error expected, since the tuning was performed using the $\Omega$ which contains three strange quarks, and thus any error due to the uncertainty of the tuning would be the largest in this case. 

\begin{figure}[!ht]\hspace*{-6cm}
\center
{\includegraphics[width=0.92\textwidth]{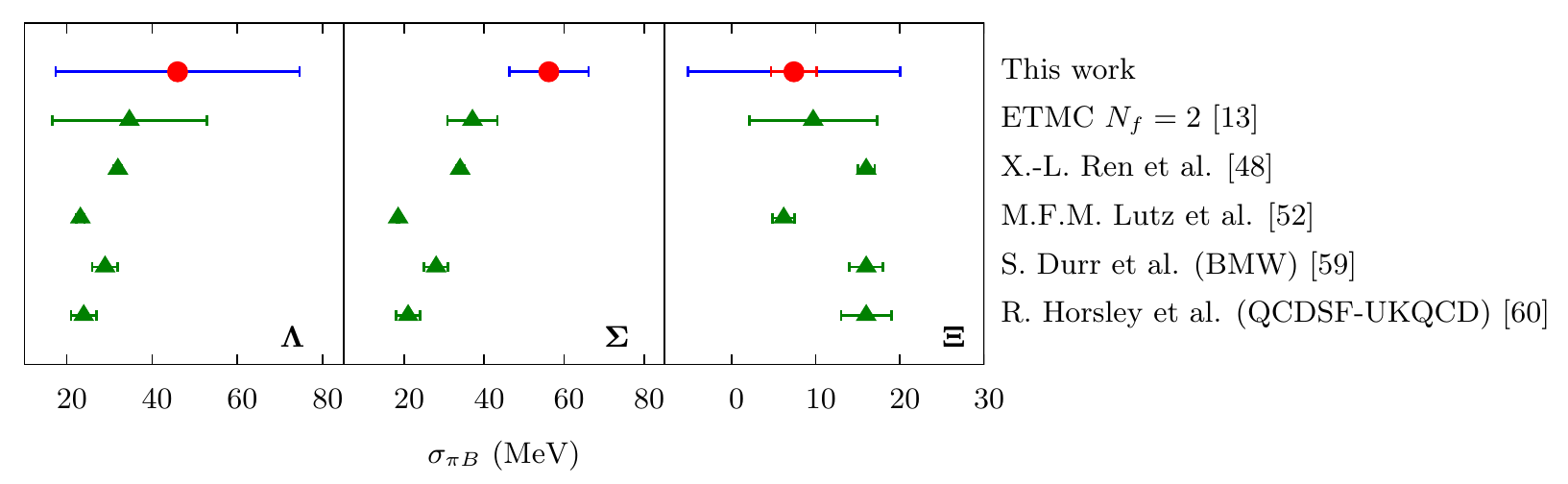}}
\caption{Comparison of the light $\sigma$-term of the spin-1/2 hyperons in MeV, extracted from the $\mathcal{O}(p^3)$  in this work with the results from other lattice calculations.
Our result shows the statistical error in red and a systematic error  in blue  taken as  the difference between the value obtained using the $\mathcal{O}(p^3)$ and $\mathcal{O}(p^4)$ expressions (Eqns. (\ref{eq:octet_expressions_lo}) and (\ref{eq:octet_expressions_nlo}) respectively) providing an estimate of the uncertainty due to the chiral extrapolation. }
\label{Fig:octet_sterms}
\end{figure}

\begin{figure}[!ht]\hspace*{-3.5cm}
\center
{\includegraphics[width=0.92\textwidth]{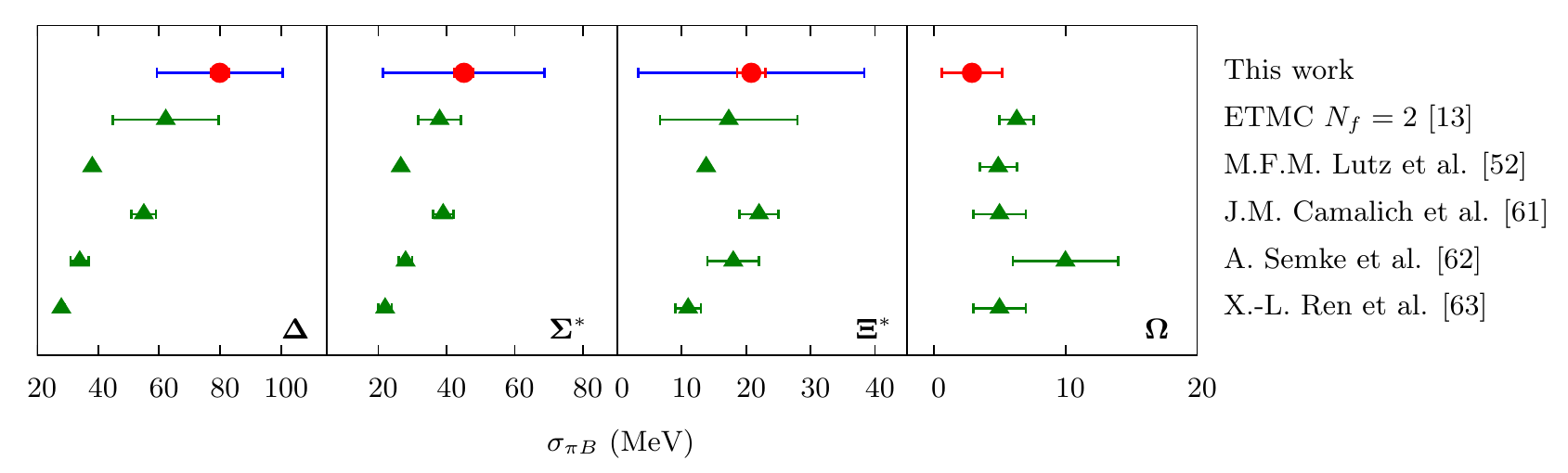}}
\caption{Comparison of the light $\sigma$-term of the spin-3/2 hyperons in MeV, extracted from the $\mathcal{O}(p^3)$ in this work with the results from other lattice calculations.  The notation is the same as that in Fig~\ref{Fig:octet_sterms}.}
\label{Fig:decuplet_sterms}
\end{figure}

As in the nucleon case, an estimate of the light $\sigma$-term of all the hyperons and charmed baryons considered in this work can be made, by taking the derivative $m_\pi^2 \partial m_B/\partial m_\pi^2$. For the octet and decuplet we calculate $\sigma_{\pi B}$ using the LO as well as the NLO expressions. It is apparent that the value extracted depends on the fitting Ansatz, and since the slope of the NLO fit is larger at the physical point, the resulting values for $\sigma_{\pi B}$ from the NLO expressions are larger, again indicating the sensitivity on the chiral extrapolations. We list the values extracted for the octet and decuplet baryons in \tbl{Table:fit_params_810}. A number of other recent works \cite{Alexandrou:2009qu,Ren:2014vea,Lutz:2014oxa,Durr:2011mp,Horsley:2011wr,MartinCamalich:2010fp,Semke:2012gs,Ren:2013oaa} have computed the light $\sigma$-terms for the octet and decuplet baryons by analyzing lattice QCD data from various collaborations. We compare our results with the results of these calculations in Figs. \ref{Fig:octet_sterms} and \ref{Fig:decuplet_sterms}. As for the case of the nucleon $\sigma$-term, we take the difference between the values obtained using 
 $\mathcal{O}(p^3)$ and  $\mathcal{O}(p^4)$  perturbation theory as an estimate
of the systematic error arising from the chiral extrapolation. This explains why our results have a larger error as compared to other groups which, typically, do not include such an estimate.  Extending this analysis we can compute the poorly known $\sigma$-terms for the charmed baryons from the fitting Ansatz of \eq{eq:charm_ansatz}. We list the resulting values in \tbl{Table:fit_params_charm}. 

It is worth mentioning that a number of analyses based on   baryon chiral perturbation theory have been carried out for the octet baryon masses and sigma terms. We refer for example to Refs.~\cite{Frink:2004ic,Frink:2005ru,Bernard:2009mw} for details.

\begin{table}[h]
\begin{center}
\renewcommand{\arraystretch}{1.25}
\renewcommand{\tabcolsep}{7.5pt}
\begin{tabular}{|l|l|l|l|l|}
\hline
\multirow{2}{*}{Baryon}    &    \multirow{2}{*}{$m_B^{(0)}$ (GeV)}    &     \multirow{2}{*}{$-4c_B^{(1)}$ (GeV$^{-1}$)}      & \multicolumn{2}{c|}{$\sigma_{\pi B}$ (MeV)}          \\
\cline{4-5}
					  &									&						  & \;\;$\mathcal{O}(p^3)$ & \;\;\; NLO                                  \\	

\hline\hline
$N$                          & 0.867(2)   &  4.574            & 64.9(1.5)  & 45.3(4.3)         \\
\hline
$\Lambda$              & 1.067(16)  &  3.544(97)     & 46.0(1.8)  & 74.5(1.8)         \\
\hline
$\Sigma^+$            & 1.110(21)  &  4.470(113)   & 55.6(2.1)  & 65.3(2.2)         \\
\hline
$\Sigma^0$             & 1.117(17)  &  4.422(95)    & 54.7(1.7)  & 64.5(1.8)         \\
\hline
$\Sigma^-$             & 1.095(18)  &  4.618(102)    & 58.3(1.9)  & 68.3(1.9)         \\
\hline
$\Xi^0$                    & 1.307(16)  &  0.433(147)    & 6.8(2.7)   & 18.9(2.7)         \\
\hline
$\Xi^-$                    & 1.312(12)  &  0.497(107)     & 8.0(2.0)   & 20.4(1.9)         \\
\hline
$\Delta$                   & 1.207(31)  &  6.496(162)     & 79.9(3.0)  &100.3(3.1)         \\
\hline
$\Sigma^*$              & 1.405(23)  &  3.603(156)     & 45.1(2.8)  & 68.6(2.7)         \\
\hline
$\Xi^*$                    & 1.535(19)  &  1.562(123)     & 20.8(2.2)  & 38.2(2.2)         \\
\hline
$\Omega$                & 1.669(19)  &  0.161(124)      & 2.9(2.3)   &                   \\
\hline
\end{tabular}
\end{center}
\caption{The mass at the chiral limit, $m_B^{(0)}$, and the fit parameter $c_B^{(1)}$ as determined from fitting to the leading one-loop order expressions for the octet and decuplet baryons at the tuned strange quark mass. Also shown in the value of the light $\sigma$-term at the physical point determined from the fits.}
\label{Table:fit_params_810}
\end{table}
\begin{table}[h]
\begin{center}
\renewcommand{\arraystretch}{1.25}
\renewcommand{\tabcolsep}{7.5pt}
\begin{tabular}{|l|l|l|l|l|}
\hline
 Baryon          &    $m_B^{(0)}$ (GeV)    &     $c_1$ (GeV$^{-1}$)    &     $c_2$ (GeV$^{-2}$)      &   $\sigma_{\pi B}$ (MeV)       \\
\hline\hline
$\Lambda_c$        & 2.272(26)  &  0.799(935)    & -0.118(1.834)      &  14.1(10.3)       \\
\hline
$\Sigma_c$   & 2.445(32)  &  0.903(1.118)  & -0.662(2.159)      & 14.0(12.4)        \\
\hline
$\Xi_c$       & 2.469(28)  &  0.233(906)    & -0.087(1.782)    & 4.6(10.0)         \\
\hline
$\Xi_c$       & 2.447(25)  &  0.855(788)    & -1.128(1.527)   & 11.4(8.8)         \\
\hline
$\Xi_c^\prime$ & 2.542(27)  &  1.242(870)    & -1.924(1.690)    &15.5(9.7)          \\
\hline
$\Omega_c$            & 2.629(22)  &  1.028(768)    & -2.017(1.507)    &  11.3(8.5)        \\
\hline
$\Xi_{cc}$     & 3.561(22)  &  0.516(725)    & -0.880(1.415)     &6.2(8.0)           \\
\hline
$\Omega_{cc}$                & 3.654(18)  &  0.341(602)    & -0.937(1.193)    & 2.8(6.6)          \\
\hline
$\Sigma_c^*$   & 2.513(38)  &  0.887(1.345)  & -0.481(2.593)  &14.4(15.0)         \\
\hline
$\Xi_c^*$     & 2.628(33)  &  0.483(1.178)  & -0.766(2.339)    & 6.0(12.9)         \\
\hline
$\Omega_c^{*}$        & 2.709(26)  &  1.408(875)    & -2.623(1.710)   &16.0(9.7)          \\
\hline
$\Xi_{cc}^*$            & 3.642(26)  &  0.703(891)    & -1.087(1.733)   &8.8(9.9)           \\
\hline
$\Omega_{cc}^{*}$               & 3.724(21)  &  0.792(719)    & -1.695(1.418)   &8.2(7.9)           \\
\hline
$\Omega_{ccc}$              & 4.733(18)  &  0.156(551)    & -0.443(1.082)     & 1.2(6.1)          \\
\hline
\end{tabular}
\end{center}
\caption{The mass at the chiral limit, $m_B^{(0)}$, and fit parameters $c_i$ as determined from fitting to the Ansatz of \eq{eq:charm_ansatz} for the charm baryons at the tuned strange and charm quark masses. Also listed is the value of the light $\sigma$-term in MeV.}
\label{Table:fit_params_charm}
\end{table}
\begin{table}[!ht]
\begin{center}
\renewcommand{\arraystretch}{1.25}
\renewcommand{\tabcolsep}{7.5pt}
\begin{tabular}{|l|l|}
\hline
 Baryon (PDG)   & \qquad  m (GeV)       \\
\hline\hline
$N$ (0.939)                        &      0.939            \\
\hline
$\Lambda$ (1.116)                  & 1.120(15)(54)(22)     \\
\hline
$\Sigma$ (1.193)                   & 1.168(32)(14)(44)     \\
\hline
$\Xi$ (1.318)                      & 1.318(19)(23)(9)      \\
\hline
$\Delta$ (1.232)                   & 1.299(30)(66)         \\
\hline
$\Sigma^*$ (1.384)                 & 1.457(22)(28)(32)     \\
\hline
$\Xi^*$ (1.530)                    & 1.558(18)(41)(19)     \\
\hline
$\Omega$ (1.672)                   & 1.672(18)             \\
\hline\hline
$\Lambda_c$  (2.286)               & 2.286(17)(10)        \\
\hline
$\Sigma_c$ (2.453)                 & 2.460(20)(20)(6)     \\
\hline
$\Xi_c$  (2.470)                   & 2.467(24)(4)(5)     \\
\hline
$\Xi_{c}^\prime$(2.575)            & 2.560(16)(22)(42)    \\
\hline
$\Omega_c^0$ (2.695)               & 2.643(14)(19)(42)    \\
\hline
$\Xi_{cc}$ (3.519)                 & 3.568(14)(19)(1)     \\
\hline
$\Omega_{cc}^{+}$                  & 3.658(11)(16)(50)    \\
\hline
$\Sigma_c^*$ (2.517)               & 2.528(25)(15)(7)     \\
\hline
$\Xi_c^*$ (2.645)                  & 2.635(20)(27)(55)    \\
\hline
$\Omega_c^{*0}$ (2.765)            & 2.728(16)(19)(26)    \\
\hline
$\Xi_{cc}^*$                       & 3.652(17)(27)(3)     \\
\hline
$\Omega_{cc}^{*+}$                 & 3.735(13)(18)(43)    \\
\hline
$\Omega_{ccc}^{++}$                & 4.734(12)(11)(9)     \\
\hline
\end{tabular}
\end{center}
\caption{Our values of the  masses of the baryons considered in this work after extrapolating to  the physical point and taking the continuum limit given in GeV, with the associated statistical error shown in the first parenthesis. The error in the second parenthesis is an estimate of the systematic error due to the chiral extrapolation and  in the third parenthesis (except for $\Delta$, which contains only light quarks) is an estimate of the systematic error due to the tuning. There are no systematic errors  for $\Omega^-$ and $\Lambda_c^+$ since these are used for the tuning of the strange and charm quark mass, respectively.}
\label{Table:phys_masses}
\end{table}


\section{Comparison with results from other collaborations}

In this section we compare our lattice results with those of other collaborations which use different discretization schemes. Having already extrapolated to the continuum, we also compare our values at the physical pion mass with the corresponding results of other collaborations and with experiment.

Several collaborations have calculated the strange spectrum. The Budapest-Marseille-Wuppertal (BMW) collaboration carried out simulations using tree level improved 6-step stout smeared $N_f=2+1$ clover fermions and a tree level Symanzik improved gauge action. The lattice spacing values used to obtain the continuum limit were $a=0.065$ fm, $0.085$ fm and $0.125$ fm. Using pion masses as low as $190$ MeV, a polynomial fit was performed to extrapolate to the physical point \cite{Durr:2008zz}. The PACS-CS collaboration obtained results using $N_f=2+1$ non-perturbatively $\mathcal{O}(a)$ improved clover fermions on an Iwasaki gauge action on a lattice of spatial length of $2.9$ fm and a value of lattice spacing $a=0.09$ fm \cite{Aoki:2008sm}.  In addition, the octet and decuplet spectrum was obtained in Ref.~\cite{Bietenholz:2011qq}, using $N_f=2+1$ SLiNC configurations. Ref.~\cite{Bali:2012ua} also includes results on the charmed baryons from an analysis on $N_f=2+1$ 2-HEX~\cite{Durr:2010aw} and SLiNC~\cite{Bietenholz:2010jr,Bietenholz:2011qq} configurations produced by the BMW-c and QCDSF collaborations respectively. Finally, we compare with the LHPC collaboration, which obtained results using a hybrid action of domain wall valence quarks on a staggered sea on a lattice of spatial length $2.5$ fm and $3.5$ fm at lattice spacing $a=0.124$ fm \cite{WalkerLoud:2008bp}.

\begin{figure}[!ht]
\center
\begin{minipage}{8cm}
{\includegraphics[width=0.9\textwidth]{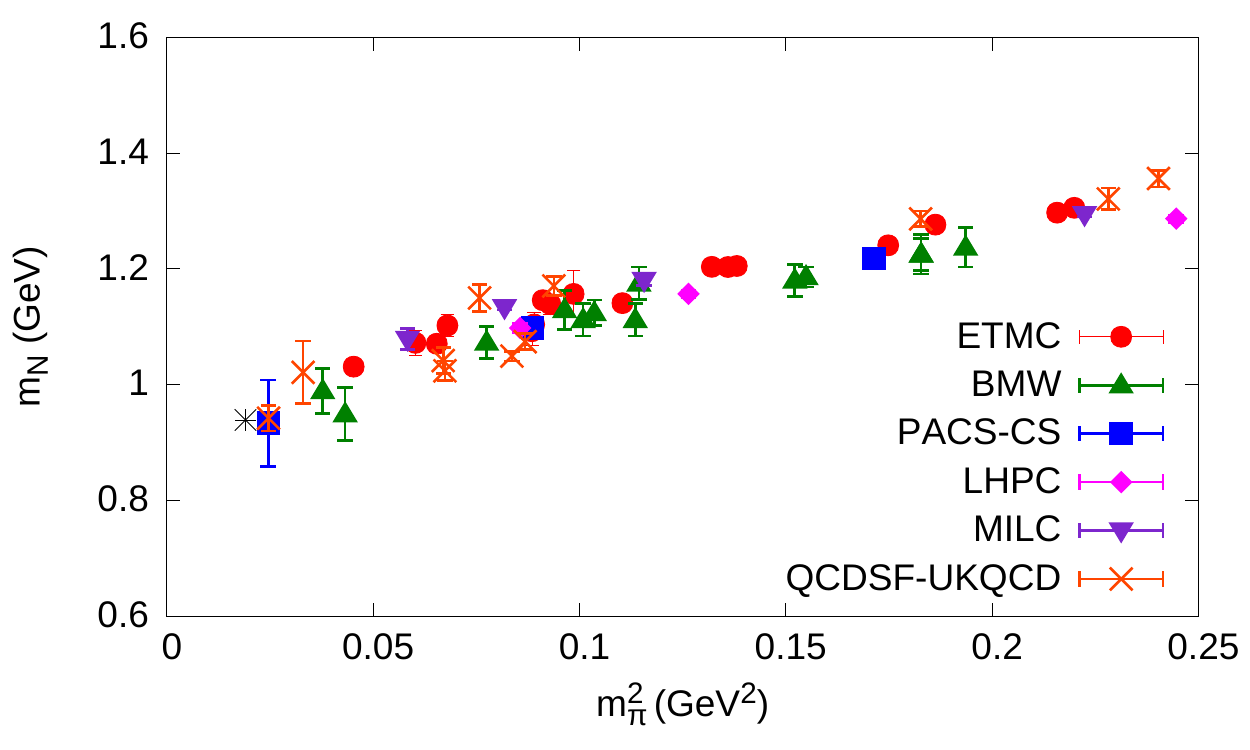}}
\end{minipage}\hfill
\begin{minipage}{8cm}
{\includegraphics[width=0.9\textwidth]{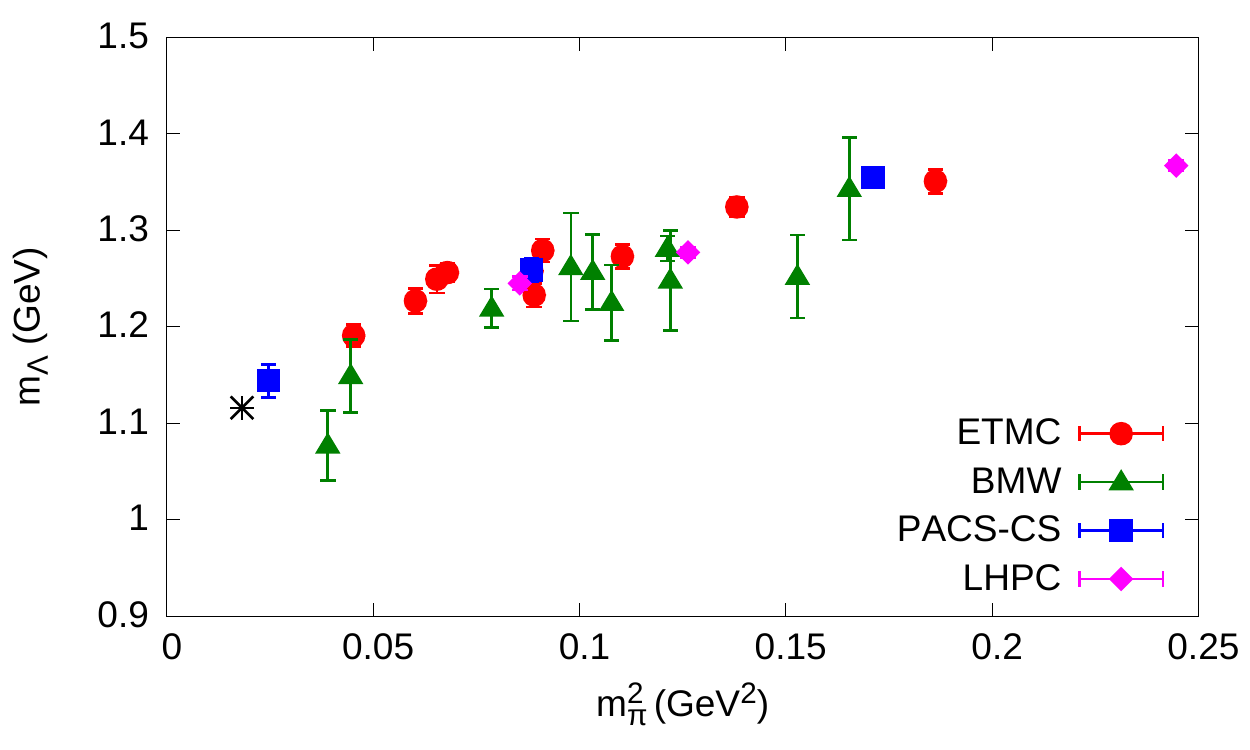}}
\end{minipage}
\begin{minipage}{8cm}
{\includegraphics[width=0.9\textwidth]{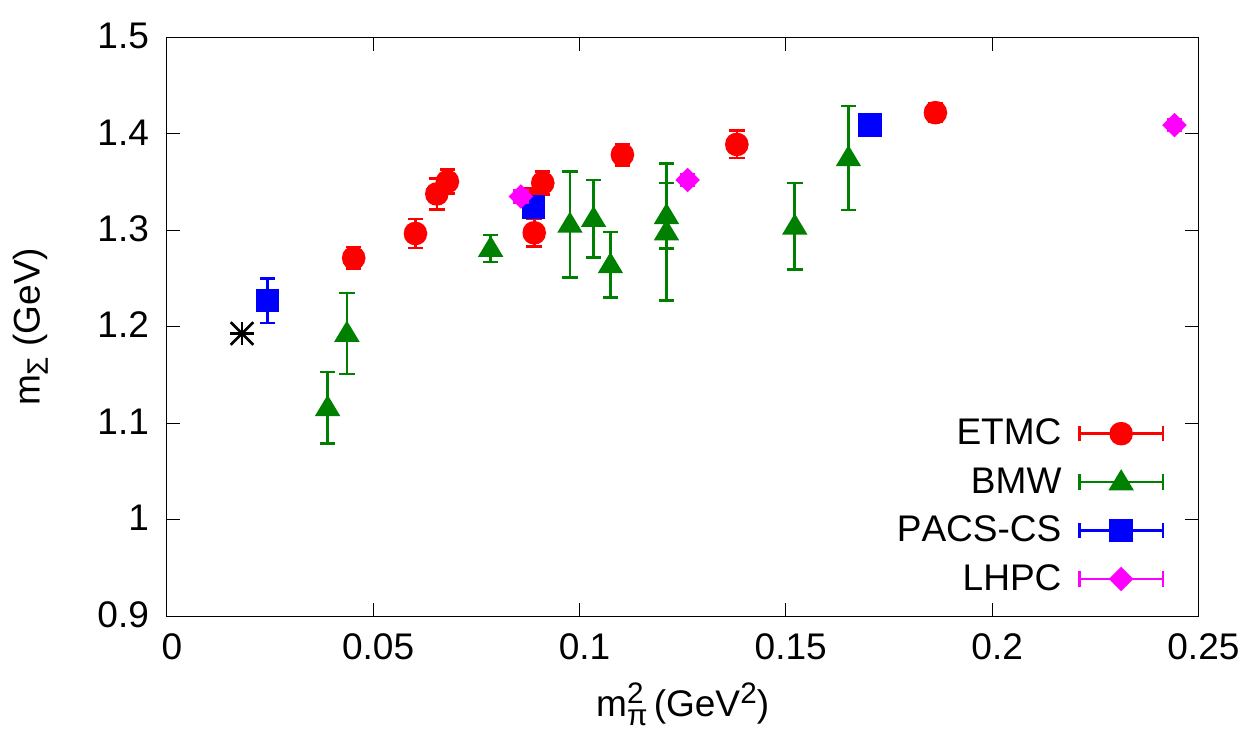}}
\end{minipage}\hfill
\begin{minipage}{8cm}
{\includegraphics[width=0.9\textwidth]{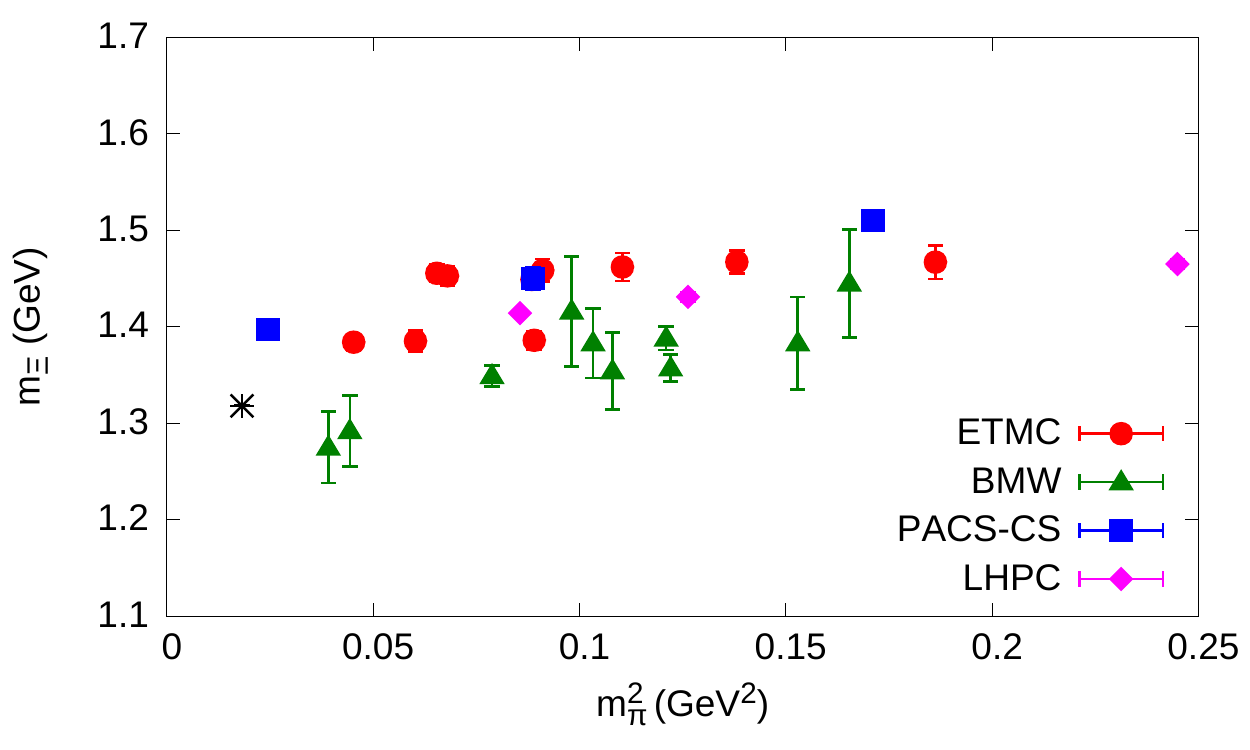}}
\end{minipage}
\caption{Comparison of lattice results of this work (red filled circles) with those from other collaborations for the octet baryons. Results using clover fermions from BMW \cite{Durr:2008zz} are shown in green triangles and from PACS-CS \cite{Aoki:2008sm} with blue squares. Domain wall valence quarks by the LHPC \cite{WalkerLoud:2008bp} are shown in magenta diamonds. In the nucleon case we additionally show results from the MILC collaboration \cite{Bernard:2001av} in purple inverted triangles and from QCDSF-UKQCD \cite{Bali:2012qs} with orange crosses. The physical point is shown with the black asterisk.}
\label{Fig:octet_comparison}
\end{figure}

\begin{figure}[!ht]
\center
\begin{minipage}{8cm}
{\includegraphics[width=0.9\textwidth]{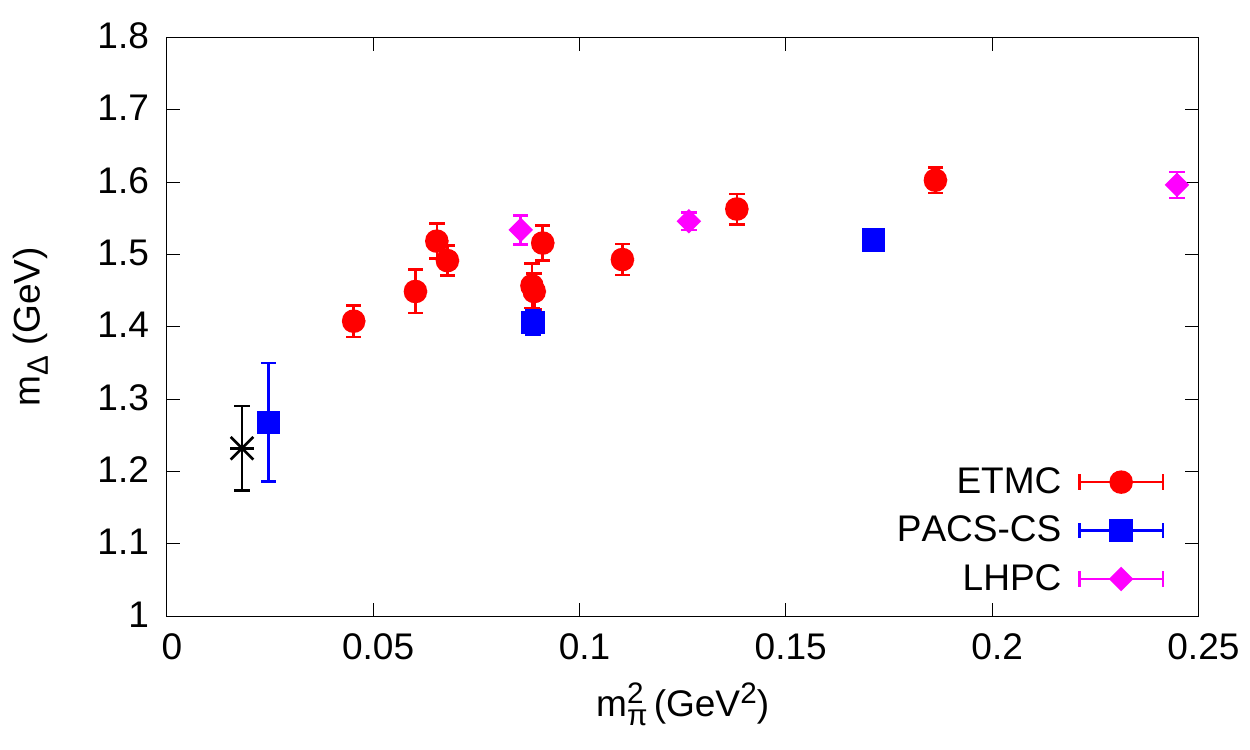}}
\end{minipage}\hfill
\begin{minipage}{8cm}
{\includegraphics[width=0.9\textwidth]{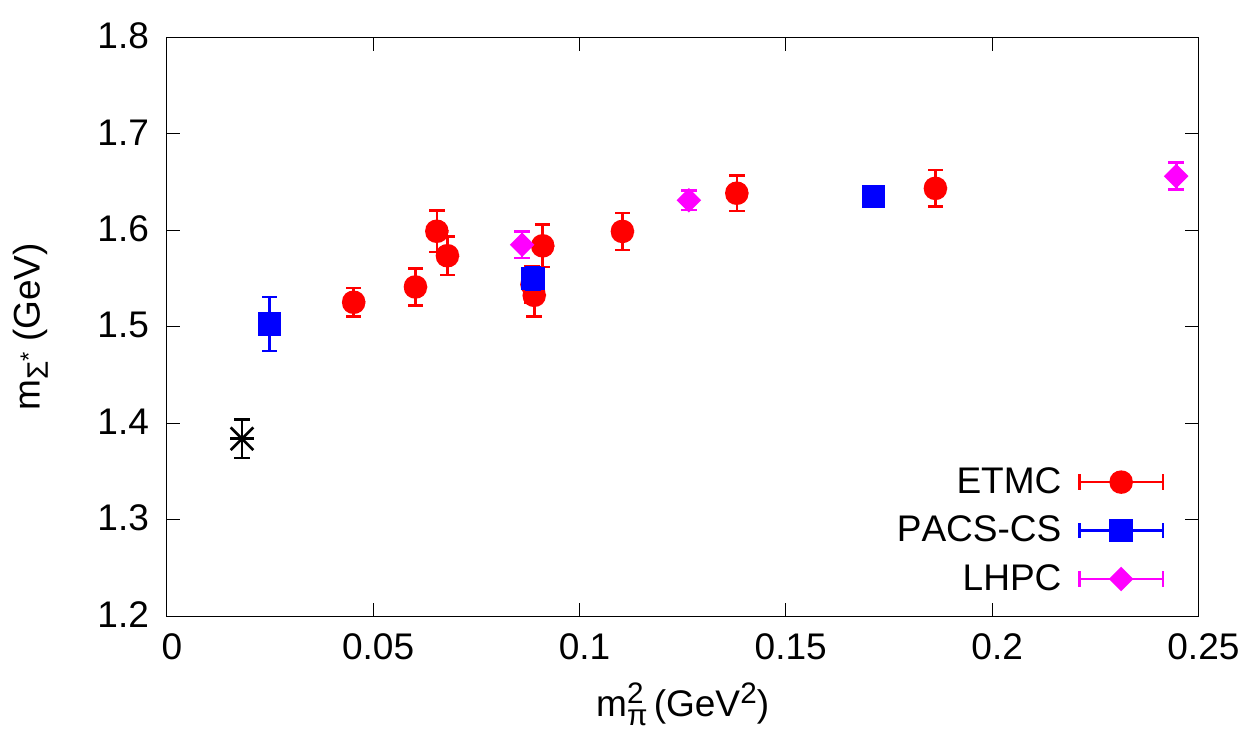}}
\end{minipage}
\begin{minipage}{8cm}
{\includegraphics[width=0.9\textwidth]{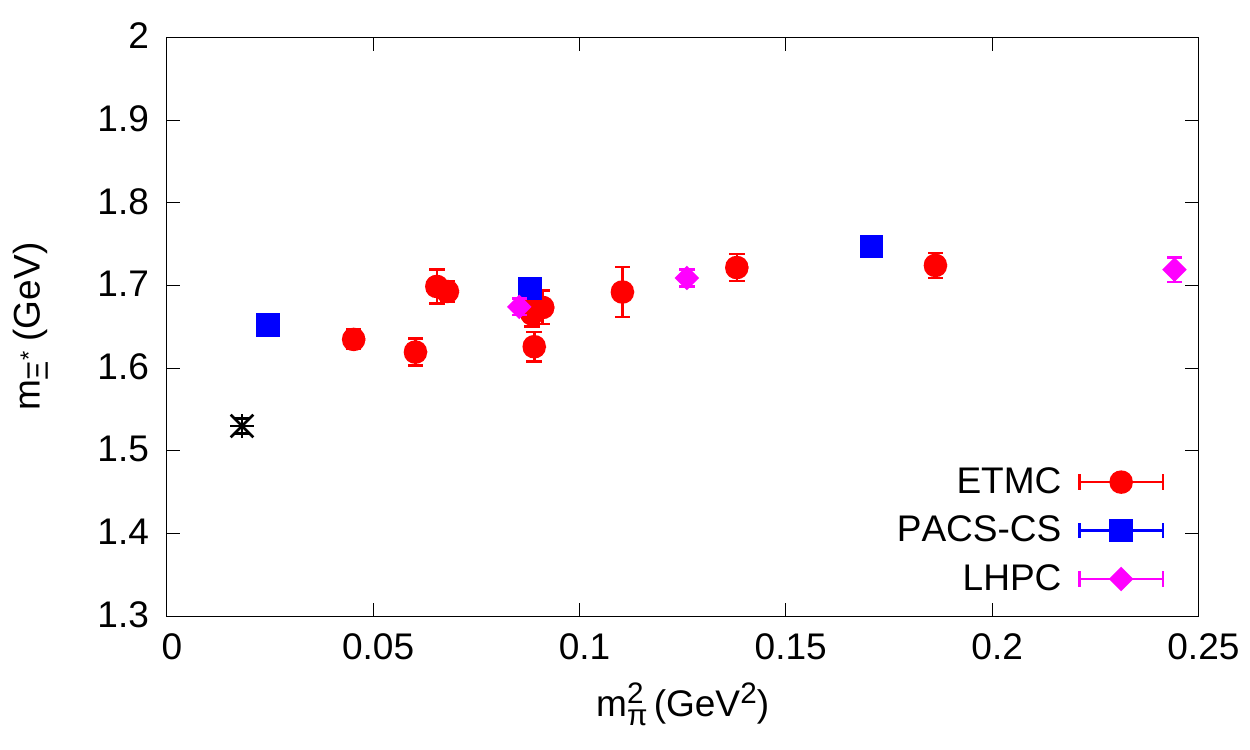}}
\end{minipage}\hfill
\begin{minipage}{8cm}
{\includegraphics[width=0.9\textwidth]{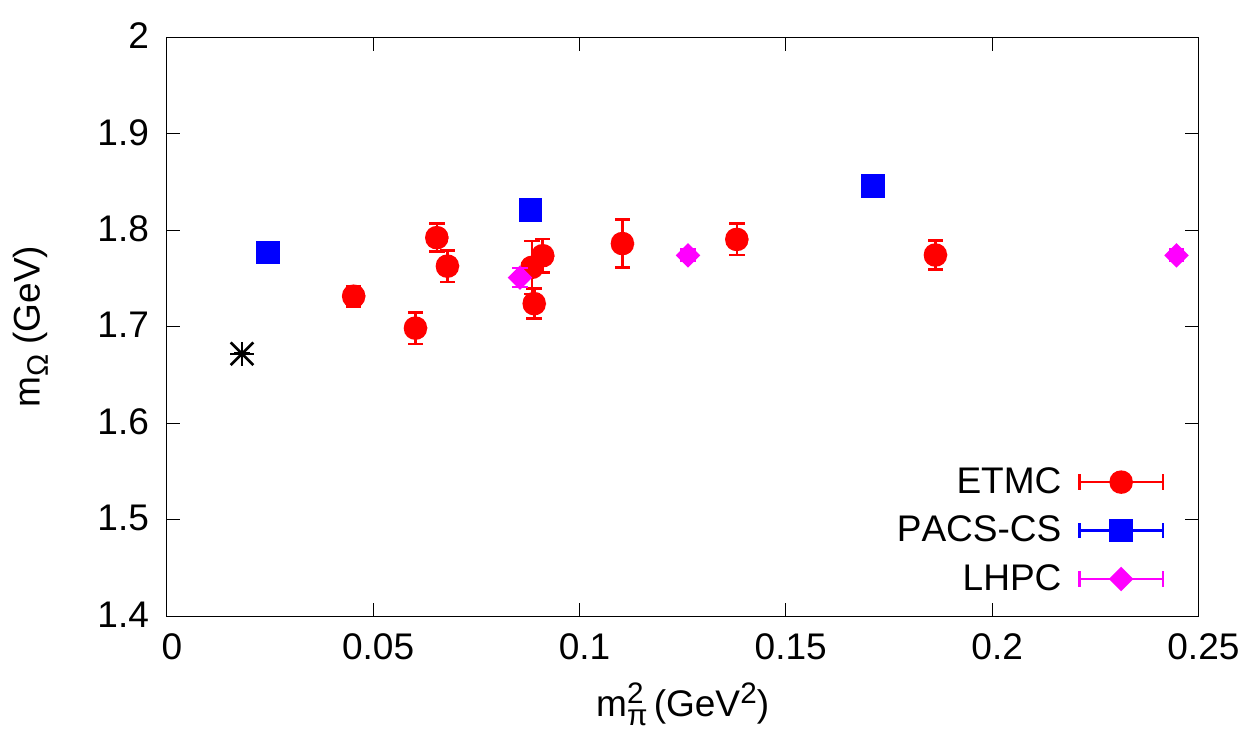}}
\end{minipage}
\caption{Comparison of the results for the decuplet baryons in this work with the results from PACS-CS using clover fermions \cite{Aoki:2008sm} and from the LHPC collaboration \cite{WalkerLoud:2008bp} using domain wall valence quarks. The notation is as in \fig{Fig:octet_comparison}.}
\label{Fig:decuplet_comparison}
\end{figure}

\begin{figure}[!ht]
\center
\begin{minipage}{8cm}
{\includegraphics[width=0.9\textwidth]{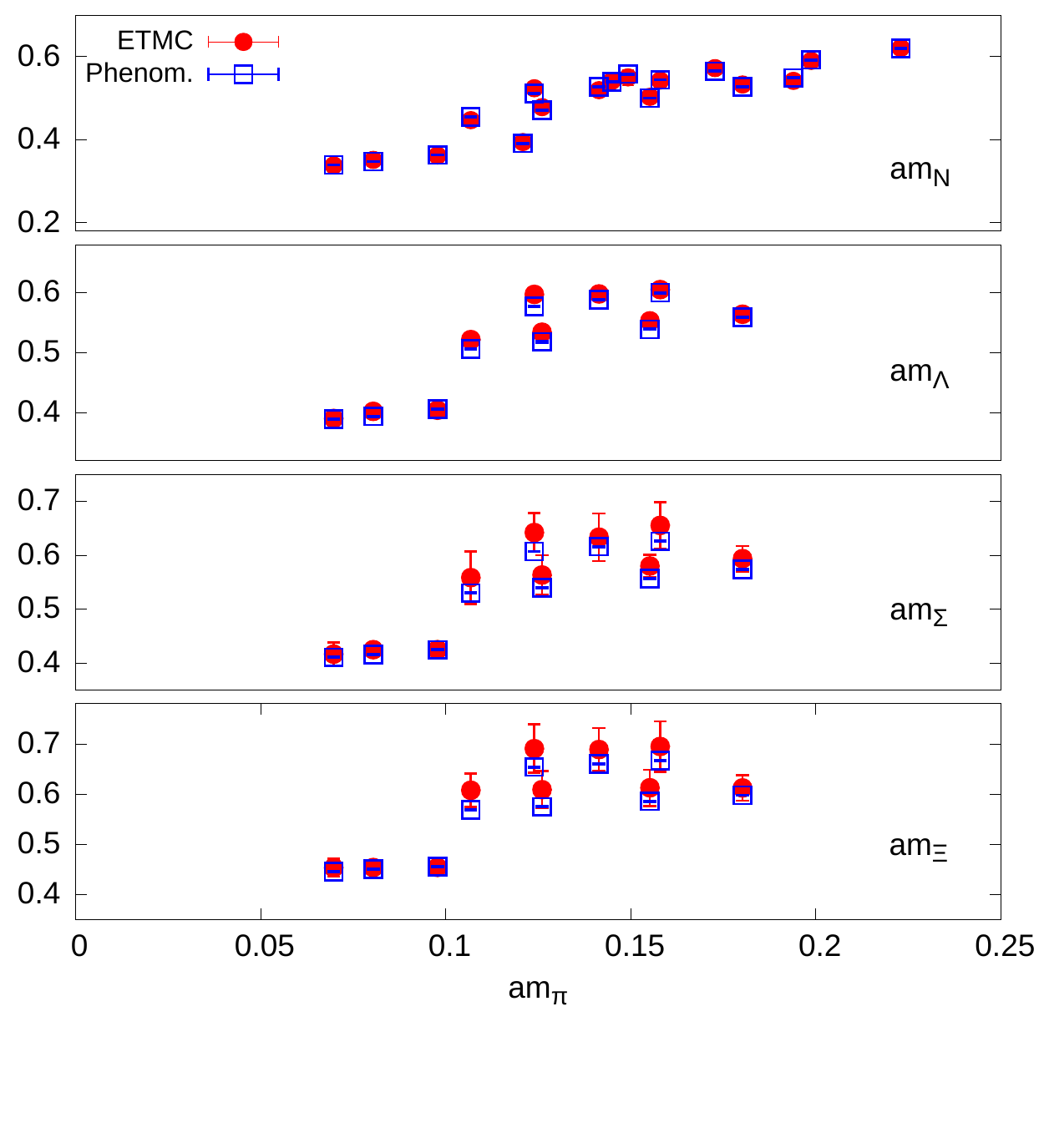}}
\end{minipage}\hfill
\begin{minipage}{8cm}
{\includegraphics[width=0.9\textwidth]{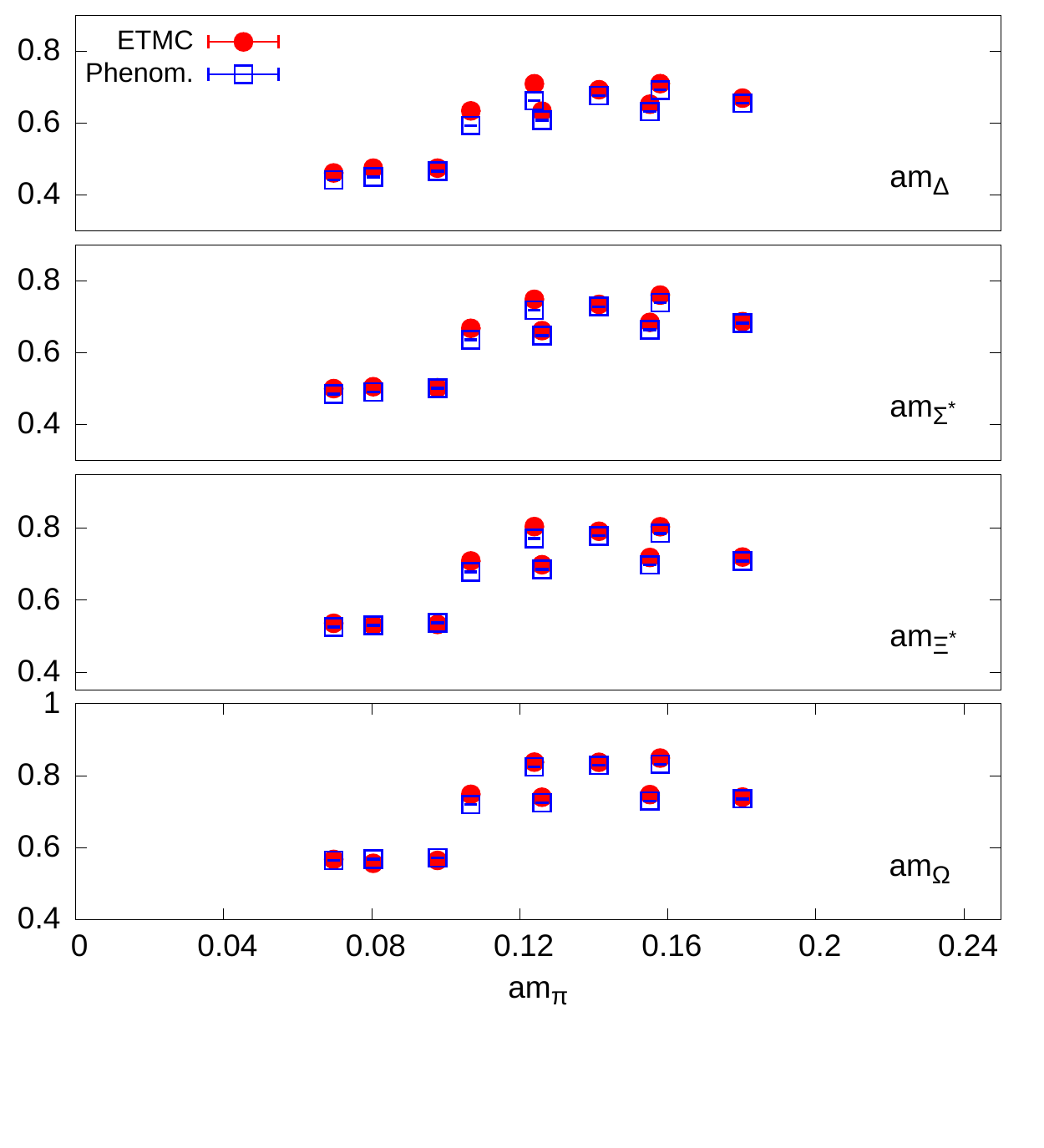}}
\end{minipage}
\caption{Comparison of the lattice results for the octet (left) and decuplet (right) baryons from this work (red circles) with the phenomenology results from Ref.~\cite{Lutz:2014oxa} (blue open squares). The results are consistent for all $\beta$ values.}
\label{Fig:phen_comparison}
\end{figure}

In \fig{Fig:octet_comparison} we compare our lattice results on the octet baryons with those of BMW, the PACS-CS and the LHPC collaborations. In the nucleon case, we furthermore compare with results from the MILC collaboration \cite{Bernard:2001av}, obtained from $N_f=2+1+1$ simulations using the one-loop Symanzik improved gauge action and an improved Kogut-Susskind quark action at a lattice spacing value $a=0.130$ fm and with results from QCDSF-UKQCD, obtained using $N_f=2$ simulations at three values of the lattice spacing, $a=0.076\;, 0.072\;, 0.060$ fm \cite{Bali:2012qs}. We note that our results shown in these plots and the results from the PACS-CS and LHPC are not continuum extrapolated, while the results from BMW are continuum extrapolated and have larger errors than the rest. Nevertheless, there is an overall agreement,  best seen in the case of the nucleon mass, which indicates that cut-off effects are small. A similar behavior is also seen in the case for the mass in the decuplet shown in \fig{Fig:decuplet_comparison}, where we compare our results with those from PACS-CS and LHPC. We stress
that these lattice results need to be  extrapolated to zero lattice spacing (continuum limit)  and therefore small deviations are to be expected the raw data. 
A  comparison is also made with recent phenomenology results on the octet and decuplet baryon masses, obtained from an analysis of lattice QCD data based on the relativistic chiral Lagrangian~\cite{Lutz:2014oxa}. As can be seen from \fig{Fig:phen_comparison}, results show an overall agreement.

\begin{figure}[!ht]
\center
{\includegraphics[width=0.5\textwidth]{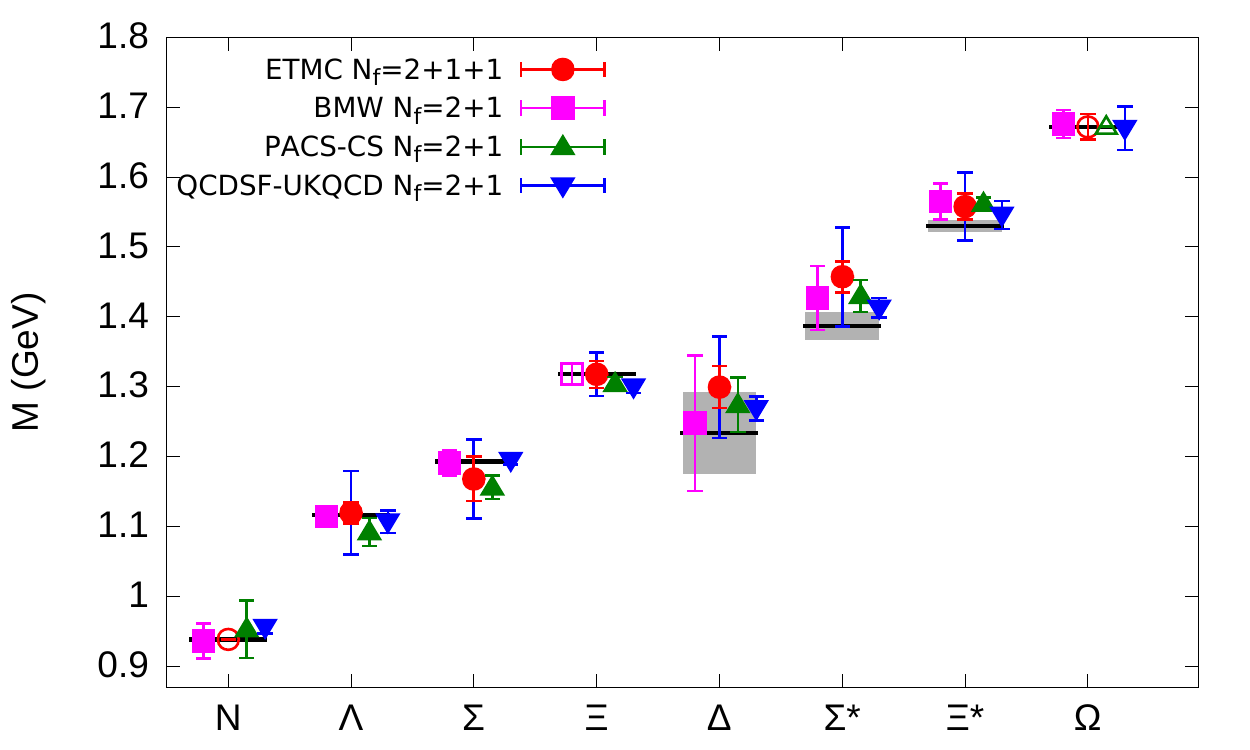}}
\caption{The octet and decuplet baryon masses obtained at the physical point and the experimental masses \cite{Hagiwara:2002fs} shown by the horizontal bands. For most baryons the band is too small to be visible. For the twisted mass results of this work (red circles) the chiral extrapolation was performed using the leading order HB$\chi$PT. In our results, the statistical error is shown in red, whereas the blue error bar includes the statistical error and the systematic errors due to the chiral extrapolation and due to the tuning added in quadrature. Results using clover fermions from BMW \cite{Durr:2008zz} are shown in magenta squares and from PACS-CS \cite{Aoki:2008sm} with green triangles. Results from QCDSF-UKQCD collaborations~\cite{Bietenholz:2011qq} using $N_f=2+1$ SLiNC configurations are also displayed in blue inverted triangles. Open symbols are used wherever the mass was used as input to the calculations.}
\label{Fig:physpoints_810}
\end{figure}

 In \fig{Fig:physpoints_810} we show the masses for the octet and decuplet baryons obtained after extrapolating to the continuum limit and to the physical pion mass. Our results are obtained using the leading order expansions from HB$\chi$PT and   the statistical error and total error are shown separately. The error in red in our results shown in Figs. \ref{Fig:physpoints_810} represents the statistical error. The total error bar, shown in blue, is obtained after adding quadratically the statistical error and the systematic errors due to the chiral extrapolation and due to the tuning.

In addition, we compare our results obtained in the charm sector with the corresponding results of other lattice calculations. Specifically, the MILC collaboration has obtained results using a clover charm valence quark in $N_f=2+1+1$ gauge configurations at three values of the lattice spacing, $a=0.09\;, 0.12\;, 0.15$ fm \cite{Na:2007pv,Na:2008hz}. Moreover, results for the charm spectrum were produced from $N_f=2+1+1$ gauge configurations at lattice spacing values $a=0.06\;, 0.09\;, 0.12$ fm using the highly improved staggered quark (HISQ) action, whereas the valence up, down and strange quark propagators were generated using the clover improved Wilson action \cite{Briceno:2012wt}. A relativistic heavy quark action was implemented for the charm quark in order to reduce discretization artifacts. In Ref. \cite{Liu:2009jc} domain wall fermions are used for the up, down and strange quarks with $N_f=2+1$ simulations using the improved Kogut-Susskind sea quarks at a lattice spacing value $a=0.12$ fm. For the charm quark the relativistic Fermilab action was adopted. Finally, the PACS-CS has obtained results in the charm sector using the relativistic heavy quark action on $N_f=2+1$ configurations with the light and strange quarks tuned to their physical masses, a lattice spacing of $a=0.09$ fm and a spatial length of $L=2.9$ fm \cite{Namekawa:2013vu}. We compare our results with those from Refs.~\cite{Na:2007pv,Na:2008hz,Briceno:2012wt,Liu:2009jc,Namekawa:2013vu}.

\begin{figure}[!ht]
\center
\begin{minipage}{7.9cm}
{\includegraphics[width=\textwidth]{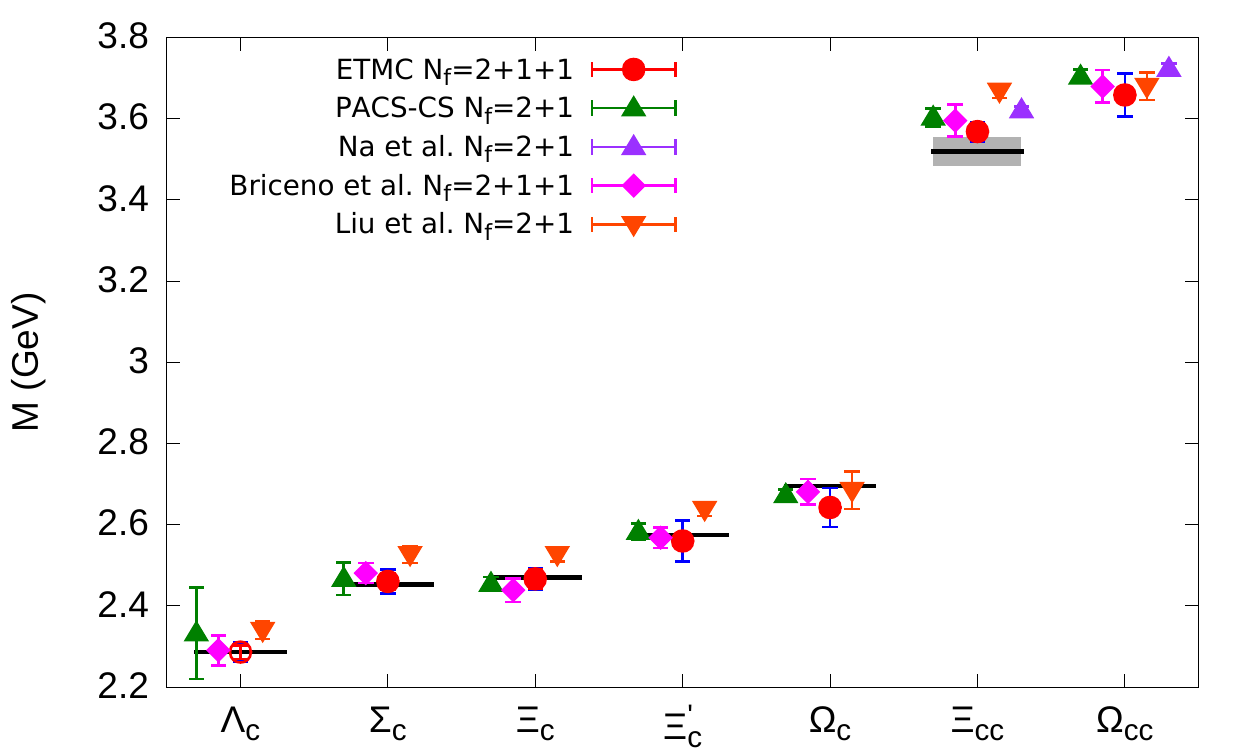}}
\end{minipage}\hfill
\begin{minipage}{7.9cm}
{\includegraphics[width=\textwidth]{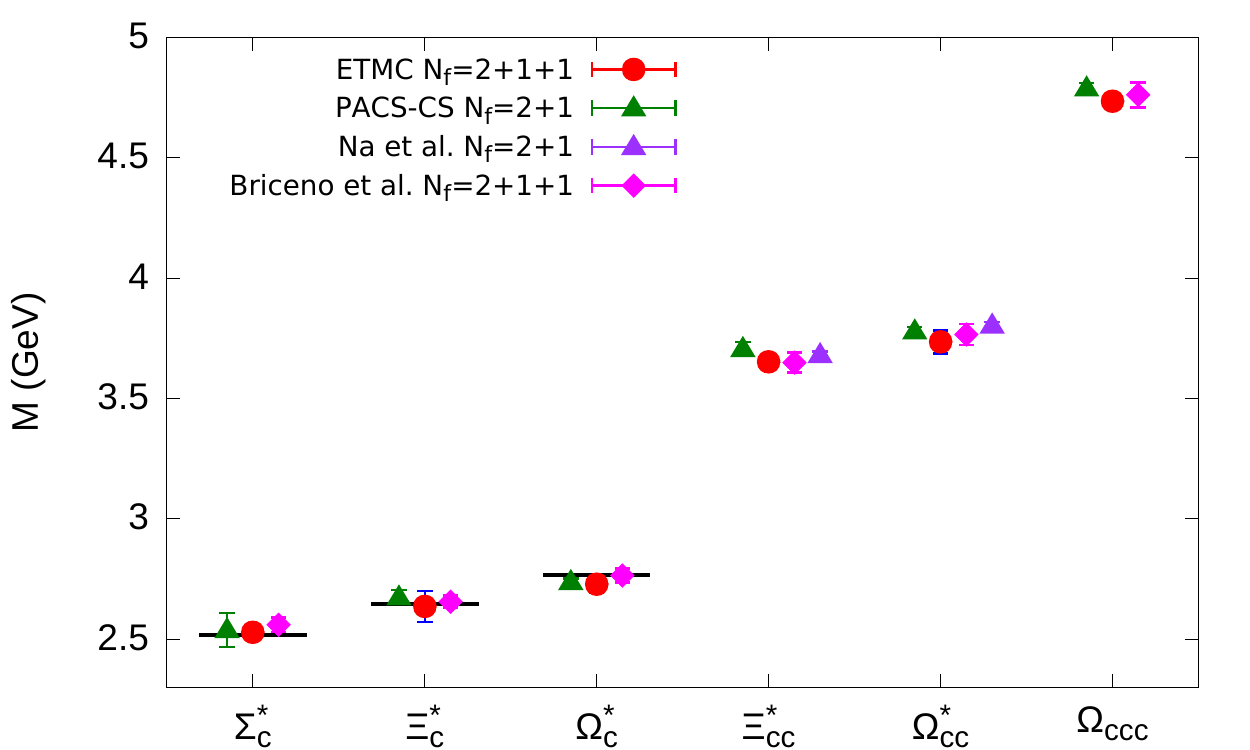}}
\end{minipage}
\caption{The masses of spin-1/2 (left) and spin-3/2 (right) charm baryons. The notation of our results (ETMC) is the same as in \fig{Fig:physpoints_810}. The experimental values are from Ref. \cite{Hagiwara:2002fs} and are shown with the horizontal bands. Included are results from various hybrid actions with staggered sea quarks from Refs. \cite{Na:2007pv,Na:2008hz} (purple triangles), \cite{Briceno:2012wt} (magenta diamonds) and \cite{Liu:2009jc} (orange inverted triangles). Results from PACS-CS~\cite{Namekawa:2013vu} are shown in green triangles.}
\label{Fig:charm_spectrum}
\end{figure}

 In \fig{Fig:charm_spectrum} we compare our continuum extrapolated results on the charmed spectrum with experiment again showing separately the statistical error and the total error. Given the agreement with the experimental values, lattice QCD thus provides predictions for the mass of the $\Xi_{cc}^*$, $\Omega_{cc}$, $\Omega^*_{cc}$ and $\Omega_{ccc}$. These predictions are consistent among lattice calculations, as shown in \fig{Fig:charm_spectrum}. We also point out that our value for $\Xi_{cc}$ is within errors with the value measured by the SELEX experiment.


\section{Conclusions}
The twisted mass formulation allowing simulations with  dynamical strange and charm quarks with their mass fixed to approximately their physical values provides a good framework for studying the baryon spectrum. A number of gauge ensembles are analyzed spanning pion masses from about 450~MeV to 210~MeV for three lattice spacings. For the strange and charm valence quarks we use the Osterwalder-Seiler formulation and tuned their mass using the mass of the $\Omega$ and $\Lambda_c$, respectively. Thus the strange and charm quarks are treated in the same manner as the light quarks. This is to be contrasted with other lattice calculations where $N_f=2+1$ staggered gauge configurations are used and the charm valence quark is introduced using a different discretization scheme such as clover or  described by a relativistic heavy quark action. A comparison  of our lattice results to other lattice calculations before extrapolations shows an overall similar tread for all lattice formulations.

Having values for the masses at three lattice spacings is crucial in order to both verify that cut-off effects are under control and to extrapolate the results to the continuum limit. We perform a continuum extrapolation to all our data and chiral extrapolate to the physical pion mass. In most cases, the largest systematic error arises because of the chiral extrapolation and the tuning of the strange and charm quark masses. We estimate the error due to the chiral extrapolation  by comparing results at different orders of the chiral expansion. The systematic error  due to tuning is estimated by varying the strange and charm quark mass within the error band of the $\Omega$ and $\Lambda_c$ masses at the physical point. 
From the chiral fits we can determine the light $\sigma$-terms for all baryons via the  Feynman-Hellmann theorem. The largest uncertainty in their determination arises from the chiral extrapolation which,
 in some cases amounts to  over 30\% error.  Therefore direct determinations of the $\sigma$-terms~\cite{Abdel-Rehim:2013wlz,Alexandrou:2014vya}
although very computer intensive can provide a valuable alternative. The values extracted for $\sigma_{\pi B}$  for all the baryons are given in Table~\ref{Table:fit_params_810}.

Our values for the baryon masses at the physical point, shown in Figs.~\ref{Fig:physpoints_810} and \ref{Fig:charm_spectrum}, reproduce the known baryon masses. For the $\Xi_{cc}$ we find a mass of 3.568(14)(19)(1)~GeV, which is higher by one standard deviation as compared 
with the value of  3.519~GeV  measured by the SELEX collaboration. Our prediction for the mass of the $\Xi_{cc}^*$ is 3.652(17)(27)(3)~GeV, for the  $\Omega^+_{cc}$ is 3.658(11)(16)(50)~GeV, for $\Omega^{*+}_{cc}$ 3.735(13)(18)(43)~GeV and for  $\Omega^{++}_{ccc}$ 4.734(12)(11)(9)~GeV.

\section*{Acknowledgments}

We would like to thank all members of the ETMC for the many valuable and constructive discussions and the very fruitful collaboration that took place during the development of this work.
 The project used computer time granted by the John von Neumann Institute for Computing (NIC) on JUQUEEN (project \texttt{hch02}) and JUROPA (project \texttt{ecy00}) at the J\"ulich Supercomputing Centre as well as by the Cyprus Institute
on the Cy-Tera machine (project \texttt{lspro113s1}), under the Cy-Tera project  (NEA $\Upsilon\Pi$O$\Delta$OMH/$\Sigma$TPATH/0308/31). We thank the staff members of these computing centers for their technical advice and support. C.K. is partly supported by the project GPUCW (T$\Pi$E/$\Pi\Lambda$HPO/0311(BIE)/09), which is co-financed by the European Regional Development Fund and the Republic of Cyprus through the Research Promotion Foundation.


\clearpage

\begin{appendices}
\section{APPENDIX: Interpolating fields for baryons}\label{App:int_fields}

In the following tables, we give the interpolating fields for the baryons used in this work in correspondence with \fig{Fig:spin12_baryons} and \fig{Fig:spin32_baryons}. Throughout, $C$ denotes the charge conjugation matrix and  spinor indices  are suppressed.
\begin{table}[!ht]
\begin{center}
\renewcommand{\arraystretch}{1.5}
\renewcommand{\tabcolsep}{5.5pt}
\small
\makebox[\textwidth]{%
\begin{tabular}{c|c|c c c c c}
\hline
\hline
\multirow{2}{*}{Charm} & \multirow{2}{*}{Strange} &  \multirow{2}{*}{Baryon} & Quark & \multirow{2}{*}{Interpolating field}  & \multirow{2}{*}{$I$} & 	\multirow{2}{*}{$I_z$}	 \\
	                     &                         &                          & content  &                                     &                      &           \\
\hline
\renewcommand{\arraystretch}{1.8}
\renewcommand{\tabcolsep}{5.8pt} 
\multirow{3}{*}{$c=2$}&	  \multirow{2}{*}{$s=0$}		&$\Xi_{cc}^{++}$        & ucc &   $    \cone{c}{u}{c} $ & 1/2 & +1/2 \\
		    		&	                     &$\Xi_{cc}^{+}$        &  dcc & $    \cone{c}{d}{c} $ & 1/2 & -1/2 \\ 
\cline{2-7}
		         &$s=1$			&$\Omega_{cc}^{+}$        & scc &   $    \cone{c}{s}{c} $  & 0 & 0\\ 
\hline\hline
\multirow{9}{*}{$c=1$}&	\multirow{4}{*}{$s=0$} & $\Lambda_c^+$  &	udc	 & $   \reci{\sqrt{6}} \eps \left[ 2\con{u}{d}{c} + \con{u}{c}{d} - \con{d}{c}{u} \right] $   & 0 & 0   \\
\cline{3-7}
  &	 	&$\Sigma_c^{++}$  &  uuc & $    \cone{u}{c}{u} $ & 1 & +1 \\
		&	  	&$\Sigma_c^{+}$   & udc & $   \reci{\sqrt{2}} \eps \left[ \con{u}{c}{d} + \con{d}{c}{u} \right] $ & 1 & 0 \\
		    	&	     	&$\Sigma_c^{0}$  &   ddc & $    \cone{d}{c}{d} $ & 1 & -1 \\						
\cline{2-7}
& \multirow{4}{*}{$s=1$}	 	&$\Xi_c^+$      & usc &	$   \cone{u}{s}{c} $ & 1/2 & +1/2 \\
 &	  &$\Xi_c^0$   & dsc & $   \cone{d}{s}{c} $ & 1/2 & -1/2 \\
\cline{3-7}		 
	&	 &$\Xi_c^{\prime +}$  &  usc & $   \reci{\sqrt{2}} \eps \left[ \con{u}{c}{s} + \con{s}{c}{u} \right] $ & 1/2 & +1/2 \\				
					 &     		&$\Xi_c^{\prime 0}$  &   dsc &$   \reci{\sqrt{2}} \eps \left[ \con{d}{c}{s} + \con{s}{c}{d} \right] $ & 1/2 & -1/2 \\
\cline{2-7}
		  & $s=2$	&$\Omega_c^{0}$        & ssc &   $    \cone{s}{c}{s} $ & 0 & 0\\
\hline\hline
\multirow{8}{*}{$c=0$}&	\multirow{2}{*}{$s=0$}	   & \emph{p}     &	uud	      &    $\cone{u}{d}{u} $      & 1/2 & +1/2     \\
	&	 	& \emph{n}     & udd	   & $   \cone{d}{u}{d} $      & 1/2 & -1/2      \\
\cline{2-7}	 
& \multirow{4}{*}{$s=1$}		&$\Lambda$     & uds     &	$   \reci{\sqrt{6}} \eps \left[ 2\con{u}{d}{s} + \con{u}{s}{d} - \con{d}{s}{u} \right] $ & 0 & 0 \\
\cline{3-7}
	   	&	&$\Sigma^{+}$  &  uus    & $    \cone{u}{s}{u} $     & 1 & +1\\
      & 	&$\Sigma^{0}$  & uds     & $   \reci{\sqrt{2}} \eps \left[ \con{u}{s}{d} + \con{d}{s}{u} \right] $& 1 & 0 \\
	   &   &$\Sigma^{-}$  &   dds   & $    \cone{d}{s}{d} $     & 1 & -1\\
\cline{2-7}
& \multirow{2}{*}{$s=2$}    &$\Xi^{0}$      & uss    &   $    \cone{s}{u}{s} $    & 1/2 & +1/2 \\
		&   &$\Xi^{-}$     &  dss   & $    \cone{s}{d}{s} $       & 1/2 & -1/2\\
\hline\hline
\end{tabular}
}
\end{center}
\caption{Interpolating fields and quantum numbers for the $20^\prime$-plet of spin-1/2 baryons.}
\label{Table:spin12_intfields}
\end{table}
%
\clearpage
\begin{table}[!ht]
\begin{center}
\renewcommand{\arraystretch}{1.8}
\renewcommand{\tabcolsep}{5.8pt}
\makebox[\textwidth]{%
\begin{tabular}{c|c|c c c c c}
\hline
\hline
\multirow{2}{*}{Charm} & \multirow{2}{*}{Strange} &  \multirow{2}{*}{Baryon} & Quark & \multirow{2}{*}{Interpolating field}  & \multirow{2}{*}{$I$} & 	\multirow{2}{*}{$I_z$}	 \\
	                     &                         &                          & content  &                                     &                      &           \\
\hline
\renewcommand{\arraystretch}{1.6}
\renewcommand{\tabcolsep}{5.5pt}
	$c=3$ &	 $s=0$	&$\Omega_{ccc}^{++}$        & ccc &   $     \conme{c}{c}{c} $ & 0 & 0 \\
\hline\hline
 \multirow{3}{*}{$c=2$}    & \multirow{2}{*}{$s=0$}	&$\Xi_{cc}^{\star ++}$        & ucc &   $     \conme{c}{u}{c} $ & 1/2 & +1/2\\
		    		&           	&$\Xi_{cc}^{\star +}$        &  dcc & $     \conme{c}{d}{c} $ & 1/2 & -1/2\\ 
\cline{2-7}
		  		&$s=1$ 	&$\Omega_{cc}^{\star +}$        & scc &   $     \conme{c}{s}{c} $ & 0 & 0\\ 
\hline\hline
\multirow{6}{*}{$c=1$}&	  \multirow{3}{*}{$s=0$}	&$\Sigma_c^{\star ++}$  & uuc &   $   \reci{\sqrt{3}} \eps \left[ \conm{u}{u}{c} + 2\conm{c}{u}{u} \right] $ & 1 & +1 \\
			 &			    &$\Sigma_c^{\star +}$   		   & udc   & $   \sqrt{\frac{2}{3}} \eps \left[ \conm{u}{d}{c} + \conm{d}{c}{u}+ \conm{c}{u}{d} \right] $ & 1 & 0 \\
		    	&		        &$\Sigma_c^{\star 0}$  &   ddc        &   $     \reci{\sqrt{3}} \eps \left[ \conm{d}{d}{c} + 2\conm{c}{d}{d} \right] $ & 1 & -1 \\					
\cline{2-7}								 
       &	  \multirow{2}{*}{$s=1$}	&$\Xi_c^{\star +}$      & usc &	$    \conme{s}{u}{c} $  & 1/2 & +1/2\\
		 &		      	&$\Xi_c^{\star 0}$   & dsc & $    \conme{s}{d}{c} $  & 1/2 & -1/2\\
\cline{2-7}
		  &$s=2$	&$\Omega_c^{\star 0}$        & ssc &   $     \conme{s}{c}{s} $ & 0 & 0  \\
\hline\hline
\multirow{10}{*}{$c=0$}&   \multirow{4}{*}{$s=0$}		& $\Delta^{++}$  & uuu & $    \conme{u}{u}{u} $   & 3/2 & +3/2         \\
	&	& $\Delta^{+}  $  &  uud & $   \reci{\sqrt{3}} \eps \left[ 2\conm{u}{d}{u} + \conm{u}{u}{d} \right] $& 3/2 & +1/2         \\
	&	& $\Delta^{0}  $  &  udd & $   \reci{\sqrt{3}} \eps \left[ 2\conm{d}{u}{d} + \conm{d}{d}{u} \right] $ & 3/2 & -1/2        \\						
 	&	& $\Delta^{-}$  & 	ddd & $    \conme{d}{d}{d} $   & 3/2 & -3/2         \\						
\cline{2-7}									 
   & \multirow{3}{*}{$s=1$} 		 &$\Sigma^{\star +}$  & uus &   $   \reci{\sqrt{3}} \eps \left[ \conm{u}{u}{s} + 2\conm{s}{u}{u} \right] $ & 1 & +1 \\
    &	&$\Sigma^{\star 0}$   		   & uds   & $   \sqrt{\frac{2}{3}} \eps \left[ \conm{u}{d}{s} + \conm{d}{s}{u}+ \conm{s}{u}{d} \right] $& 1 & 0 \\
	 & &$\Sigma^{\star -}$  &   dds        &   $     \reci{\sqrt{3}} \eps \left[ \conm{d}{d}{s} + 2\conm{s}{d}{d} \right] $ & 1 & -1\\
\cline{2-7}
& \multirow{2}{*}{$s=2$}		&$\Xi^{\star 0}$        & uss &   $     \conme{s}{u}{s} $ & 1/2 & +1/2\\
 		&	&$\Xi^{\star -}$        &  dss & $     \conme{s}{d}{s} $ & 1/2 & -1/2\\
\cline{2-7}
& $s=3$		& $\Omega^-$  & sss &$   \conme{s}{s}{s}$ &0&0 \\
\hline\hline
\end{tabular}
}
\end{center}
\caption{Interpolating fields and quantum numbers for the 20-plet of spin-3/2 baryons.}
\label{Table:spin32_intfields}
\end{table}

\begin{table}[hp]
\begin{center}
\renewcommand{\arraystretch}{1.6}
\renewcommand{\tabcolsep}{5.8pt}
\makebox[\textwidth]{%
\begin{tabular}{c|c|ccccc}
\hline
\hline
\multirow{2}{*}{Charm} & \multirow{2}{*}{Strange} &  \multirow{2}{*}{Baryon} & Quark & \multirow{2}{*}{Interpolating field}  & \multirow{2}{*}{$I$} & 	\multirow{2}{*}{$I_z$}	 \\
	                     &                         &                          & content  &                                     &                      &           \\
\hline
\renewcommand{\arraystretch}{1.6}
\renewcommand{\tabcolsep}{5.5pt}
&&\multicolumn{5}{c}{Spin-1/2 baryons} \\	 
\hline

\multirow{2}{*}{$c=1$}& \multirow{2}{*}{$s=1$}	&$\Xi_c^+$ & usc &	$   \reci{\sqrt{6}} \eps \left[ 2\con{s}{u}{c} + \con{s}{c}{u} - \con{u}{c}{s} \right] $ & 1/2 & +1/2 \\
		    &           	&$\Xi_c^0$   & dsc & $   \reci{\sqrt{6}} \eps \left[ 2\con{s}{d}{c} + \con{s}{c}{d} - \con{d}{c}{s} \right]$ & 1/2 & -1/2 \\
\hline\hline
&&\multicolumn{5}{c}{Spin-3/2 baryons} \\
\hline
\multirow{2}{*}{$c=0$}& \multirow{2}{*}{$s=2$}		&$\Xi^{\star 0}$        & uss &   $   \reci{\sqrt{3}} \eps \left[ 2\conm{s}{u}{s} + \conm{s}{s}{u} \right] $ & 1/2 & +1/2\\
		            &       			&$\Xi^{\star -}$        &  dss &  $   \reci{\sqrt{3}} \eps \left[ 2\conm{s}{d}{s} + \conm{s}{s}{d} \right] $ & 1/2 & -1/2\\
\hline								 
\multirow{3}{*}{$c=1$}& \multirow{2}{*}{$s=1$}	&$\Xi_c^{\star +}$      & usc &	$   \sqrt{\frac{2}{3}} \eps \left[ \conm{u}{s}{c} + \conm{s}{c}{u}+ \conm{c}{u}{s} \right] $  & 1/2 & +1/2\\
		          &		      	&$\Xi_c^{\star 0}$      & dsc  &    $   \sqrt{\frac{2}{3}} \eps \left[ \conm{d}{s}{c} + \conm{s}{c}{d}+ \conm{c}{d}{s} \right] $  & 1/2 & -1/2\\
\cline{2-7}
		         &$s=2$	&$\Omega_c^{\star 0}$        & ssc &    $   \reci{\sqrt{3}} \eps \left[ 2\conm{s}{c}{s} + \conm{s}{s}{c} \right] $  & 0 & 0  \\
\hline              
\multirow{3}{*}{$c=2$}& \multirow{2}{*}{$s=0$}	&$\Xi_{cc}^{\star ++}$        & ucc &   $   \reci{\sqrt{3}} \eps \left[ 2\conm{c}{u}{c} + \conm{c}{c}{u} \right] $ & 1/2 & +1/2\\
		    		&           	&$\Xi_{cc}^{\star +}$        &  dcc & $   \reci{\sqrt{3}} \eps \left[ 2\conm{c}{d}{c} + \conm{c}{c}{d} \right] $ & 1/2 & -1/2\\ 
\cline{2-7}
		  		&$s=1$ 	&$\Omega_{cc}^{\star +}$        & scc &   $   \reci{\sqrt{3}} \eps \left[ 2\conm{c}{s}{c} + \conm{c}{c}{s} \right] $ & 0 & 0\\ 	
\hline\hline
\end{tabular}
}
\end{center}
\caption{Additional interpolating fields for spin-1/2 and spin-3/2 baryons. There are two of the spin-1/2 baryons and eight of the spin-3/2 baryons.}
\label{Table:alt_intfields}
\end{table}

\newpage
\section{APPENDIX: Lattice results}\label{App:numerical_results}

In the tables below we list the baryon masses in lattice units and the continuum extrapolated values in physical units. The masses in physical units are in GeV and are converted from lattice units using the lattice spacing values extracted from the nucleon in this work, \eq{eq:lat_spacings}. The masses for the nucleon, $\Omega$ and $\Lambda_c^+$ are listed in Tables \ref{Table:nucleon_masses} and \ref{Table:omega_lambdac_masses}.

\begin{table}[h]
\begin{center}
\renewcommand{\arraystretch}{1.25}
\renewcommand{\tabcolsep}{5pt}
\begin{tabular}{l|llllll}
\hline\hline
$\; a\mu_l$ & $\quad\; am_\Lambda$  & $\quad\; am_\Sigma$ & $\quad\; am_\Xi$ & $\quad\; am_\Delta$ & $\quad\; am_{\Sigma^*}$  & $\quad\; am_{\Xi^*}$  \\
\hline\hline
\multicolumn{7}{c}{$\beta=1.90$} \\
\hline
0.0030 & 0.5972(46) &  0.6420(60) &   0.6906(50) &  0.7090(100)  &   0.7481(95) &   0.8046(61)     \\
0.0040 & 0.5978(46) &  0.6335(52) &   0.6888(38) &  0.6924(145)  &   0.7339(89) &   0.7918(73)     \\
0.0050 & 0.6051(60) &  0.6552(52) &   0.6949(69) &  0.7097(101)  &   0.7600(91) &   0.8044(144)    \\
\hline\hline
\multicolumn{7}{c}{$\beta=1.95$}\\
\hline
0.0025 & 0.5217(59) &  0.5586(66) &   0.6077(38) &  0.6340(100)  &   0.6677(89) &   0.7093(87)     \\
0.0035 & 0.5341(50) &  0.5633(50) &   0.6090(48) &  0.6329(102)  &   0.6614(92) &   0.6987(84)     \\
0.0055 & 0.5529(43) &  0.5800(60) &   0.6126(50) &  0.6525(88)   &   0.6841(77) &   0.7189(68)     \\
0.0075 & 0.5640(52) &  0.5937(39) &   0.6125(72) &  0.6691(74)   &   0.6862(80) &   0.7199(62)     \\
\hline\hline
\multicolumn{7}{c}{$\beta=2.10$}\\
\hline
0.0015 & 0.3904(37) &  0.4167(37) &   0.4537(28) &  0.4614(71)   &   0.5000(48) &   0.5359(39)     \\
0.0020 & 0.4021(43) &  0.4250(49) &   0.4540(35) &  0.4749(98)   &   0.5052(63) &   0.5308(53)     \\
0.0030 & 0.4041(40) &  0.4253(46) &   0.4543(32) &  0.4749(81)   &   0.5024(71) &   0.5330(58)     \\
\hline\hline
\end{tabular}
\caption{Octet and decuplet baryon masses in lattice units with the associated statistical error.}
\label{Table:octet_decuplet_lresults}
\end{center}
\vspace*{-.0cm}
\end{table}


\begin{table}[h]
\begin{center}
\renewcommand{\arraystretch}{1.25}
\renewcommand{\tabcolsep}{5pt}
\begin{tabular}{l|llllll}
\hline\hline
$\; a\mu_l$ & $\quad\; m_\Lambda$  & $\quad\; m_\Sigma$ & $\quad\; m_\Xi$ & $\quad\; m_\Delta$ & $\quad\; m_{\Sigma^*}$  & $\quad\; m_{\Xi^*}$  \\
\hline\hline
\multicolumn{7}{c}{$\beta=1.90$} \\
\hline
0.0030 & 1.2329(394)   & 1.3103(435) & 1.3331(356)  & 1.4909(834) & 1.5669(678)  & 1.6139(539)   \\
0.0040 & 1.2343(394)   & 1.2924(431) & 1.3294(349)  & 1.4560(863) & 1.5372(674)  & 1.5869(545)   \\
0.0050 & 1.2496(402)   & 1.3381(431) & 1.3422(369)  & 1.4923(835) & 1.5920(675)  & 1.6133(604)   \\
\hline\hline
\multicolumn{7}{c}{$\beta=1.95$}\\
\hline
0.0025 & 1.2314(364)   & 1.3067(399) & 1.3632(312)  & 1.5178(749) & 1.5938(608)  & 1.6379(504)   \\
0.0035 & 1.2610(356)   & 1.3180(385) & 1.3662(320)  & 1.5152(750) & 1.5787(610)  & 1.6126(502)   \\
0.0055 & 1.3063(351)   & 1.3580(393) & 1.3748(322)  & 1.5621(740) & 1.6332(598)  & 1.6609(487)   \\
0.0075 & 1.3328(358)   & 1.3909(378) & 1.3746(345)  & 1.6019(731) & 1.6382(600)  & 1.6633(483)   \\
\hline\hline
\multicolumn{7}{c}{$\beta=2.10$}\\
\hline
0.0015 & 1.1798(287)   & 1.2522(308) & 1.3272(250)  & 1.4074(598) & 1.5222(470)  & 1.5973(380)   \\
0.0020 & 1.2157(294)   & 1.2775(324) & 1.3282(258)  & 1.4484(632) & 1.5380(486)  & 1.5819(395)   \\
0.0030 & 1.2216(291)   & 1.2783(320) & 1.3290(253)  & 1.4484(609) & 1.5294(497)  & 1.5885(402)   \\
\hline\hline
\end{tabular}
\caption{Octet and decuplet baryon masses in physical units with the associated statistical error.}
\label{Table:octet_decuplet_presults}
\end{center}
\vspace*{-.0cm}
\end{table}


\begin{table}[h]
\begin{center}
\renewcommand{\arraystretch}{1.25}
\renewcommand{\tabcolsep}{5pt}
\begin{tabular}{l|llllll}
\hline\hline
$\; a\mu_l$ & $\quad\; am_{\Sigma_c}$  & $\quad\; am_{\Xi_c}$ & $\quad\; am_{\Xi_c^\prime}$ & $\quad\; am_{\Omega_c^0}$ & $\quad\; am_{\Xi_{cc}}$  & $\quad\; am_{\Omega_{cc}^+}$  \\
\hline\hline
\multicolumn{7}{c}{$\beta=1.90$} \\
\hline
0.0030 &   1.2543(72) & 1.2611(46) & 1.3028(53) & 1.3575(46) &  1.8187(48) &  1.8704(38)   \\
0.0040 &   1.2448(53) & 1.2580(62) & 1.2983(50) & 1.3506(37) &  1.8166(42) &  1.8694(33)    \\
0.0050 &   1.2696(55) & 1.2599(61) & 1.3185(49) & 1.3655(47) &  1.8303(44) &  1.8781(37)    \\
\hline\hline
\multicolumn{7}{c}{$\beta=1.95$}\\
\hline
0.0025 &   1.0896(55) & 1.0900(43) & 1.1388(42) & 1.1764(41) &  1.5684(34) &  1.6099(29)    \\
0.0035 &   1.0927(49) & 1.0920(41) & 1.1322(43) & 1.1726(39) &  1.5684(32) &  1.6077(27)    \\
0.0055 &   1.1091(51) & 1.1027(37) & 1.1440(44) & 1.1788(39) &  1.5782(36) &  1.6138(33)    \\
0.0075 &   1.1112(43) & 1.1024(36) & 1.1412(37) & 1.1691(37) &  1.5739(34) &  1.6065(34)    \\
\hline\hline
\multicolumn{7}{c}{$\beta=2.10$}\\
\hline
0.0015 &   0.8348(35) & 0.8362(25) & 0.8682(27) & 0.9010(23) &  1.2136(25) &  1.2449(19)    \\
0.0020 &   0.8384(64) & 0.8419(33) & 0.8735(35) & 0.9000(30) &  1.2078(31) &  1.2414(21)    \\
0.0030 &   0.8376(49) & 0.8410(26) & 0.8741(33) & 0.9028(28) &  1.2139(25) &  1.2438(19)    \\
\hline\hline
\end{tabular}
\caption{Charm spin-1/2 baryon masses in lattice units with the associated statistical error.}
\label{Table:charm12_lresults}
\end{center}
\vspace*{-.0cm}
\end{table}


\begin{table}[h]
\begin{center}
\renewcommand{\arraystretch}{1.25}
\renewcommand{\tabcolsep}{5pt}
\begin{tabular}{l|llllll}
\hline\hline
$\; a\mu_l$ & $\quad\; m_{\Sigma_c}$  & $\quad\; m_{\Xi_c}$ & $\quad\; m_{\Xi_c^\prime}$ & $\quad\; m_{\Omega_c^0}$ & $\quad\; m_{\Xi_{cc}}$  & $\quad\; m_{\Omega_{cc}^+}$  \\
\hline\hline
\multicolumn{7}{c}{$\beta=1.90$} \\
\hline
0.0030 & 2.5020(560) & 2.4921(374) &  2.5890(412) & 2.6663(350) & 3.5829(344) & 3.6631(268)   \\
0.0040 & 2.4820(551) & 2.4856(384) &  2.5796(410) & 2.6518(345) & 3.5784(340) & 3.6611(265)   \\
0.0050 & 2.5342(552) & 2.4896(384) &  2.6221(410) & 2.6831(350) & 3.6072(341) & 3.6794(267)   \\
\hline\hline
\multicolumn{7}{c}{$\beta=1.95$}\\
\hline
0.0025 & 2.5042(492) & 2.4865(334) &  2.6102(363) & 2.6713(311) & 3.5687(300) & 3.6461(235)   \\
0.0035 & 2.5114(489) & 2.4912(332) &  2.5946(364) & 2.6623(310) & 3.5687(299) & 3.6408(234)   \\
0.0055 & 2.5509(490) & 2.5168(330) &  2.6228(364) & 2.6771(310) & 3.5921(301) & 3.6554(238)   \\
0.0075 & 2.5558(485) & 2.5161(329) &  2.6160(360) & 2.6538(309) & 3.5818(300) & 3.6378(239)   \\
\hline\hline
\multicolumn{7}{c}{$\beta=2.10$}\\
\hline
0.0015 & 2.4816(387) & 2.4746(261) &  2.5766(286) & 2.6585(242) & 3.5867(239) & 3.6686(186)   \\
0.0020 & 2.4927(421) & 2.4921(269) &  2.5927(294) & 2.6557(249) & 3.5690(245) & 3.6581(188)   \\
0.0030 & 2.4902(401) & 2.4891(262) &  2.5944(292) & 2.6643(247) & 3.5877(239) & 3.6652(186)   \\
\hline\hline
\end{tabular}
\caption{Charm spin-1/2 baryon masses in physical units with the associated statistical error.}
\label{Table:charm12_presults}
\end{center}
\vspace*{-.0cm}
\end{table}


\begin{table}[h]
\begin{center}
\renewcommand{\arraystretch}{1.25}
\renewcommand{\tabcolsep}{5pt}
\begin{tabular}{l|llllll}
\hline\hline
$\; a\mu_l$ & $\quad\; am_{\Sigma_c^*}$  & $\quad\; am_{\Xi_c^*}$  & $\quad\; am_{\Omega_c^{*0}}$ & $\quad\; am_{\Xi_{cc}^*}$  & $\quad\; am_{\Omega_{cc}^{*+}}$ & $\quad\; am_{\Omega_{ccc}^{++}}$ \\
\hline\hline
\multicolumn{7}{c}{$\beta=1.90$} \\
\hline
0.0030  & 1.2828(103) & 1.3333(78)  &  1.3780(58)  &  1.8464(71)    &    1.8941(47)  &      2.3788(37)    \\
0.0040  & 1.2812(76)  & 1.3337(57)  &  1.3846(48)  &  1.8407(100)   &    1.9034(38)  &      2.3845(48)    \\
0.0050  & 1.3057(65)  & 1.3543(57)  &  1.3953(51)  &  1.8665(52)    &    1.9092(41)  &      2.3857(42)    \\
\hline\hline
\multicolumn{7}{c}{$\beta=1.95$}\\
\hline
0.0025  & 1.1296(90)  & 1.1757(52)  &  1.2049(46)  &  1.6084(54)    &    1.6400(41)  &      2.0486(29)    \\
0.0035  & 1.1295(53)  & 1.1588(63)  &  1.1999(46)  &  1.6037(45)    &    1.6394(35)  &      2.0537(27)    \\
0.0055  & 1.1435(63)  & 1.1767(54)  &  1.2028(51)  &  1.6153(42)    &    1.6451(36)  &      2.0578(29)    \\
0.0075  & 1.1471(54)  & 1.1608(64)  &  1.2016(43)  &  1.6107(39)    &    1.6386(38)  &      2.0570(28)    \\
\hline\hline
\multicolumn{7}{c}{$\beta=2.10$}\\
\hline
0.0015  & 0.8591(41)  & 0.8951(32)  &  0.9239(28)  &  1.2380(26)    &    1.2669(21)  &      1.5958(20)    \\
0.0020  & 0.8612(73)  & 0.8928(53)  &  0.9277(30)  &  1.2377(40)    &    1.2702(26)  &      1.5928(20)    \\
0.0030  & 0.8596(55)  & 0.8909(44)  &  0.9296(29)  &  1.2384(33)    &    1.2665(26)  &      1.5946(16)    \\
\hline\hline
\end{tabular}
\caption{Charm spin-3/2 baryon masses in lattice units with the associated statistical error.}
\label{Table:charm32_lresults}
\end{center}
\vspace*{-.0cm}
\end{table}


\begin{table}[h]
\begin{center}
\renewcommand{\arraystretch}{1.25}
\renewcommand{\tabcolsep}{5pt}
\begin{tabular}{l|llllll}
\hline\hline
$\; a\mu_l$ & $\quad\; m_{\Sigma_c^*}$  & $\quad\; m_{\Xi_c^*}$  & $\quad\; m_{\Omega_c^{*0}}$ & $\quad\; m_{\Xi_{cc}^*}$  & $\quad\; m_{\Omega_{cc}^{*+}}$ & $\quad\; m_{\Omega_{ccc}^{++}}$ \\
\hline\hline
\multicolumn{7}{c}{$\beta=1.90$} \\
\hline
0.0030 & 2.5529(709) & 2.6263(552)  & 2.7461(402)  & 3.6555(497) & 3.7362(335) &  4.7432(263)   \\
0.0040 & 2.5496(694) & 2.6271(541)  & 2.7599(396)  & 3.6435(518) & 3.7556(330) &  4.7552(270)   \\
0.0050 & 2.6012(689) & 2.6704(541)  & 2.7824(397)  & 3.6978(486) & 3.7680(332) &  4.7576(266)   \\
\hline\hline
\multicolumn{7}{c}{$\beta=1.95$}\\
\hline
0.0025 & 2.5928(631) & 2.6778(479)  & 2.7677(354)  & 3.6756(436) & 3.7361(298) &  4.7049(231)   \\
0.0035 & 2.5927(607) & 2.6373(487)  & 2.7557(353)  & 3.6642(430) & 3.7347(293) &  4.7171(229)   \\
0.0055 & 2.6261(612) & 2.6803(481)  & 2.7626(357)  & 3.6921(428) & 3.7481(294) &  4.7268(231)   \\
0.0075 & 2.6349(607) & 2.6422(488)  & 2.7600(351)  & 3.6810(427) & 3.7327(295) &  4.7250(230)   \\
\hline\hline
\multicolumn{7}{c}{$\beta=2.10$}\\
\hline
0.0015 & 2.5515(482) & 2.6459(377)  & 2.7459(277)  & 3.6679(336) & 3.7469(230) &  4.7443(183)   \\
0.0020 & 2.5581(517) & 2.6388(398)  & 2.7576(279)  & 3.6669(349) & 3.7568(234) &  4.7350(184)   \\
0.0030 & 2.5530(495) & 2.6329(387)  & 2.7632(278)  & 3.6694(342) & 3.7457(234) &  4.7406(180)   \\
\hline\hline
\end{tabular}
\caption{Charm spin-3/2 baryon masses in physical units with the associated statistical error.}
\label{Table:charm32_presults}
\end{center}
\vspace*{-.0cm}
\end{table}

\clearpage
\section{APPENDIX: HB$\chi$PT next-to-leading order expressions for the octet and decuplet baryons}\label{App:chiral_expressions_nlo}

For the octet baryons the NLO expressions read
\bea \label{eq:octet_expressions_nlo}
	m_\Lambda^{\rm NLO} (m_\pi) &=& m_\Lambda^{(0)} - 4c_\Lambda^{(1)}m_\pi^2 - \frac{g_{\Lambda\Sigma}^2}{(4\pi f_\pi)^2} \mathcal{F}(m_\pi,\Delta_{\Lambda\Sigma},\lambda) - \frac{4g_{\Lambda\Sigma^*}^2}{(4\pi f_\pi)^2} \mathcal{F}(m_\pi,\Delta_{\Lambda\Sigma^*},\lambda) \nonumber\\
	m_\Sigma^{\rm NLO} (m_\pi)    &=& m_\Sigma^{(0)} - 4c_\Sigma^{(1)}m_\pi^2 - \frac{2g_{\Sigma\Sigma}^2}{16\pi f_\pi^2} m_\pi^3  - \frac{g_{\Lambda\Sigma}^2}{3(4\pi f_\pi)^2} \mathcal{F}(m_\pi,-\Delta_{\Lambda\Sigma},\lambda) - \frac{4g_{\Sigma^*\Sigma}^2}{3(4\pi f_\pi)^2} \mathcal{F}(m_\pi,\Delta_{\Sigma\Sigma^*},\lambda) \nonumber\\
	m_\Xi^{\rm NLO} (m_\pi)  &=& m_\Xi^{(0)} - 4c_\Xi^{(1)}m_\pi^2 - \frac{3g_{\Xi\Xi}^2}{16\pi f_\pi^2} m_\pi^3  - \frac{2g_{\Xi^*\Xi}^2}{(4\pi f_\pi)^2}\mathcal{F}(m_\pi,\Delta_{\Xi\Xi^*},\lambda)
\eea
and for the decuplet baryons
\bea \label{eq:decuplet_expressions_nlo}
	m_\Delta^{\rm NLO} (m_\pi) &=& m_\Delta^{(0)} - 4c_\Delta^{(1)}m_\pi^2 - \frac{25}{27}\frac{g_{\Delta\Delta}^2}{16\pi f_\pi^2} m_\pi^3 - \frac{2g_{\Delta N}^2}{3(4\pi f_\pi)^2} \mathcal{F}(m_\pi,-\Delta_{N\Delta},\lambda)  \nonumber\\
	m_{\Sigma^*}^{\rm NLO} (m_\pi) &=& m_{\Sigma^*}^{(0)} - 4c_{\Sigma^*}^{(1)}m_\pi^2 -\frac{10}{9}  \frac{g_{\Sigma^*\Sigma^*}^2}{16\pi f_\pi^2} m_\pi^3 - \frac{2}{3(4\pi f_\pi)^2} \left[g_{\Sigma^*\Sigma}^2\mathcal{F}(m_\pi,-\Delta_{\Sigma\Sigma^*},\lambda) + g_{\Lambda\Sigma^*}^2 \mathcal{F}(m_\pi,-\Delta_{\Lambda\Sigma^*},\lambda) \right]  \nonumber\\
	m_{\Xi^*}^{\rm NLO} (m_\pi) &=& m_{\Xi^*}^{(0)} - 4c_{\Xi^*}^{(1)}m_\pi^2 - \frac{5}{3} \frac{g_{\Xi^*\Xi^*}^2}{16\pi f_\pi^2} m_\pi^3 -\frac{g_{\Xi^*\Xi}^2}{(4\pi f_\pi)^2} \mathcal{F}(m_\pi,-\Delta_{\Xi\Xi^*},\lambda) \nonumber\\
	m_{\Omega}^{\rm NLO} (m_\pi)       &=& m_{\Omega}^{(0)}        - 4c_{\Omega}^{(1)}m_\pi^2\;.
\eea
The non-analytic function $\mathcal{F}(m,\Delta,\lambda)$ is of the form \cite{Tiburzi:2005na}
\be \label{eq:F_non_analytic}
\mathcal{F}(m,\Delta,\lambda) = (m^2-\Delta^2)\sqrt{\Delta^2-m^2+i\epsilon}\; \log\left(\frac{\Delta-\sqrt{\Delta^2-m^2+i\epsilon}}{\Delta +\sqrt{\Delta^2-m^2+i\epsilon}} \right) - \frac{3}{2} \Delta\; m^2 \log\left( \frac{m^2}{\lambda^2}\right) - \Delta^3 \log \left(\frac{4\Delta^2}{m^2} \right)
\ee
depending on the threshold parameter $\Delta_{XY} = m_Y^{(0)}-m_X^{(0)}$ and on the scale $\lambda$ of chiral perturbation theory, fixed to $\lambda = 1$ GeV. For $\Delta >0$ the real part of the function $\mathcal{F}(m,\Delta,\lambda)$ has the property
\be\label{eq:F_property}
\mathcal{F}(m,-\Delta,\lambda) = \left\lbrace 
\begin{array}{ll}
-\mathcal{F}(m,\Delta,\lambda) & m< \Delta \\
-\mathcal{F}(m,\Delta,\lambda) + 2\pi (m^2-\Delta^2)^{3/2} & m>\Delta 
\end{array} \right.
\ee
which corrects a typo in the sign of the second term in Ref. \cite{WalkerLoud:2008bp}. 

A noticeable result of this expansion is the absence of a cubic term in the expressions for the $\Lambda$ and $\Omega$ baryons given in Eqns. (\ref{eq:octet_expressions_nlo}) and (\ref{eq:decuplet_expressions_nlo}). In the case of $\Omega$ it follows from the absence of light valence quarks. However, the absence of a cubic term in the NLO expression for $\Lambda$, although a consequence of $\chi$PT, is nevertheless a questionable result, since it relies on the assumption that $m_\pi \ll M_\Sigma - M_\Lambda$. In the limit $\Delta \rightarrow 0$ the non-analytic function of \eq{eq:F_non_analytic} becomes
\be \label{eq:Flimit}
\mathcal{F}(m_\pi,\Delta\rightarrow 0,\lambda) = \pi m_\pi^3\;,
\ee
which generates a cubic term for the $\Lambda$ and slightly modifies the existing one for $\Sigma$. The corresponding expressions are given by
\begin{eqnarray} \label{eq:LambaSigma_limit}
m_\Lambda (m_\pi) &=& m_\Lambda^{(0)} - 4c_\Lambda^{(1)}m_\pi^2 - \frac{g_{\Lambda\Sigma}^2}{16\pi f_\pi^2} m_\pi^3 \nonumber\\
m_\Sigma (m_\pi) &=& m_\Sigma^{(0)}   - 4c_\Sigma^{(1)}m_\pi^2    - \frac{2g_{\Sigma\Sigma}^2+g_{\Lambda\Sigma}^2/3}{16\pi f_\pi^2} m_\pi^3
\end{eqnarray}

\end{appendices}

\clearpage
\bibliographystyle{./apsrev}               
\bibliography{references}

\end{document}